\documentclass[11pt]{article}
\pdfoutput=1

\usepackage{jheppub} 

\usepackage{amsmath,amssymb,epsfig,amsfonts}
\usepackage{graphicx,subfigure}
\usepackage{epsf}
\usepackage{makeidx}
\usepackage[all]{xy}
\usepackage{slashed}
\usepackage{verbatim} 
\usepackage{float}
\restylefloat{table}
\usepackage[all]{xy}
\usepackage{pdflscape}
\usepackage{tikz}
\usepackage{array}
\usepackage{youngtab}
\usepackage{multirow}
\usepackage{color}
\usepackage{ulem}
\usepackage{cleveref}

\usepackage{mathtools}

\usepackage{tensor}

\makeatletter

\newcommand{\AdS}{\hbox{AdS}}
\newcommand{\Y}{\mathcal{Y}}

\newcommand{\Xtwo}{\mathcal{S}^\tau_{4}}

\newcommand{\be}{\begin{equation}}
\newcommand{\ee}{\end{equation}}
\newcommand{\ba}{\begin{aligned}}
\newcommand{\ea}{\end{aligned}}

\newcommand{\A}{A}
\newcommand{\B}{B}
\newcommand{\imab}{\Im[\A^{*}\B]}
\newcommand{\bbcw}{z}

\newcommand{\tM}{\widetilde{\mathcal{M}}}
\newcommand{\Kthreebase}{B_2}
\newcommand{\alphaalpha}{\alpha}

\newcommand{\nn}{\nonumber}

\newcommand{\Yodd}{\mathfrak{Y}}
\newcommand{\podd}{\mathfrak{p}}
\newcommand{\qodd}{\mathfrak{q}}
\newcommand\F{\mathbb{F}}

\newcommand\D{\mathcal{D}}

\newcommand{\newbase}{\mathcal{Z}_4}
\newcommand\T{\mathbb{T}}
\newcommand\diff{\mathrm{d}}

\newcommand{\dd}{\mathrm{d}}
\newcommand{\me}{\mathrm{e}}
\newcommand{\ii}{\mathrm{i}}

\newcommand{\vol}{\mathrm{vol}}

\newlength{\sswidth}

\newcommand{\M}{\mathcal{M}}

\newcommand{\lb}{\left(}
\newcommand{\rb}{\right)}

\newcommand{\ff}{F^{(2)}}

\newcommand{\alphai}{\alpha}

\newcommand{\C}{\mathbb{C}}
\renewcommand{\P}{\mathbb{P}}

\newcommand{\cN}{\mathcal{N}}

\newcommand{\bea}{\begin{eqnarray}}
\newcommand{\eea}{\end{eqnarray}}

\newcommand{\R}{{\mathbb R}}
\newcommand{\Z}{{\mathbb Z}}

\newcommand{\CFT}{{\rm CFT}}

\def\Im{\mathop{\mathrm{Im}}\nolimits}

\def\Tr{\mathop{\mathrm{Tr}}\nolimits}

\def\unit{{1\kern-.65ex {\rm l}}}
\def\1{{1\kern-.65ex {\rm l}}}

\def\ls{{\ell_{\rm s}}}
\def\lp{{\ell_{\rm p}}}
\def\gs{{g_{\rm s}}}

\newcount\hour \newcount\minute
\hour=\time \divide \hour by 60
\minute=\time
\def\now{%
\ifnum \hour<13
  \ifnum \hour=0 \advance \hour by 12 \number\hour:\else \number\hour:\fi%
     \ifnum \minute<10 0\fi%
     \number\minute%
\ A.M.%
\else \advance \hour by -12 \number\hour:%
  \ifnum \minute<10 0\fi%
  \number\minute%
  \ P.M.%
\fi%
}

\makeatother

\begin{document}

\baselineskip=18pt  
\numberwithin{equation}{section}  
\allowdisplaybreaks  



\thispagestyle{empty}


\vspace*{3mm} 
\begin{center}
{\huge F-theory and $\AdS_3/\CFT_2$ $(2,0)$}\\

 \vspace*{1.4cm}
 
{Christopher Couzens$\,^1$,   Dario Martelli$\,^1$, and\, Sakura Sch\"afer-Nameki$\,^2$}\\

 \vspace*{.8cm} 
{\it $^1$ Department of Mathematics, King's College London, \\
  The Strand, London, WC2R 2LS,  UK}\\
  \bigskip
{\it $^2$ Mathematical Institute, University of Oxford \\
Woodstock Road, Oxford, OX2 6GG, UK}\\
\bigskip

{\tt {kcl.ac.uk}: christopher.couzens, dario.martelli}\\
{\tt {gmail:  sakura.schafer.nameki}}
\end{center}
\vspace*{.2cm}


\noindent
We continue to develop the program initiated in \cite{Couzens:2017way} of studying supersymmetric AdS$_3$ backgrounds of F-theory and their holographic dual 2d  superconformal field theories, 
which are dimensional reductions of theories with varying coupling. Imposing 2d $\mathcal{N}=(0,2)$ supersymmetry, we derive the general conditions on the geometry for Type IIB $\AdS_3$ solutions with varying axio-dilaton and five-form flux. 
Locally the compact part of spacetime  takes the form of a circle fibration over an eight-fold $\mathcal{Y}_8^\tau$, which
 is elliptically fibered over a base $\widetilde{\M}_{6}$.  We construct two classes of  solutions given in terms of a product ansatz $\widetilde{\M}_{6} = \Sigma \times \mathcal{M}_4$, 
where $\Sigma$ is a complex curve  and $\mathcal{M}_4$ is locally a K\"ahler surface. 
 In the first class  $\mathcal{M}_4$ is  {globally a  K\"ahler surface}  and we take the elliptic fibration to vary non-trivially over either of these two factors, where in both cases the metrics on the total space of the elliptic fibrations are not Ricci-flat. 
In the second class  the metric on the total space of the elliptic fibration over either curve or surface are Ricci-flat. 
This results in solutions of the type AdS$_3\times K3 \times \M_5^\tau$, dual to 2d $(0,2)$ SCFTs, and  $\AdS_3\times S^3/\Gamma \times CY_3$, dual to 2d $(0,4)$ SCFTs, respectively.
In all cases we compute the charges for the dual field theories with varying coupling and find agreement with the holographic results.
We also show that  solutions with enhanced 2d $\mathcal{N}=(2,2)$ supersymmetry must have constant axio-dilaton.  Allowing the internal geometry  to be non-compact leads to the most general class of Type IIB $\AdS_5$ solutions with varying axio-dilaton, i.e. F-theoretic solutions, that are dual to 4d $\mathcal{N}=1$ SCFTs.

\newpage


\tableofcontents


\section{Introduction}

Twenty years after holography was uncovered in string theory, it still provides us with surprising and deep results about strongly coupled superconformal field theories (SCFTs) and quantum gravity in anti-de Sitter ($\AdS$) spacetimes. 
Progress is as far ranging as finding new supergravity solutions, matching with dual field theory observables, as well as performing precision tests of the duality in particular regimes. 
In the present paper we expand this AdS/CFT dictionary towards theories with spacetime varying coupling constant, a program initiated in \cite{Couzens:2017way}. The main goal is the construction of Type IIB solutions, where the axio-dilaton $\tau$ varies over parts of the spacetime, including monodromies in the $SL_2\mathbb{Z}$ duality group of Type IIB. 
In this sense these are $\AdS$ solutions in F-theory \cite{Vafa:1996xn}. In a brane realisation, the non-trivial monodromies arise through the presence of non-perturbative $(p,q)$ 7-branes, which contribute a new sector to the field theory duals. 

There are a multitude of motivations for studying field theories with varying coupling, e.g. field theories arising within F-theory such as D3-branes  and duality defects in SCFTs. Often the field theory side is somewhat difficult to study due to the genuinely non-perturbative effects. For example,  in F-theory D3-branes wrapped on cycles inside  the compactification geometry give rise to a varying complexified coupling $\tau$.  A field theoretic description of these is available for abelian theories \cite{Martucci:2014ema, Assel:2016wcr, Haghighat:2015ega, Lawrie:2016axq}, but remains elusive for the non-abelian generalization. Some special cases of S-duality twists can be studied along the lines of \cite{Ganor:2010md}, but do not correspond to varying axio-dilaton configurations.
In this context the holographic dual can shed some light upon some of the physical properties of these theories. 
In this paper we determine several new classes of solutions in Type IIB supergravity, which have a varying axio-dilaton profile, dual to both 2d and 4d SCFTs. A summary of the solutions is given in Table \ref{starwars}. In all cases we will explore both the solutions, as well as the holographic dual field theories and corroborate the duality by comparing central charges and other characteristics. Furthermore, 
 we determine dual M-theory solutions, which support some of the assumptions made in the F-theory setting.

\begin{table}\centering
\begin{tabular}{|c|c|c|c|c|c|c|}\hline
Dim& SUSY & $d\tau \not= 0$ &Type IIB/F-theory &  Field Theory & Sect.  \cr \hline\hline
2 & $(0,4)$ & $\checkmark$ & $\AdS_3 \times S^3 \times \Y_6^\tau$ & D3s on $C \subset \Y_6^\tau$ & \cite{Couzens:2017way} \cr \hline\hline
2 &$(2,2)$ & $\times$ &   $\AdS_3\times \M_7$  &  Example in \cite{Benini:2013cda} &  \ref{sec:AdS322} \cr \hline \hline
4 &$1 $ & $\checkmark$ & $\AdS_5 \times {(S^1 \rightarrow \mathcal{T}^\tau_{6} )} $ & D3s at conical singularity &  \ref{sec:AdS5}\cr \hline\hline
2&$(0,2)$ & $\checkmark$  & $\AdS_3   \times  (S^1 \rightarrow (\mathcal{S}^\tau_{4} \times \mathcal{M}_4))$ & Section \ref{sec:UTFTSurface} &  \ref{sol1deriv}\cr\hline
2&$(0,2)$ & $\checkmark$   & $\AdS_3   \times  (S^1 \rightarrow (\mathcal{T}^\tau_{6} \times \Sigma))$ &  Section \ref{sec:UTFTThreeFold} &  \ref{sec:ThreeFold}\cr\hline
2& $(0,2)$ & $\checkmark$ & $\AdS_3   \times K3^\tau  \times  \Yodd^{\podd,\qodd}$ & Section \ref{sec:BaryonicFT}  &  \ref{sec:BT}\cr\hline
\end{tabular}
\caption{\label{starwars} 
Summary of various $\AdS_3$ and $\AdS_5$ solutions in Type IIB/F-theory with five-form flux, including supersymmetries, the geometry of the solution and the dual field theory. The spaces with a superscript $\tau$ are elliptically fibered, however only in the case of $\mathcal{Y}_6^\tau$ and $K3^\tau$  they have Ricci-flat metrics. In general they are elliptic fibrations with non-trivial Ricci curvature. }
\end{table}


In \cite{Couzens:2017way} this approach was initiated by studying the F-theory solutions dual to 2d $(0,4)$ SCFTs, which were shown to be of the type $\AdS_3 \times S^3/\Gamma \times \Y_6^\tau$ where $\Y_6^\tau$ is an elliptically fibered Calabi--Yau three-fold, and the complex structure of the elliptic fiber $\mathbb{E}_\tau$ is identified with the axio-dilaton $\tau_{IIB}$. 
This provides a generalization of the known solutions with $(4,4)$ supersymmetry where $\Y_6= \Y_4 \times \mathbb{E}_\tau$ with $\Y_4 =$ K3 or $T^4$ and the axio-dilaton is  constant. 
The discrete subgoup $\Gamma$ of $SU(2)$ can be modded out, whilst retaining $(0,4)$ supersymmetry.
In fact, these solutions were shown to be the most general ones dual to 2d $(0,4)$ SCFTs, supported by five-form flux. 

The dual field theories are closely related to the MSW string \cite{Maldacena:1997de}, and have a dual description in terms of D3-branes wrapped on a curve inside  the base K\"ahler $B_2$ of the elliptically fibered Calabi--Yau three-fold:  $C\subset B_4\subset \Y_6^\tau$. The varying axio-dilaton induces a varying coupling of the 4d $\mathcal{N}=4$ Super-Yang Mills theory on the D3-brane, along the curve $C$. The resulting theory is supersymmetric when a particular topological twist, the so-called topological duality twist \cite{Martucci:2014ema, Assel:2016wcr}, is applied along the curve $C$ \cite{Haghighat:2015ega, Lawrie:2016axq, Lawrie:2016rqe}. The dual M-theory setup is the MSW string wrapped on an elliptic surface, and a dual M-theory solution confirms the F-theoretic results in \cite{Couzens:2017way}, including the holographic comparison of the central charges to leading and sub-leading orders. In this context this comparison is non-trivial due to the presence of duality defects, i.e. 7-branes, which in the M-theory picture have a geometric origin in resolution cycles of the singular elliptic fibers.

A richer class of theories in 2d preserves only $(0,2)$ supersymmetry, where the central charge is not determined by the UV spectrum of the field theory, but due to mixing of the $U(1)_R$ symmetry with {global} symmetries along the RG flow, one needs to invoke $c$-extremization to compute the central charges \cite{Benini:2012cz} (see also \cite{Silverstein:1994ih}). 

Holographically the constant axio-dilaton case supported by only five-form flux was studied in \cite{Kim:2005ez}, where it was shown that the internal space locally admits a circle fibration, realising the $U(1)_R$ symmetry in the dual field theory. 
A related analysis appeared in \cite{Donos:2008ug}, which again has trivial $\tau$ but allows for {a particular} three-form flux on the internal manifold $\M_{7}$. 
Examples of solutions were obtained in  \cite{Gauntlett:2006ns,Benini:2015bwz}, again for constant $\tau$, where starting with the general framework of \cite{Kim:2005ez}, the 6d K\"ahler 
base  is assumed to be a direct product  $C_g\times \M_4$, with $C_g$ a genus-$g$ constant curvature Riemann surface, and $\M_4$ a locally K\"ahler space equipped with a metric admitting an $SU(2)\times U(1)$ isometry.

In the present paper we generalise these results  to allow for varying axio-dilaton $\tau$, and determine the geometric constraints on the supergravity solutions preserving $(0,2)$ in the dual 2d SCFT. {Locally} the F-theory 
solution {takes the form} $\AdS_3\times (S^1\rightarrow \mathcal{Y}_8^\tau)$, where $\mathcal{Y}^\tau_8$ is elliptically fibered and K\"ahler, however \emph{not Calabi--Yau}. 
{The $U(1)$ isometry of the $S^1$ corresponds to the R-symmetry of the dual $(0,2)$ SCFTs and} the curvature of $ \mathcal{Y}_8^\tau$ has to satisfy (\ref{Master8d}). 
This is of course a formidable equation to solve and in this paper we will focus on two classes of solutions to this equation, which result from specialisations of $\mathcal{Y}^\tau_8$. Let us denote the base of the elliptic fibration by $\widetilde{\M}_{6}$, which is (locally) a K\"ahler three-fold. 

The first class of solutions, which will be  {referred to as} \emph{universal twist solutions}, arise from making a product ansatz for the base in terms of a complex curve and surface  
\be
\ba
{\mathbb{E}_\tau} \ \hookrightarrow\  & \ \mathcal{Y}_8^\tau  \cr 
& \,  \downarrow \cr 
 &  \widetilde{\mathcal{M}}_6 =\Sigma \times \mathcal{M}_4
\ea
 \quad \hbox{where} \quad 
\left\{ \ba
\mathcal{Y}_8^\tau &= (\mathbb{E}_\tau \rightarrow \Sigma) \times \mathcal{M}_4 \equiv \mathcal{S}^{\tau}_4 \times \mathcal{M}_4 \cr 
\mathcal{Y}_8^\tau &= (\mathbb{E}_\tau \rightarrow \mathcal{M}_4) \times \Sigma   \equiv \mathcal{T}^{\tau}_6 \times \Sigma
\ea\right.
\label{babel}
\ee
Thus $\mathcal{Y}_8^\tau$, either has an elliptic surface $\mathcal{S}^{\tau}_4$ or an elliptic three-fold $\mathcal{T}^{\tau}_6$ as a factor. 
The key is that none of the factors appearing in $\mathcal{Y}_8^\tau$ are Ricci-flat. This is to be contrasted with the standard flat-space analysis, which results in an elliptic Calabi-Yau compactification. Here, due to the fluxes, the supersymmetry implies a non-trivial condition on the curvature of the factors in $\mathcal{Y}_8^\tau$, which is not simply Ricci-flatness. These solutions will be discussed in section \ref{sec:Universal02}. 
Their field theory duals have a characterization in terms of a topological duality twisted reduction of 4d $\mathcal{N}=1$ SCFTs. 

The second class of solutions, that are the subject of section \ref{sec:BT} can be summarised in terms of an ansatz 
\be
\mathcal{Y}_8^\tau =       (\mathbb{E}_\tau \rightarrow \mathbb{P}^1) \times \mathcal{M}_4   =  K3^\tau \times \mathcal{M}_4 \,,
\ee
whereby the base factors as $\widetilde{\mathcal{M}}_6 = \mathbb{P}^1 \times \mathcal{M}_4$ and the elliptic fibration over the $\mathbb{P}^1$ is Ricci-flat\footnote{Likewise we can consider a factorization $\mathcal{Y}_8^\tau = CY_3^\tau \times \Sigma$, with a Calabi-Yau three-fold factor, and as we shall show in section \ref{sec:Recovery}, this gives rise to the already known solutions in \cite{Couzens:2017way} with $(0,4)$ supersymmetry in 2d.}. The dual field theories are obtained by turning on  baryonic flux, and we shall refer to these as the \emph{baryonic flux solutions}. 
We find that this yields a class of geometries that are closely related to the $Y^{p,q}$ Sasaki--Einstein manifolds \cite{Gauntlett:2004yd}  as well as the constant axio-dilaton $\AdS_3$ solutions obtained in 
\cite{Gauntlett:2006ns, Donos:2008ug} and interpreted in \cite{Benini:2015bwz} as gravity duals to certain topologically twisted compactification of  the $Y^{p,q}$ quiver theories   \cite{Benvenuti:2004dy}.
The F-theory solution, which we find, including the geometrized axio-dilaton, is $\AdS_3\times K3^\tau \times \mathfrak{Y}^{\podd, \qodd}$, or as a Type IIB solution $\AdS_3 \times \mathbb{P}^1 \times \mathfrak{Y}^{\podd, \qodd}$, 
 where -- similarly to     \cite{Gauntlett:2004yd} --   the latter is a circle fibration over the Hirzebruch surface $\mathbb{F}_0 = \mathbb{P}^1 \times \mathbb{P}^1$,  equipped with a non-direct product metric. 
 The constant $\tau$ precursor of the baryonic flux solutions  were conjectured in \cite{Benini:2015bwz} to be gravity   dual to the $Y^{p,q}$ field theories compactified on a $T^2$, with a twisting by the baryonic 
symmetry of these theories. Upon a careful analysis of the geometry {that we include in Appendix \ref{app:Ypqstuffstuff}}, we have uncovered some puzzling features of this duality, that  deserve a separate investigation.

Another interesting class of SCFTs in 2d have $(2,2)$ supersymmetry, which are amenable to localization computations \cite{Benini:2012ui, Doroud:2012xw}. Recent progress on constructions of dual pairs, using orbifolds of the NS-NS-flux supported AdS$_3 \times S^3 \times T^4$ solutions with 2d $(4,4)$  supersymmetry were obtained in \cite{Datta:2017ert, Eberhardt:2017uup}. 
We find that by turning on the axio-dilaton as well as requiring 2d  $\mathcal{N}=(2,2)$ supersymmetry in the dual SCFT, no compact solutions exist. However, we determine the general constraints for these Type IIB solutions with constant axio-dilaton, showing in particular that they realise geometrically the $U(1)_R\times U(1)_R$ $R$-symmetry of the dual SCFTs. We will  recover some known solutions, but we have refrained from exploring the general conditions further.  {Interestingly}, if we relax the assumption that the internal space  is compact, the axio-dilaton can vary and we find $\AdS_5 \times \M_7^\tau$, 
where {$S^1 \hookrightarrow \M_7^{\tau}  \rightarrow \mathcal{T}^\tau_{6}$ with  $\mathcal{T}^\tau_{6}$  the same elliptic three-fold  geometries that appeared in the AdS$_3$ solutions in (\ref{babel}). As we will discuss 
 in the paper this motivates the conjecture that the 2d SCFTs dual to the second type of universal twist solutions are the compactification of the 4d SCFTs dual to this class of AdS$_5$ solutions.}
 In fact it is not too difficult to show that this is the most general $\AdS_5$ solution, dual to SCFTs preserving $\mathcal{N}=1$ in 4d, supported with only five-form flux. This makes contact with the results in \cite{Kehagias:1998gn}, however, we shall provide an intrinsically F-theoretic interpretation in terms of elliptically fibered Calabi--Yau four-fold compactifications. This surprising appearance of duals to 4d SCFTs,
{generalising known constant but non-perturbative $\tau$ solutions of \cite{Aharony:1998xz, Ahn:1998tv,  Fayyazuddin:1998fb, Kruczenski:2003pv,  Aharony:2007dj}}, is  welcome and will be studied elsewhere in more detail \cite{toappearAdS}.

The plan of the paper is as follows: in section \ref{sec:AdS302} we begin by developing the general setup for solutions in Type IIB with five-form flux and varying axio-dilaton, requiring only $(0,2)$ supersymmetry in 2d. The ``master equation'' derived in this section will underlie most of the remaining part of the paper.
We then analyse in section \ref{sec:22} the special case of $(2,2)$ supersymmetry and find, as already alluded to, that there are no varying axio-dilaton solutions, unless the internal space decompactifies. The resulting solutions have $\AdS_5$ factors and varying $\tau$, as explained in section \ref{sec:AdS5}, and a proof that these are the most general such $\AdS_5$ solutions dual to 4d $\mathcal{N}=1$ is provided in appendix \ref{app:AdS5}. In section \ref{sec:NewSol} we derive two new classes of $(0,2)$ F-theory solutions: the universal twist and baryonic twist solutions. The holographic central charges are computed in section \ref{sec:HolC}.
The dual field theories are discussed in section \ref{sec:FT}, where a brief review of the duality twist is included. Dual M-theory solutions
 are discussed in section \ref{sec:M}, before we conclude with a summary and outlook in section \ref{sec:DO}. 
Several appendices supplement the content of the main body of the paper.

\section{$\AdS_3$ Solutions in F-theory dual to 2d $\mathcal{N}=(0,2)$}
\label{sec:AdS302}

The starting point of our analysis is a comprehensive exploration of the conditions of Type IIB/F-theory supergravity which yield $\AdS_3$ solutions with at least  2d $(0,2)$ supersymmetry and vanishing three-form fluxes. The main difference to earlier results in \cite{Kim:2005ez} is that we allow the axio-dilaton $\tau$ to have a non-trivial dependence on spacetime.  
This generalises our earlier work in \cite{Couzens:2017way}, where we found the most general  solutions in F-theory with only five-form flux dual to 2d $(0,4)$ SCFTs. The requirement for 2d $(0,2)$ supersymmetry leads us to both recover the earlier $(0,4)$ results, as well as new classes of solutions dual to $(0,2)$ SCFTs. The ``master equation'' (\ref{Master}), which constrains the internal geometry, yields potentially more solutions, whose exploration we leave for the future. This equation has a reformulation in terms of an F-theoretic setting, where the axio-dilaton becomes part of the compactification geometry.

\subsection{Killing Spinor Equations}

The starting point is the 10d Type IIB supergravity Killing spinor equations\footnote{Our conventions will be those of \cite{Gauntlett:2005ww, Couzens:2017way}.}, i.e. the requirement that the supersymmetry transformations of the fermions vanish identically
\bea
\delta\psi_{M}&=& \mathcal{D}_{M}\epsilon+\frac{\ii}{192}\Gamma^{P_{1}...P_{4}}F_{MP_{1}...P_{4}}\epsilon \cr 
&& \ \qquad -\frac{1}{96}\left(\tensor{\Gamma}{_{M}^{P_{1}..P_{3}}}G_{P_{1}..P_{3}}-9\Gamma^{P_{1}P_{2}}G_{MP_{1}P_{2}}\right)\epsilon^{c} =0~,\label{susy1}\\   
\delta \lambda&=& \ii \Gamma^{M}P_{M}\epsilon^{c}+\frac{\ii}{24}\Gamma^{P_{1}..P_{3}}G_{P_{1}...P_{3}}\epsilon=0\label{susy2}~.
\eea
The covariant derivative $\mathcal{D}$ is with respect to both Lorentz transformations and local $U(1)_D$ transformations, where the duality $U(1)_D$ has gauge field depending on the variation of the axio-dilaton $\tau=\tau_{1}+ i\tau_{2}$ 
\be\label{ALMOSTTHERE}
Q=-\frac{1}{2 \tau_{2}}\dd \tau_{1}~.
\ee
In addition we define the combination 
\be
P=\frac{\ii}{2 \tau_{2}}\dd \tau~.
\ee
The Killing spinors have $U(1)_D$ charge $1/2$, $P$ has charge 2 and $G$ has charge 1.
This $U(1)_D$ symmetry will play a key role in the following as it encodes the varying axio-dilaton. 
In supergravity the gauge symmetry itself is well-known and we summarise some of the salient points in appendix \ref{app:SL2}. We will return to this in section \ref{sec:FT} in the field theory analysis.

The equations of motion consist of the Einstein equation 
\bea
R_{MN}=2P_{(M}P^{*}_{N)}+\frac{1}{96}F_{MP_{1}..P_{4}}\tensor{F}{_{N}^{P_{1}..P_{4}}}+\frac{1}{8}\lb 2\tensor{G}{_{(M}^{P_{1}P_{2}}}G^{*}_{N)P_{1}P_{2}}-\frac{1}{6}g_{MN}G^{P^{1}..P_{3}}G^{*}_{P_{1}..P_{3}}\rb~~~
\eea
and the flux equations of motion and Bianchi identities
\be
\ba
\mathcal{D}*G &=P\wedge *G^{*}-\ii F\wedge G~\,,\quad 
\mathcal{D}*P=-\frac{1}{4}G\wedge *G~,\quad 
F=*F \,, \cr 
\mathcal{D}P&=0 \,,\quad 
\mathcal{D}G=-P\wedge G^{*} \,,\quad 
\dd F=\frac{\ii}{2}G\wedge G^{*} \,.
\ea
\ee

\subsection{AdS$_{3}$ Ansatz and $(0,2)$ Supersymmetry}

In this paper we consider the most general class of bosonic, minimally supersymmetric Type IIB supergravity solutions with $SO(2,2)$ symmetry and vanishing three-form flux $G=0$. 
As in \cite{Kim:2005ez,Couzens:2017way} the 10d metric will be taken in Einstein frame to be a warped product of the form\footnote{For the entirety of the paper, we will indicate by subscripts of spaces always the real dimension.}
\be
\dd s^{2}=\me^{2 \Delta}\lb \dd s^{2}(\text{AdS}_{3})+\dd s^{2}(\M_{7})\rb\label{metricform}~,
\ee
where $\dd s^{2}($AdS$_{3})$ is the metric on AdS$_{3}$, with Ricci tensor $R_{\mu\nu}=-2 m^{2}g_{\mu\nu}$ and  $\dd s^{2}(\M_{7})$ is the metric on an arbitrary internal seven-dimensional manifold $\M_{7}$. We take $\Delta\in\Omega^{(0)}(\M_{7},\mathbb{R}),~P\in \Omega^{(1)}(\M_{7},\mathbb{C}),~\tau\in \Omega^{(0)}(\M_{7},\mathbb{C})$ and the five-form flux to be of the form
\be
F^{(5)}=(1+*)\dd \vol(\text{AdS}_{3})\wedge F^{(2)}~,
\ee
with $F^{(2)}\in\Omega^{(2)}(\M_{7},\mathbb{R})$ in order to preserve the $SO(2,2)$ symmetry of AdS$_3$. The Bianchi identity for $F^{(5)}$ implies 
\be
\dd F^{(2)}=0 \,,\qquad \dd\hat{*}_{7}F^{(2)}= 0\label{d*F} \,,
\ee
where $\hat{*}_{7}$ is the hodge star on the unwarped metric $\dd s^{2}(\M_{7})$. 
We use the spinor ansatz developed in appendix A of \cite{Couzens:2017way}
\be\label{KSansatz}
\epsilon=\psi_{1}\otimes \me^{\Delta/2} \xi_{1}\otimes \theta+\psi_{2}\otimes \me^{\Delta/2}\xi_{2}\otimes \theta \,      , 
\ee
where $\psi_{i}$ are Majorana Killing spinors on AdS$_{3}$ and satisfy 
\be
\nabla_{\alphai}\psi_{i}=\frac{\alphai_{i} m}{2}\rho_{\alpha}\psi_{i}\label{AdSKSeq} \,  .
\ee
The chirality of the spinor of the dual SCFT is determined by the choice of $\alphai_{i}=\pm 1$. The spinors $\psi_{i}$ are taken to be independent Killing spinors on AdS$_3$, whilst the $\xi_{i}$ are Dirac spinors on $\M_{7}$. 
Each independent Dirac spinor $\xi_{1/2}$ will give 2 (anti-) chiral supercharges on the boundary SCFT. To preserve $(0,2)$ supersymmetry we take $\xi_{2}$ to vanish. We shall also be interested in preserving $(2,2)$ supersymmetry in section \ref{sec:22} in which case both spinors are kept, but with opposite values for $\alpha$.

The reduced supersymmetry equations for the spinors on $\M_{7}$ are, as in \cite{Couzens:2017way}, obtained by inserting the ansatz \eqref{KSansatz} into the $10d$ supersymmetry equations, (\ref{susy1}) and (\ref{susy2}), 
\bea
\gamma^{\mu}P_{\mu}\xi_{j}^{c}
&=&0\label{dilalg}~,\\
\lb \frac{1}{2} \partial_{\mu}\Delta\gamma^{\mu}-\frac{\ii \alphai_{j}m}{2}+\frac{ \me^{-4\Delta}}{8}\slashed{F}^{(2)}\rb \xi_{j}
&=&0   \label{gravalg}~, \\
\lb\D_{\mu} +\frac{\ii \alphai_{j}m}{2}\gamma_{\mu}-\frac{\me^{-4 \Delta}}{8}F^{(2)}_{\nu_{1}\nu_{2}}\tensor{\gamma}{_{\mu}^{\nu_{1}\nu_{2}}}\rb\xi_{j}
&=&0 \,. \label{gravdif}
\eea


\subsection{Constraints on the Geometry}\label{Constraints}

In this section we investigate the torsion conditions arising from imposing the minimal amount of supersymmetry, namely $\mathcal{N}=(0,2)$ in 2d. This amount of supersymmetry is preserved by the existence of a single Dirac spinor on $\M_{7}$, and signifies that the internal 7d space admits an $SU(3)$ structure. In 7d an $SU(3)$ structure implies the existence of a real vector which foliates the space with the transverse 6d space admitting a canonical $SU(3)$ structure. In the following we show that the  transverse 6d space is conformally K\"ahler and the existence of a supersymmetric solution is determined by a single partial differential equation, similar to the equation found in \cite{Kim:2005ez}, for the  K\"ahler metric on the 6d space. The remaining geometry is fixed by the choice of this K\"ahler metric. 

The torsion conditions for preserving $\cN=(0,4)$ supersymmetry were computed in appendix C of \cite{Couzens:2017way} and may be specialised to preserve $\cN=(0,2)$  by setting, without loss of generality, $\xi_{2}=0$. We present the non-trivial torsion conditions in the present case below\footnote{We refine the notation of \cite{Couzens:2017way} for ease of reading. By setting $\xi_{2}=0$ the bilinears with a `2' index are set to zero, and it therefore becomes superfluous to keep the `11' subscript on the non-zero bilinears; apart from removing this labelling the names of the bilinears are otherwise kept the same. We also set the parameter $\alpha$ in \cite{Couzens:2017way} to 1 without loss of generality in the following. As explained previously this parameter takes values $\pm1$ and is related to the chirality of the preserved supersymmetry for the dual SCFT.} and the general equations in appendix \ref{sec:torsion}.
Supersymmetry implies both differential and algebraic constraints on the fluxes and bilinears. The independent differential conditions satisfied by the bilinears are
\begin{align} \label{02Cond} 
\dd S&=0\\
\me^{-4 \Delta}\dd \lb \me^{4 \Delta}K\rb &=-2\ii m  U-\me^{-4 \Delta}\ff~,\label{dK}\\
\dd \lb \me^{4 \Delta}U \rb&=0~,\label{dU}\\
\me^{-6 \Delta}\D\lb \me^{6 \Delta}Y\rb &=2 m  * Y~,\\
\me^{-6 \Delta}\D \lb \me^{6 \Delta}*Y\rb &=0~,\label{d*Y}\\
4 \dd \Delta \wedge *Y&=-\ii \me^{-4 \Delta}\ff\wedge Y~,\\
\me^{-8 \Delta}\dd \lb \me^{8 \Delta}*U\rb&= 2 \ii m *K~.\label{d*U}
\end{align}
Again, as in \cite{Couzens:2017way} the scalar $S$ can be set to 1 by a constant rescaling of the Killing spinor. 

To proceed we introduce an orthonormal frame for the metric and by a suitable frame rotation we may set $K$ to be parallel to the vielbein $e^{7}$. In this frame the remaining bilinears become 
\begin{align} 
K &=- e^7~, \\
U &= -\ii(e^{12} + e^{34} + e^{56}) ~,\\
X &= U \wedge K ~,\\
Y &=(e^{1}-\ii e^{2})\wedge (e^{3}-\ii e^{4})\wedge (e^{5}-\ii e^{6})~ .
\end{align}
A 2d SCFT with $\mathcal{N}=(0,2)$ supersymmetry has a $U(1)_R$ R-symmetry, which by the AdS/CFT dictionary is dual on the gravity side to a Killing vector generating a $U(1)$ isometry of the full solution. From the torsion conditions it follows that $K$ defines such a Killing vector and thus is identified with the R-symmetry of the putative dual SCFT. It is useful to introduce coordinates adapted to this Killing vector (and dual one-form) 
\be
K^{\#}=2m\partial_{\psi}~,\qquad 
K=\frac{1}{2m}(\dd \psi+\rho)~,
\ee
so that the 7d metric can be written as follows 
\be
\dd s^{2}=\frac{1}{4m^2}(\dd \psi+\rho)^{2}+\dd s^{2}(\M_{6})~.
\ee
Observe from \eqref{dU} that the bilinear $U$ is conformally closed and this motivates us to define the following conformally rescaled forms
\be
J=\ii m^{2} \me^{4 \Delta}U \,,\qquad  \bar{\Omega}= m^{3}\me^{6 \Delta}Y \,.
\ee
These new forms define a canonical $SU(3)$ structure on $\widetilde{\M}_{6}$ whose metric is conformally related to $\M_{6}$ by
\be
\dd s^{2}({\M_{6}})=\frac{\me^{-4 \Delta}}{m^{2}}\dd s^{2}({\widetilde{\M}_{6}})~.
\ee
They satisfy the $SU(3)$ structure algebraic conditions
\be 
J \wedge \Omega = 0 \,,\qquad 
\Omega \wedge \bar\Omega =- \frac{8\ii}{6}J \wedge J \wedge J = -8\ii~\dd \vol(\widetilde{\M}_6)
\ee
and in addition the differential conditions
\be\label{DOmdJ}
\bar{\D}\Omega=-2\ii m K \wedge \Omega \,,\qquad 
\dd J=0 \,,
\ee
which imply integrability of the complex structure defined by $\Omega$ and that $\widetilde{\M}_{6}$ is K\"ahler. 
Finally, we should extract the conditions of the varying axio-dilaton on the metric. 
From the supersymmetry equation \eqref{dilalg} 
\be
\tensor{J}{^{\mu}_{\nu}}P^{\mu}=\ii P^{\mu}\,,\qquad P_{\mu}K^{\mu} =0 \,,\label{Phol02}
\ee
i.e. $P$ varies holomorphically on $\widetilde{\M}_{6}$ and the Killing vector $K$ is a symmetry of $\tau$, $\mathcal{L}_{K}\tau=0$.

Due to the foliation of the space by the Killing vector we may decompose the exterior derivative as
\be
\dd =\dd \psi \wedge \partial_{\psi}+\dd_{6}~.
\ee
With this splitting of the exterior derivative \eqref{DOmdJ} becomes
\begin{align}
\partial_{\psi}\Omega&= -\ii  \Omega~,\label{psiOmega}\\
\dd_{6}\Omega&=-\ii (Q+ \rho)\wedge \Omega~.\label{d6Omega}
\end{align}
Equation \eqref{psiOmega} may be solved by extracting a suitable $\psi$ dependent phase from $\Omega$. This phase will play no role in the following analysis and will be assumed to have been extracted. Subsequently, (\ref{d6Omega}) implies
\be
\mathfrak{R}=-(\dd Q+ \dd\rho)~,\label{Ricciform}
\ee
where $\mathfrak{R}$ is the Ricci form on $\widetilde{\M}_{6}$. The Ricci tensor on $\widetilde{\M}_6$ is given in terms of the Ricci form as
\be 
R_{\mu\nu}=-\tensor{J}{_{\mu}^{\rho}}\mathfrak{R}_{\rho\nu}~.
\ee
The flux is fixed by equation \eqref{dK} to be 
\be
 m\ff=-2   J-\frac{1}{2}\dd (\me^{4 \Delta}(\dd \psi+\rho))~.\label{ffdef1}
 \ee
Notice that  the flux has legs along the Killing direction and may be decomposed such that 
\be
\hat{F}^{(2)}=\ff+ \dd \me^{4\Delta} \wedge K~,
\ee
has no legs along the Killing direction\footnote{The explicit $K$ factor cancels out with that in $\ff$.}. By contracting the indices of the Ricci-form with the complex structure one finds the Ricci scalar for $\widetilde{\M}_{6}$ to be\footnote{In deriving this result it is necessary to use the algebraic equation
$F^{(2)}_{\mu \nu} J^{\mu \nu} =\hat{F}^{(2)}_{\mu\nu}J^{\mu\nu}= -\frac{8}{m}$,
which is obtained from the supersymmetry equation \eqref{dilalg}.
}
\be
R=2|P|^{2}+8 \me^{-4 \Delta}~.\label{Rexpression}
\ee
By imposing equations \eqref{dK}, \eqref{DOmdJ}, \eqref{Ricciform} it follows that equations \eqref{d*Y}-\eqref{d*U} are immediately satisfied.


\subsection{Sufficiency of the Conditions}

So far supersymmetry has implied that the solution satisfies \eqref{DOmdJ}, \eqref{Phol02}, \eqref{Ricciform}, \eqref{ffdef1} and  \eqref{Rexpression}. We show in this section that this set of equations in addition to imposing the equation of motion for $\ff$ are both necessary and sufficient conditions for a bosonic supersymmetric solution. As we show, the equation of motion for $\ff$ may be rephrased as a differential condition on the K\"ahler metric of the 6d space. We proceed by first considering the equations of motion before proving that there exists a globally defined Killing spinor satsifying the Killing spinor equations \eqref{dilalg}-\eqref{gravdif}. 

\subsubsection{Equations of Motion}

Recall that the equation of motion for the five-form flux is equivalent to the two equations in \eqref{d*F} for the two-form $\ff$. Using equation \eqref{ffdef1} as the definition of $\ff$ it is clear after using \eqref{DOmdJ} that it is closed. Supersymmetry does not however impose the equation of motion for the flux, $\dd * \ff=0$, which must be imposed in addition. One may understand this equation as giving a ``master equation'' for the K\"ahler base which generalises the one found in \cite{Kim:2005ez} to include varying axio-dilaton, 
\be\label{Master}
{\square}_{6}(R-2|P|^{2})-\frac{1}{2}R^{2}+R_{\mu\nu}R^{\mu\nu}+2 |P|^{2}R-4 R_{\mu\nu}P^{\mu}P^{*\nu}=0~.
\ee
A discussion of its derivation is given in appendix \ref{app:Mast}. We conclude that both the equation of motion for the five-form flux and the self-duality constraint are satisfied. The Bianchi identity for $P$ is implied by construction whilst its equation of motion reduces to $\tau$ being harmonic on the K\"ahler manifold. As $\tau$ is holomorphic it follows that it is also harmonic and therefore the flux equations of motion and Bianchi identities are satisfied. 

By using the analysis of \cite{Gauntlett:2005ww} and some case dependent algebra we may show that the Einstein equation is satisfied. Integrability of the Killing spinor equations and use of the flux equations of motion and Bianchi identities implies 
\be
E_{MN}\Gamma^{N} \epsilon=0
\ee
where $E_{MN}=0$ is equivalent to Einstein's equation and $\epsilon$ is the 10d Killing spinor. One may construct a null vector bilinear, $\widehat{K}\equiv \bar{\epsilon}\,\Gamma_{(1)}\epsilon$, which implies that the metric admits a frame such that it takes the form
\be
\dd s^{2}=2e^{+}e^{-} +e^{a}e^{a}~,
\ee
with $\widehat{K}=e^{+}$ and $a=1,..,8$. The argument of \cite{Gauntlett:2002fz} shows that the only component of $E_{MN}$ which may be non-zero is $E_{++}$. For this class of solutions $E_{++}$ lies along AdS$_{3}$ and by explicit computation one finds that the Ricci-tensor on the warped AdS$_3$ satisfies 
\be
R_{\mu\nu}=\lb -2 m^{2}+ 8 \nabla_{\mu} \Delta \nabla^{\mu}\Delta - \square \Delta \rb g_{\mu\nu}~.
\ee
It follows that $E_{++}\propto g_{++}$ which therefore vanishes and we conclude supersymmetry implies the Einstein equation. We determine that all the equations of motion are satisfied by supersymmetry and equation \eqref{Master}.


\subsubsection{Supersymmetry}

We now show that any solution satisfying the necessary conditions presented above admits a globally defined Killing spinor satisfying \eqref{susy1} and \eqref{susy2}. By construction it follows that any global solution to the 7d Killing spinor equations \eqref{dilalg}-\eqref{gravdif} may be uplifted to a global Killing spinor in 10d satisfying both \eqref{susy1} and \eqref{susy2}. Preserving supersymmetry is therefore equivalent to proving that equations \eqref{dilalg}-\eqref{gravdif} admit a globally defined Killing spinor. We shall construct such a spinor by making use of the canonical spin$^c$ structure that every K\"ahler manifold admits.

We begin by defining the notation and vielbein we shall be using in the following. Recall that the metric takes the form
\be
 \dd s^{2}(\M_{7})= \frac{\me^{-4 \Delta}}{m^{2}} \dd s^{2} (\widetilde{\M}_{6})+\frac{1}{4m^{2}}(\dd \psi+ \rho)^{2}= \dd s^{2}(\widetilde{\M}_{6}) + (e^{7})^{2} \,,
\ee
where, in keeping with the frame in section \ref{Constraints}, $e^7=-\frac{1}{2m} (\dd \psi+\rho)$. The flat index for the vielbein on $\widetilde{\M}_{6}$ will be taken from the middle of the Latin alphabet, $i,j,k$ and run from $1,\dots ,6$ whereas the curved index on $\widetilde{\M}_{6}$ will be from the middle of the Greek alphabet, $\mu, \nu,\sigma$, finally the seven-dimensional indices will be from the beginning of the respective alphabets. The K\"ahler two-form on $\widetilde{\M}_{6}$, written in terms of the vielbeine, is $j=e^{12}+e^{34}+e^{56}$, which in general is only conformally closed, whilst the closed two-form on $\widetilde{\M}_{6}$ is denoted $J$.

On any K\"ahler manifold there exists a spin$^c$ structure that admits a section $\eta$ satisfying the spin$^c$ Killing spinor equation
\be
\lb\widetilde{\nabla}_{\mu}+\frac{\ii }{2} \widehat{P}_{\mu}\rb \eta=0\,,
\label{SpincKSE}
\ee
where $\widehat{P}$ is the one-form Ricci potential of the K\"ahler metric. For a 6d space, if one takes the spinor $\eta$ to satisfy the projection conditions
\be
\gamma_{12}\eta=\gamma_{34}\eta=\gamma_{56}\eta=-\ii \eta~ \,,
\label{Projcond}
\ee
it is easy to see that the term arising from the spin-connection precisely cancels the contribution from $\widehat{P}$ and therefore any constant section $\eta$, subject to the projection conditions, solves \eqref{SpincKSE}. Clearly this spinor is globally defined on $\widetilde{\M}_{6}$, and we may use it to  to construct a Killing spinor satisfying the 7d supersymmetry equations. On $\widetilde{\M}_{6}$, equation \eqref{SpincKSE} reads
\be 
\lb \widetilde{\mathcal{D}}_{\mu}-\frac{\ii}{2} \rho_{\mu}\rb \eta=0~.\label{SpincKSEred}
\ee
The spin connection on $\M_{7}$ is found to be
\begin{align}
\omega^{jk}&= \widetilde{\omega}^{jk}- 2(\partial^{k}\Delta e^{j}- \partial^{j}\Delta e^{k})-\frac{1}{4m} \left[ \frak{R}^{jk} -\ii (P^{j} P^{* k}- P^{*j} P^{k})\right] e^{7}~,\\
\omega^{7 j}&=\frac{1}{4m}\left[\tensor{\mathfrak{R}}{^{j}_{k}}  e^{k} -\ii ( P^{j} P^{*}-P^{*j} P)\right]~,
\end{align}
and the flux is 
\be
m \ff= -2 m^{2} \me^{4 \Delta}j+ \frac{\me^{4 \Delta}}{2} (\mathfrak{R}+\dd Q) +4 m \me^{4 \Delta} \dd \Delta \wedge e^{7}~.
\ee
By inserting the above spin connection, \eqref{Ricciform}, \eqref{ffdef1} and \eqref{Rexpression} into \eqref{gravdif}, and computing along the Killing direction and along $\widetilde{\M}_{6}$, respectively, yields 
\be\ba
0=&\lb \nabla_{\psi}+\frac{\ii m}{2} -\frac{\me^{-4 \Delta}}{8} F_{a b}\tensor{\gamma}{_{\psi}^{ab}}\rb \xi =\lb \partial_{\psi}-\frac{\ii}{2}\rb \xi \cr 
0=&\lb \mathcal{D}_{\mu} +\frac{\ii m}{2} \gamma_{\mu} -\frac{\me^{- 4\Delta}}{8}F_{ab} \tensor{\gamma}{_{\mu}^{ab}}\rb \xi= \lb \widetilde{\mathcal{D}}_{\mu} -\frac{\ii}{2} \rho_{\mu} \rb \xi~.
\ea\ee
We may solve both equations by taking the Killing spinor to be
\be
\xi= \me^{\frac{\ii}{2}\psi} \eta~.
\ee
Notice that the functional dependence on $\psi$  is consistent with \eqref{psiOmega}.

It remains to show that the algebraic conditions \eqref{dilalg} and \eqref{gravalg} are satisfied. Using the holomorphicity of $P$ one finds that the dilatino equation, \eqref{dilalg}, vanishes upon application of the projection conditions \eqref{Projcond}. The algebraic gravitino equation becomes
\begin{align}
0=\lb \frac{1}{2} \partial_{\mu}\Delta\gamma^{\mu}-\frac{\ii \alphai_{j}m}{2}+\frac{ \me^{-4\Delta}}{8}\slashed{F}^{(2)}\rb \xi=\lb \frac{\ii m}{ 4} + \frac{1}{16 m}\lb \frac{1}{2}\mathfrak{R}_{ij}-\ii P_{i}P^{*}_{j}\rb\gamma^{ij}\rb \xi\,,
\end{align}
which vanishes after some gamma matrix algebra and application of \eqref{Rexpression} and \eqref{Projcond}. We conclude that supersymmetry is preserved if we satisfy \eqref{DOmdJ}, \eqref{Phol02}, \eqref{Ricciform}, \eqref{ffdef1}, \eqref{Rexpression} and \eqref{Master}.

\subsection{Summary of Conditions}

Let us summarise the necessary and sufficient conditions for a supersymmetric solution with at least $\mathcal{N}=(0,2)$ supersymmetry, metric of the form \eqref{metricform}, arbitrary five-form flux, $F$ and varying axio-dilaton, $\tau$ all preserving the isometries of AdS$_3$. We have shown that the metric of the solution takes the form
\be\label{generalmetric}
\dd s^{2}= \me^{2 \Delta}\left[ \dd s^{2}(\text{AdS}_{3})+ \frac{1}{m^{2}}\lb \frac{1}{4} (\dd \psi+\rho)^{2}+\me^{-4 \Delta}\dd s^{2}(\widetilde{\M}_{6})\rb\right]\,,
\ee
where $\dd s^{2}(\widetilde{\M}_{6})$ is a K\"ahler metric satisfying the ``master equation'' \eqref{Master}. The remaining geometry is determined in terms of the metric on $\dd s^{2}(\widetilde{\M}_{6})$ to be
\begin{align}
\me^{-4 \Delta}&=\frac{1}{8}( R-2 |P|^{2})~,\label{warpdef}\\
\dd\rho&=-(\dd Q +\mathfrak{R})~,\label{rhodef}
\end{align}
 and the flux is given by
\be
\ba
F&=(1+*) \dd \vol(\text{AdS}_{3})\wedge \ff\cr
 m\ff&=-2   J-\frac{1}{2}\dd (\me^{4 \Delta}(\dd \psi+\rho))~.\label{ffdef}
\ea
\ee
The axio-dilaton $\tau$ is a holomorphic function on $\widetilde{\M}_{6}$, and when it is constant, the above conditions consistently reduce to those in \cite{Kim:2005ez}. As shown in the previous subsection these conditions are necessary and sufficient for the existence of a supersymmetric solution.

\subsection{F-theoretic Formulation}\label{F-theory reformulation}

The condition on the curvature and axio-dilaton  (\ref{Master}) has again a nice geometrised form which will allow a re-interpretation  of the Type IIB supergravity equations with varying $\tau$ in terms of an F-theory model, where the axio-dilaton $\tau$ is identified with the complex structure of an elliptic curve. The varying of the complex structure, which is compatible with the $SL_2\mathbb{Z}$ duality group action on Type IIB string theory, is then encoded in a geometric elliptic fibration in a putative 12d space. 

The geometry that incorporates the axio-dilaton in terms of an elliptic fibration over the Type IIB spacetime $\widetilde{\M}_{6}$ is a K\"ahler four-fold, with metric 
\be\label{F-theory metric}
\dd s^{2}(\mathcal{Y}_{8}^{\tau})=\frac{1}{\tau_{2}}\lb (\dd x +\tau_{1}\dd y)^{2}+\tau_{2}^{2}\dd y^{2}\rb +\dd s^{2}(\widetilde{\M}_{6})~,
\ee
whose Ricci-form is written in terms of that of $\widetilde{\M}_{6}$,  $\mathfrak{R}^{(\tM)}$, as 
\be
\mathfrak{R}^{(\mathcal{Y})}=\mathfrak{R}^{(\tM)}-\ii P\wedge P^{*}~.
\ee
It is clear from this expression that the Ricci-form has legs only along $\widetilde{\M}_{6}$ and therefore
\begin{align}
R^{(\mathcal{Y})}_{\mu\nu}&=R^{(\tM)}_{\mu\nu}-2 P_{(\mu}P^{*}_{\nu)}~,\\
R^{(\mathcal{Y})}&=R^{(\tM)}-2|P|^{2}~.
\end{align}
Using the above expressions in \eqref{Master} and that the coordinates of the auxiliary elliptic fibration generate Killing directions of the full solution we find
\be
\ba\label{Master8d}
0&=\square_{\tM}(R^{(\tM)}-2|P|^{2})-\frac{1}{2}(R^{(\tM)})^2+R^{(\tM)}_{\mu\nu}R^{(\tM)\mu\nu}+2 |P|^{2}R^{(\tM)}-4 R^{(\tM)}_{\mu\nu}P^{\mu}P^{*\nu}\cr 
&=\square_{\mathcal{Y}} R^{(\mathcal{Y})}-\frac{1}{2} (R^{(\mathcal{Y})})^2+R^{(\mathcal{Y})}_{ij}R^{(\mathcal{Y})\, ij}~.
\ea\ee
This is the ``master equation'' presented in \cite{Kim:2005ez} in two more dimensions. Solving \eqref{Master} is equivalent to solving \eqref{Master8d} and imposing that the 8d K\"ahler metric for $\mathcal{Y}_8^{\tau}$ is elliptically fibered. The condition is thus {\it not} that this space is Calabi-Yau, but a more refined condition, which only in special cases will be shown to reduce to containing Ricci-flat elliptic fibrations. 
Alternatively, the geometry may also be specified in terms of the metric on $\mathcal{Y}_8^{\tau}$ using \eqref{warpdef} and \eqref{rhodef} as 
\be\label{warp8d}
R^{(\mathcal{Y})}=8\me^{-4 \Delta}~, \qquad 
\dd \rho=- \mathfrak{R}^{(\mathcal{Y})} \,   . 
\ee
Note that solutions to this equation will  also automatically give rise to supersymmetric solutions of eleven dimensional supergravity of the form $\AdS_2\times \M_9$  \cite{Kim:2006qu}, 
 where $\M_9$ is locally a circle fibration over $\mathcal{Y}_8^{\tau}$. We thus obtain a 1--1 correspondence of F-theory  AdS$_3$ solutions and elliptically fibered M-theory $\AdS_2$ solutions. We shall {discuss} this point later in section \ref{sec:M}.

\section{$\AdS_3$ with 2d $\mathcal{N}=(2,2)$ and $\AdS_5$ with Varying $\tau$} 
\label{sec:22}

Before entering an extensive analysis of new solutions with $\mathcal{N}=(0,2)$ supersymmetry, it is worthwhile singling out the special case of $\mathcal{N}=(2,2)$ supersymmetry.  Again we consider only five-form flux in the present setup, and analyse the general torsion conditions on the geometry. There are two main conclusions:
the first is that there are no $\AdS_3$ solutions which preserve $(2,2)$ and have non-constant axio-dilaton. We again provide the constraints on the geometry, and show how known solutions for constant axio-dilaton $\tau$ such as $\AdS_3 \times S^3 \times CY_2$ are recovered from this in appendix \ref{app:Recover22}. 
Interestingly, requiring the axio-dilaton to vary, implies that the solution is non-compact, and in fact becomes $\AdS_5$. These solutions are in fact the most general solutions of this type, and we conclude this section by showing the classification of the constraints on Type IIB $\AdS_5$ solutions with varying axio-dilaton and vanishing three-form fluxes, preserving $\mathcal{N}=1$ supersymmetry in the dual 4d gauge theory.



 \subsection{Torsion Conditions}
 
The torsion conditions for $\mathcal{N}=(2,2)$ supersymmetry may be extracted from \cite{Couzens:2017way} by specialising the $\alpha$ parameters to be $\alpha_{1}=-\alpha_{2}=1$ for the two spinors $\xi_{i}$. The existence of two non-vanishing Dirac Killing spinors on $\M_{7}$ implies that $\M_{7}$ supports an $SU(2)$ structure. In 7d an $SU(2)$ structure is determined by three independent vectors which specify a dreibein for a 3d space $\M_3$ and a transverse 4d space $\M_{4}$, admitting a canonical $SU(2)$ structure
\be
\M_{7} =  \M_{3}\rtimes\M_{4} \, .
\ee
 In 4d an $SU(2)$ structure is determined by the existence of a real two-form, $j$ of maximal rank and a holomorphic two-form $\omega$, satisfying the algebraic conditions
\begin{align}
j \wedge \omega&=0~,\\
\omega\wedge \bar{\omega}&=2 j \wedge j~.
\end{align}
From \eqref{dilalg} and \eqref{gravalg} we may find various algebraic conditions that the bilinears must satisfy.  From \eqref{gravalg} we have
\be
(\alphai_{i}+\alphai_{j})A_{ij}=0 \quad \Rightarrow \quad A_{11}=A_{22}=0 \,   .
\ee
From \eqref{dilalg} we find
\be
A_{ij}^{*}P=0~,\label{APeq}
\ee 
and therefore for $\tau$ to vary we require $A_{12}=0$. We may then split the cases into those with varying $\tau$ and those where $\tau$ is fixed to be constant or equivalently to the cases of vanishing $A_{12}$ or non-trivial $A_{12}$ respectively. Using the results of \cite{Couzens:2017way} we see that both $K_{11}$ and $K_{22}$ are Killing vectors. In addition one finds from \eqref{gravalg} the two equations
\begin{align}
i_{K_{ij}}\dd \Delta&= -\frac{\ii m}{2}(\alphai_{i}-\alphai_{j}) S_{ij}~,\label{iKijdDelta}\\
S_{ij}\dd \Delta +\frac{\ii m}{2}(\alphai_{i}-\alphai_{j})K_{ij}&=\frac{\me^{-4 \Delta}}{4} i_{K_{ij}}F~,\label{iKijF}
\end{align}
which may be used to show that the vectors $K_{11}$ and $K_{22}$ are also symmetries of both the warp factor and flux. They correspond to the left and right moving R-current in the putative dual SCFT. In all cases the scalars $S_{11}$ and $S_{22}$ are constant and the spinors may be normalised such that both of these scalars are unity. This concludes the general analysis, and we must now specialise to one of the two cases. 
Requiring that the solution space transverse to the $\AdS_3$ is compact implies that $\tau$ is constant. This case will be discussed in section \ref{sec:AdS322}, supplmented with details in appendix \ref{app:22}.  
If we relax the compactness condition then we find $\AdS_5$ with varying $\tau$. This allows us to classify all F-theoretic $\AdS_5$ solutions in Type IIB with five-form flux in section \ref{sec:AdS5}.


\subsection{Constant $\tau$: $\AdS_3$ Duals to $\mathcal{N}=(2,2)$}
\label{sec:AdS322}

In this section we provide the necessary conditions for the existence of a compact internal manifold that allows for 2d $(2,2)$ supersymmetry, further discussion can be found in appendix \ref{app:22} where the conditions are derived and known solutions in the literature are recovered. The analysis of the torsion conditions shows that for a compact internal space constant $\tau$ is a necessary condition. In this section we consider the case where the scalar bilinear $A_{12}$ is non-trivial. The conditions for the existence of a solution are reminiscent of the conditions found in \cite{Gauntlett:2004zh} for AdS$_5$ solutions in M-theory.  
Locally the internal metric takes the form
\be
m^{2}\dd s^{2}( \M_{7})=  (1- y\me^{-4 \Delta}) (\dd \psi_{1}+\sigma_{1})^{2}+ y \me^{-4 \Delta} \dd \psi_{2}^{2}+\frac{\me^{-4 \Delta}}{4y(1- y \me^{-4 \Delta})}\dd y^{2}+ \me^{-4 \Delta} g^{(4)}(y,x)_{ij}\dd x^{i}\dd x^{j} \,,\label{22metric}
\ee
where both $\psi_{1}$ and $\psi_{2}$ are Killing vectors and generate the expected $U(1)\times U(1)$ symmetry that is dual to the R-symmetry on the field theory side.\footnote{As explained in appendix \ref{app:22} the Killing directions dual to the left and right moving R-symmetries are linear combinations of these two Killing vectors, they correspond to the diagonal and anti-diagonal.} For fixed $y$ the metric $g^{(4)}$ is K\"ahler with K\"ahler form $J_4$ satisfying
\be\label{final12}
\partial_{y}J_{4}=\frac{1}{2}\dd_{4}\sigma_{1}\,,\qquad 
\partial_{y}\log \sqrt{g} =-\frac{4 y \me^{-4 \Delta}}{1- y \me^{-4 \Delta}} \partial_{y}\Delta~\, ,\ee
where
\be\ba\label{Fluxdef}
\sigma_{1}&= -\widehat{P}_{4} +\frac{2y \me^{-4 \Delta}}{1- y \me^{-4 \Delta}} \dd_{4}^{c}\Delta \cr 
m \ff &=-\lb \frac{}{}  (\me^{4 \Delta}-y) \dd \sigma_{1}+2 J_{4} +4 \me^{4 \Delta} \dd \Delta \wedge (\dd \psi_{1}+\sigma_{1})\rb   \,. 
\ea\ee
 $\widehat{P}_{4}$ is the Ricci-form potential for the metric $g^{(4)}$. The complex structure $\tensor{J}{_{i}^{j}}$ is independent of $y$ and this allows us to rewrite \eqref{final12} as
\be
(\dd_{4}\sigma_{1})^{+}=-\frac{4 y \me^{-4 \Delta}}{1- y \me^{-4 \Delta}}\partial_{y}\Delta J_{4}\label{d4sigma}~,
\ee
where $(\dd_{4}\sigma_{1})^{+}$ is the self-dual part of the two-form $\dd_{4} \sigma_{1}$. 

Examples of geometries of this type were found in \cite{Kim:2005ez,Benini:2013cda} but is is likely that there exist other interesting solutions. Since the case of constant axio-dilaton is not the focus of this paper, we leave a 
further analysis of these solutions and their duals  for the future. We shall proceed in the next section with the analysis  of non-trivially varying $\tau$ by relaxing the condition of compactness of the internal 7d manifold.


\subsection{Varying $\tau$: $\AdS_5$ Duals to 4d $\mathcal{N}=1$}
\label{sec:AdS5}

In the previous section the requirement for 2d $\mathcal{N}=(2,2)$ supersymmetry and compactness of the solution led to the result that $\tau$ is constant. We shall now relax the latter condition and find that there are non-trivial varying $\tau$ solutions which are $\AdS_5$ duals to $\mathcal{N}=1$ in 4d. In the following we will provide the derivation of this starting from the present setup of $\AdS_3$ solutions. We supplement this with an analysis from a direct $\AdS_5 \times \M_5^\tau$ ansatz which shows that these are in fact all such varying $\tau$ $\AdS_5$ solutions with five-form flux.

The details of the derivation are provided in appendix \ref{app:AdS5}. The solution is given in terms of the metric 
\be\ba
\dd s^{2}=&~\me^{2 \Delta}\lb \dd s^{2}(\text{AdS}_{3})+\frac{\me^{-2 \Delta}}{m^{2}(1-\me^{-2\Delta})}\dd \Delta^{2}+\frac{ 4(1-\me^{-2 \Delta})}{m^{2}}\dd \varphi^{2}\rb +\frac{1}{m^{2}}\left [ \frac{}{} \lb \dd \psi+ \sigma\rb^{2}+\dd s^{2}(\widetilde{\M}_{4})\right]\cr 
=& ~\dd s^{2}(\text{AdS}_{5}) +\frac{1}{m^{2}}\left [ \frac{}{} \lb \dd \psi+ \sigma\rb^{2}+\dd s^{2}(\widetilde{\M}_{4})\right] \, ,\label{AdS5metric}
\ea\ee
where $\widetilde{\M}_4$ is a K\"ahler surface. The axio-dilaton varies holomorphically over $\widetilde{\M}_{4}$ {and obeys the following curvature condition
\be
\mathfrak{R}_{4}=6 J_4- \dd Q~.\label{Ricciformeqqe}
\ee
In particular, to find solutions we should solve this equation. Notice that for constant $\tau$ this reduces to the K\"ahler--Einstein condition.}

As an F-theory background, this can written as
\begin{align}
\dd s^{2}  & =   \dd s^{2}(\text{AdS}_{5}) +   \dd s^{2}(\M_7^\tau)   \nonumber\\
 & =  \dd s^{2}(\text{AdS}_{5}) +\frac{1}{m^2}\lb \dd \psi+ \sigma\rb^{2}+\dd s^{2}(\widetilde{\M}_{4})+ \frac{1}{\tau_{2}}(\dd x +\tau_{1}\dd y)^{2}+\tau_{2} \dd y^{2}~,
\end{align}
where $ S^1 \hookrightarrow \M_7^\tau  \rightarrow \mathcal{T}^\tau_{6}$, with the elliptically fibered three-fold $\mathbb{E}_\tau  \hookrightarrow \mathcal{T}^\tau_{6}\rightarrow \widetilde{\mathcal{M}}_4$, which is \emph{not Calabi-Yau}.
There is however a nice reformulation of these solutions in term of  an elliptically fibered Calabi--Yau four-fold.
The {compact part of the} geometry has an obvious {relation} with the metrics on Sasaki--Einstein solutions and in fact 
may be shown to be the link of the conical base of an elliptically fibered Calabi--Yau four-fold. Specifically, that  the metric 
\be
\dd s^{2}(\Y_4)=\frac{1}{\tau_{2}}(\dd x +\tau_{1}\dd y)^{2}+\tau_{2} \dd y^{2}+\dd r^{2}+r^{2}\lb \lb \dd \psi+ \sigma\rb^{2}+\dd s^{2}(\widetilde{\M}_{4})\rb
\ee
is both Ricci-flat and K\"ahler, where the elliptic fiber varies over the K\"ahler manifold $\widetilde{\M}_{4}$. 

For constant $\tau$ the fibration is trivial and we reduce to the usual Sasaki--Einstein solutions, which can be written as the link of a Calabi--Yau three-fold cone. Including varying $\tau$ the solution remains Sasakian, however the Calabi--Yau condition {of the 6d cone} is now relaxed. In fact as we briefly show in appendix \ref{app:AdS5}, this set of solutions is the most general with $\mathcal{N}=1$ supersymmetry in the dual 4d theory, with five-form flux and vanishing three-form fluxes. A more detailed study of these solutions {and their holographic interpretation} will appear in \cite{toappearAdS}.

The starting point in appendix \ref{app:AdS5} is the ansatz $\AdS_5 \times {\M}_5$ for the geometry, now allowing  $\tau$ to vary. 
We rename the K\"ahler form on the transverse space $\widetilde{\M}_{4}$ to be $J_{T}$, and the holomorphic two-form $\Omega_{T}$. Furthermore we introduce both a K\"ahler two-form and maximal holomorphic form in complex dimension $3$ and $4$ and we shall distinguish them by labeling with their complex dimension.
Let us first construct the conical metric $C(\M_5^\tau)=\Y_6$ 
\be
\dd s^{2}(\Y_6)=\dd r^{2}+ r^{2} \lb  (\dd \psi +\sigma)^{2}+\dd s^{2}(\widetilde{\M}_{4})\rb~.
\ee
The  K\"ahler two-form is
\bea
J_{3}&=& r \dd r \wedge (\dd \psi+\sigma) + r^{2}J_{T}~,
\eea
and from \eqref{dsigma} it follows that it is closed $\dd J_{3}= 0$.
We have proven our first assertion that the solution is Sasakian. We next check the Ricci tensor of the solution. As the metric is complex this is most easily achieved by computing the exterior derivative of the holomorphic three-form of the cone,
\bea
\Omega_{3}&=& r^{2}(\dd r +\ii  r K)\wedge \Omega_{T}~.
\eea
Using \eqref{Omegapsi} results in
$\dd \Omega_{3}= -\ii Q \wedge \Omega_{3}$, from which we may abstract the Ricci form of the  K\"ahler cone to be
\bea
\mathfrak{R}_{Y}&=& -\dd Q~.
\eea
Recall that $\tau$ varies holomorphically over $\widetilde{\M}_{4}$,  so that $\mathfrak{R}_Y$ can be reinterpreted as the Ricci form for the base $\Y_6$ of an elliptically fibered Calabi--Yau four-fold, $X_{F}$, 
\be\label{CY4Metric}
\dd s^{2}(X_{F})=\frac{1}{\tau_{2}}\lb (\dd x +\tau_{1}\dd y)^{2}+\tau_{2}^{2}\dd y^{2}\rb+\dd r^{2}+ r^{2} \lb  (\dd \psi +\sigma)^{2}+\dd s^{2}({\widetilde{\M}_{4}})\rb~.
\ee
For trivial $\tau$ this reduces to the Sasaki--Einstein case as necessary. Notice firstly that the Calabi--Yau is non-compact and that the elliptic fibration depends only on $\widetilde{\M}_{4}$. Moreover the Reeb vector is fibered over $\widetilde{\M}_{4}$.
The full solution is 
\begin{align}
\dd s^{2}&=\dd s(\text{AdS}_{5})+\frac{1}{m^{2}}\lb(\dd\psi+\sigma)^{2}+\dd s^{2}(\widetilde{\M}_{4})\rb\nonumber\\
&=\frac{1}{m^{2}}\lb r^{2}\dd s^{2}(\mathbb{R}^{1,3})+\frac{1}{r^{2}}\dd s^{2}(\Y_6)\rb
\end{align}
with self-dual five-form flux
\be
F=4m (\dd\vol(\text{AdS}_{5})+\dd\vol(\M_5^\tau))~, 
\ee
where 
\be
\dd s^{2}(\Y_6)=\dd r^{2}+r^{2} \lb  (\dd \psi +\sigma)^{2}+\dd s^{2}(\widetilde{\M}_{4})\rb
\ee
is the metric on the K\"ahler three-fold which is the base of an elliptically fibered Calabi-Yau four-fold with elliptic fiber varying over the K\"ahler manifold $\widetilde{\M}_{4}\subset X$. 

{Of course} the {five-form} flux needs to be quantized through  {the unique five-cycle in the ten-dimensional geometry} i.e. we must impose 
\be
\frac{1}{(2\pi \ls)^4 \gs}\int_{\M_5^\tau}F\in \mathbb{Z} \,,
\ee
which with the above form for the flux results in
\be
\frac{1}{(2\pi \ls)^4 \gs} \int _{\M_5^\tau} \frac{4}{m^4}\dd \vol(\M_5^\tau)= \frac{4 \vol(\M_5^\tau)}{(2 \pi m \ls)^4 \gs}=N~.\label{NAdS5}
\ee
The integer $N$ is interpreted as the number of D3-branes {as usual}.
From this it follows straightforwardly that the leading order holographic central charge {of the dual 4d SCFT} is given by
\be\label{4dcentral}
a^{4d}= \frac{\pi}{8G_{N}^{(10)}}\int_{\M_5^\tau}\me^{3 \Delta} \dd \vol(\M_5^\tau)= \frac{N^2 \pi^3}{4 \vol(\M_5^\tau)}  \,,
\ee
{exactly as in the constant $\tau$, Sasaki--Einstein case.} 

However, due to the relation (\ref{Ricciformeqqe}), the volume receives corrections with respect to the constant $\tau$ case. In particular, using    $6 J_4 = \dd Q+ \mathfrak{R}_{4}$ 
, we have 
\be
  \vol(\M_5^\tau) = \int_{\M_5^\tau}  (\diff \psi +\sigma) \wedge \frac{J_4\wedge J_4}{2}  
   =  \frac{\pi^3 \ell}{9} \int_{\M_{4}}       c_{1}(\M_{4})^2 -2 c_{1}(\M_{4})\wedge  c_{1}(\mathcal{L}_{D})+c_{1}(\mathcal{L}_{D})^2~,  
\ee
where the first term is the result of the volume for quasi-regular Sasaki--Einstein manifolds.

These solutions and their dual field theories will be further investigated in \cite{toappearAdS}. 
In the present paper they will make a re-appearance in the so-called universal twist solutions in section \ref{sec:Universal02}, which are  holographic dual of topologically twisted compactifications of the 4d $\mathcal{N}=1$ SCFTs dual to the above $\AdS_5$ solutions.

We conclude that that there are no  compact AdS$_{3}$ solutions with varying axio-dilaton dual to $(2,2)$ SCFTs, and this is supported by the absence of any non-chiral field theories from wrapped D3-branes in F-theory \cite{Lawrie:2016axq}. Nevertheless this analysis has led us to the exciting direction of $\AdS_5$ solutions in F-theory. 


\section{New $\cN=(0,2)$ Solutions with Varying $\tau$}
\label{sec:NewSol}

In this section we  turn to exploring solutions to the ``master equation'' \eqref{Master} or equivalently \eqref{Master8d}, for duals to 2d (0,2) SCFTS,
which incorporates a varying Type IIB axio-dilaton in terms 
of an elliptically fibered {local} K\"ahler four-fold $\mathcal{Y}_8^{\tau}$.
We will study two new classes of solutions, which result from different specialisations of the K\"ahler four-fold.

The first type of solution is a specialisation of the F-theoretic reformulation in section \ref{F-theory reformulation}, where $\tM_{6}$ is a direct product of a complex curve and a complex surface, 
\be\label{642}
\tM_6 = \Sigma \times \M_4 \,, 
\ee
such that the elliptic fibration is only non-trivial over one of these subspaces, i.e. there are two cases 
\be\label{TwoUniCases}
\ba
\hbox{Elliptic Surface:} \qquad &
\mathcal{Y}_8^\tau  = (\mathbb{E}_\tau \rightarrow \Sigma) \times \M_ 4  = \mathcal{S}^\tau_{4} \times \M_4  \cr 
\hbox{Elliptic Three-fold:}\qquad &\mathcal{Y}_8^\tau = 
\Sigma \times (\mathbb{E}_\tau \rightarrow  \M_ 4) = \Sigma \times \mathcal{T}^\tau_{6} \,.
\ea
\ee
where none of the factors has a  Ricci-flat metric.
This class will correspond in the dual field theory to ``universal twist solutions'', which generalise to varying $\tau$ the universal twist solutions in \cite{Benini:2015bwz}, that were originally found in \cite{Gauntlett:2006qw}. 
We will see that they are dimensional reductions with topological duality twist of  4d $\mathcal{N}=1$ SCFTs {with rational R charges}, with varying coupling. In this class of solutions we do not assume that the elliptic fibration over $\Sigma$ or $\M_4$ are Ricci-flat. In fact the ``master equation'' implies that they are not. These solutions will be studied in section \ref{sec:Universal02}. 

Another class of solutions can be obtained by a similar splitting, however we now require that the factor with the non-trivial elliptic fibration is Ricci-flat, i.e. has a Calabi-Yau $(4-s)$-fold factor {$\Y^\tau_{2(4-s)}$}
\be\label{M8ansatz}
\mathcal{Y}_{8}^\tau = \Y^\tau_{2(4-s)}\times \M_{2s}\,.
\ee
Clearly $\M_{2s}$ has to be K\"ahler as well and only the  values $s=1,2$ are interesting\footnote{Note that $s=0$ is ruled out because the Ricci scalar of $\mathcal{Y}_{8}^\tau$, and therefore also the warp factor $\me^{-4\Delta}$, vanishes in this case. $s=3$ corresponds to $\mathcal{Y}_{8}= T^2 \times \M_{6}$, which has constant axio-dilaton  \cite{Kim:2005ez}.}. 
Inserting the direct product metric into (\ref{Master8d}) one immediately finds that the K\"ahler  metric on  $\M_{2s}$ must again obey the same equation originally found in \cite{Kim:2005ez}, namely
\be
\label{Master2s}
\square_{\M_{2s}}  R^{(\M_{2s})}-\frac{1}{2} R^{(\M_{2s})2}+R^{(\M_{2s})}_{ij}R^{(\M_{2s})ij} =0~.
\ee
We shall first consider the case when $s=1$ where $\mathcal{Y}_8^\tau$ is the direct product of an elliptically fibered Calabi--Yau three-fold and a Riemann surface before considering the $s=2$ case. As we shall show the former recovers the $(0,4)$ solutions determined in \cite{Couzens:2017way} whilst the latter gives rise to a new class of strictly $(0,2)$ supersymmetic solutions. These solutions will be the subject of section \ref{sec:BT}.


\subsection{Universal Twist Solutions}
\label{sec:Universal02}

In this section we begin with the product ansatz in (\ref{642})
\be
\dd s^{2}(\widetilde{\M}_{6})= \dd s^{2}(\Sigma) +\dd s^{2} (\M_{4})\,,
\ee
where $\Sigma$ is a complex curve and $\M_{4}$ a K\"ahler surface. It is most convenient to express our ansatz in the reformulation of section \ref{F-theory reformulation}. The Ricci-form of the 8d space $\mathcal{Y}_{8}^\tau$, which is the elliptic fibration over $\widetilde{\mathcal{M}}_6$ is
\be
\mathfrak{R}_{\mathcal{Y}}= k_{1} J_{\M_4}+ k_{2} J_{\Sigma}\,,
\ee
where $k_{1}$ and $k_{2}$ are constants. {We consider} the two cases oulined in (\ref{TwoUniCases}):  $\tau$ varies non-trivially only over the curve $\Sigma$ giving an elliptic surface, or  $\tau$ varies non-trivially only over $\M_{4}$ giving an elliptic three-fold. Though the supergravity solutions are distinct much of the  analysis  will be similar, and therefore it will be useful to keep the {discussion} as general as possible. Inserting the above ansatz into the `master equation' \eqref{Master8d} the necessary condition is 
\be\label{k1k2cond}
k_{1}( k_{1}+2  k_{2})=0
\qquad \hbox{and} \qquad 
R_{\mathcal{Y}}= 4 k_{1}+2 k_{2}~.
\ee
Clearly to solve \eqref{k1k2cond} either $k_{1}=0$ or $k_{1}=-2 k_{2}$. The former recovers the $(0,4)$ solution discussed in \cite{Couzens:2017way}\footnote{We will recover these (0,4) solutions in a slightly different way in section \ref{sec:BT}.}. We therefore consider the latter solution in the remainder of this section. Evaluated on such a solution the Ricci scalar is $R_{\mathcal{Y}}=-6 k_{2}$ and thus the positivity constraint of the Ricci scalar implies that $k_{2}<0$. The overall scale of the K\"ahler metric on $\widetilde{\M}_{6}$ may be removed by a coordinate change, thus without loss of generality we may set $k_{2}$ to be any negative value, for convenience we choose $k_{2}=-3$.  The 10d solution in Einstein frame is
\begin{align}
\dd s^{2}&= \frac{2}{3}\lb \dd s^{2}(\text{AdS}_{3}) + \frac{9}{4}\lb \frac{1}{9}( \dd \chi +\rho)^{2}+ \dd s^{2}(\M_{4})+\dd s^{2}(\Sigma)\rb \rb~,\\
\me^{-4 \Delta}&= \frac{9}{4}~,\\
\ff&=-\frac{2}{3} ( 4 J_{\Sigma}+J_{\M_4})~,\\
F&= -\frac{3}{4} (\dd \chi+\rho)\wedge J_{\M_{4}}\wedge (2 J_{\M_{4}}+J_{\Sigma})-\frac{2}{3}\dd \vol (\text{AdS}_{3})\wedge (4 J_{\Sigma}+J_{\M_{4}})~,\\
\rho&=- 6\mathcal{A}_{\M_4}+3 \mathcal{A}_{\Sigma}~,\\
\dd \mathcal{A}_{i}&= J_{i}~.
\end{align}
The Ricci form on $\mathcal{Y}_{8}^\tau$ becomes $\mathfrak{R}_{\mathcal{Y}}= 6 J_{\M_4}- 3 J_{\Sigma}$, in matrix block form this is
\be\label{blocchi}
\mathfrak{R}_{\mathcal{Y}} = 
\left(\begin{array}{c|c|c}
- 3 J_{\Sigma}  & & \\\hline 
&  0 & \\\hline
& & \ 6 J_{\M_4} \ 
\end{array}\right) \,.
\ee
In the above we have not specified over which factor in $\widetilde{\M}_{6}$,  $\tau$ varies non-trivially. In the following we shall consider the two cases in which $\tau$ varies non-trivially only over $\Sigma$, giving an elliptic surface, or over $\M_{4}$, giving an elliptic three-fold. We are not aware of any existence results for metrics on either the elliptic surface or the elliptic three-fold with the specific conditions imposed on the curvatures (in particular let us re-emphasise that these are not Ricci-flat). We will assume that such metrics exist on these spaces with the Ricci-form given as above. It would indeed be of great interest to develop the mathematics that shows the existence of such metrics. Of course the bases of these elliptic fibrations will have singularities at points where $\tau$ becomes singular, but by assumption they will be otherwise smooth. In the following sections we analyse the two distinct types of solutions discussed above. {The consistency of the holographic computations using these solutions with the proposed field theory duals corroborates our conjecture that these metrics exist.}


\subsubsection{Elliptic Surface Case}\label{sol1deriv}

Let us first consider the case where $\tau$ varies non-trivially only over $\Sigma$. 
{We require the metric on $\mathcal{Y}_{8}^\tau$  to factorise as 
\be
\dd s^2 (\mathcal{Y}_{8}^\tau) = \dd s^2 (\Xtwo) + \dd s^2 (\M_4) \,,
\ee
where $\mathbb{E}_\tau \hookrightarrow \mathcal{S}_4^\tau \rightarrow \Sigma$ is an elliptic surface with section, over $\Sigma$.
The Ricci curvature then factorises into two $4\times 4$ blocks, and (\ref{blocchi}) reads
\be\label{blocchidue}
\left(
\begin{array}{cc|cc}
\mathfrak{R}_{\Xtwo} && & \\\hline
& && \mathfrak{R}_{\M_4}  \\
\end{array}
\right)
= 
\left(
\begin{array}{cc|c|cc}
- 3 J_{\Sigma}  &&& & \\\hline
& & 0& & \\\hline 
& &&& 6 J_{\M_4} \\
\end{array}
\right) \,.
\ee
To solve this equation we therefore have that the metric on $\M_{4}$ is K\"ahler--Einstein with Ricci-form $\mathfrak{R}_{\M_{4}}= 6 J_{\M_{4}}$, and 
we require the existence of a metric on the elliptic surface  $\Xtwo$ to satisfy 
\be
\mathfrak{R}_{\Xtwo}= -3 J_{\Sigma}\quad \Longleftrightarrow \quad \mathfrak{R}_{\Sigma}  +\dd Q =-3 J_{\Sigma} ~.\label{RSigma}
\ee}
{Notice that the K\"ahler--Einstein metric  on $\M_{4}$ has the  normalisation of the  base of a Sasaki--Einstein manifold. In fact the one form dual to the Reeb vector field of the Sasaki--Einstein manifold is given by $-\frac{1}{3}(\dd\chi +\rho)$ at fixed coordinate on $\Sigma$. We conclude that at fixed coordinate on $\Sigma$ the $U(1)$ fibration over $\M_{4}$ is a (quasi-regular) Sasaki--Einstein manifold. }

 Solutions of this form, where $\Sigma$ is the constant curvature Riemann surface $\mathbb{H}^{2}$ have been studied in \cite{Gauntlett:2006qw}, however there are some  differences once $\tau$ is allowed to vary non-trivially over $\Sigma$. Topologically the 7d internal {space} is a $U(1)$ fibration over $\M_{4}\times \Sigma$. Such fibrations are regular if the first Chern class of the bundle is integral over all two-cycles in $H_{2}(\M_{4}\times \Sigma,\Z)$. Let the period of $\chi$ be $2 \pi \ell$, then for a regular $U(1)$ fibration we require 
\be
{1\over 2 \pi} \frac{1}{\ell} \dd \rho  = {1\over  2 \pi \ell} (- 6 J_{\mathcal{M}_4} + 3 J_{\Sigma})\in H^{2}(\M_{4}\times \Sigma, \Z)~.
\ee
This may be rephrased  in terms of the elliptic surface $\mathcal{S}^\tau_4$ with base $\Sigma$ as
\be
c_{1}(U(1))=-\frac{1}{\ell} ( c_{1}(\M_{4})   + c_1 (\mathcal{S}_4^\tau)|_\Sigma)\in H^{2}(\M_{4}\times \Sigma, \Z)~.
\ee
Notice that the non-trivial elliptic fibration implies that the quantisation condition differs to that in \cite{Gauntlett:2006qw}. Concretely we have used the first Chern class of the elliptic surface $\mathcal{S}^\tau_4$ to rewrite the condition on $c_{1}(U(1))$.
 A convenient basis for $H_{2}(\widetilde{M}_{6})$ is furnished by the set $\{ \Sigma, \Sigma_{\alpha}\}$ where $\{\Sigma_{\alpha}\}$ is a basis of $H_{2}(\M_{4},\Z)$. Then  $c_{1}(U(1))$ being integer implies
\be\ba
\frac{1}{2\pi}\int_{\Sigma} c_{1}(U(1))&= -\frac{1}{\ell} (2(g-1)+\text{deg}(\mathcal{L}_{D}) )\in \Z \cr 
\frac{1}{2\pi} \int _{\Sigma_{a}} c_{1}(U(1)) &=- \frac{\tilde{m} n_{\alpha}}{\ell} \in \mathbb{Z}\,,
\ea\ee
where $\tilde{m}$ is the Fano-index of $\M_{4}$, see appendix B of \cite{Gauntlett:2006qw} for a  review of properties of 4d  K\"ahler--Einstein spaces, and $n_{\alpha}$ are relatively prime integers. The period $\ell$ of $\chi$ must be a divisor of both $\tilde{m}$ and $(2(g-1)+\text{deg}(\mathcal{L}_{D}))$ and consequently it has maximal value 
$\ell = \text{gcd}\{ \tilde{m},(2(g-1)+\text{deg}(\mathcal{L}_{D}))\}$. Recall that this construction only works for the regular and quasi-regular Sasaki--Einstein metrics.

\subsubsection*{Flux Quantisation}

The cycles of interest are the compact five-cycles of the geometry, of which there are two classes. The first is the five-cycle given at fixed $\Sigma$ coordinates, which is a Sasaki--Einstein (SE) manifold. The second class of five-cycles, {which we denote  $D_\alpha$}, are obtained as $U(1)$ fibrations over  $\Sigma_\alpha\times \Sigma$, where  $\Sigma_{\alpha}\in H_{2}(\M_{4},\Z)$. For the former we find
\begin{align}
N(\hbox{SE}_{5}) = \frac{1}{(2 \pi \ls)^4 \gs}\int_{SE_{5}} F 
= \frac{9}{(2 \pi \ls m)^4 \gs}\vol({SE}_{5})~,\label{universalNquant}
\end{align}
where the volumes are computed with  the canonical Sasaki--Einstein metrics, which have Ricci-tensor satisfying $R_{\mu\nu}=4 g_{\mu\nu}$. As it is necessary for the fibration to be quasi-regular we may rewrite this quantisation condition as
\be
N(SE_{5})=\frac{\ell M }{ 2^4 3 \pi m^{4}\ls^{4}\gs}\in \Z~,
\ee
where the integer $M$ is the topological invariant 
\be
M=\int_{\M_{4}} c_1 (\M_4)\wedge c_1(\M_4)~.
\ee
For the five-cycles $D_\alpha$ the condition is
\begin{align}
N( D_{\alpha})&=-\frac{\ell \tilde{m} n_{\alpha} (2(g-1)+\text{deg}(\mathcal{L}_{D}))}{2^4 3 \pi m^4 \ls^4 \gs}\in \Z~.
\end{align}
Quantisation of the flux such that the above integers are minimal implies
\be
n=\frac{\tilde{m}\ell h}{2^4 3 \pi m^4\ls^{4} \gs}, \qquad 
h= \hbox{gcd}\lb \frac{M}{\tilde{m}}, 2(g-1)+\text{deg}(\mathcal{L}_{D})\rb \,,
\ee 
from which we obtain
\be
N(SE_{5})\equiv N = \frac{n M}{\tilde{m}h}~,~~~~N(D_{\alpha})= \frac{n n_{\alpha}}{h} (2(g-1)+\text{deg}(\mathcal{L}_{D}))~.\label{QuantSigma}
\ee
In comparing with the field theory results we shall identify the integer $N$ as the number of D3-branes in  the setup.
Notice that the above analysis is a generalisation to that performed in \cite{Gauntlett:2006qw}, corresponding to $\text{deg}(\mathcal{L}_{D})=0$.

\subsubsection{Elliptic Three-fold Case}

\label{sec:ThreeFold}

{Consider now the case where $\tau$ varies non-trivially only over $\M_{4}$,
 so that  the metric on $\mathcal{Y}_{8}^\tau$   factorises as
\be
\dd s^2 (\mathcal{Y}_{8}^\tau) = \dd s^2 (\Sigma) + \dd s^2 (\mathcal{T}_{6}^\tau)~,
\ee
where $\mathbb{E}_\tau  \hookrightarrow \mathcal{T}_{6}^\tau \rightarrow \M_4$ is the elliptic three-fold.
The  Ricci curvature of this metric now factorises in one $2\times 2$ block and one $6\times 6 $ block, and (\ref{blocchi}) reads
\be\label{blocchitre}
\left(
\begin{array}{cc|cc}
\mathfrak{R}_{\Sigma} & && \\\hline
& && \mathfrak{R}_{\mathcal{T}^\tau_6}  \\
\end{array}
\right)
= 
\left(
\begin{array}{cc|c|cc}
- 3 J_{\Sigma}  &&& & \\\hline
&  &0& & \\\hline
& & && \ 6 J_{\M_4} \ \\
\end{array}
\right) \,.
\ee
The upper block of this equation implies that the metric on the Riemann surface has constant curvature $\mathfrak{R}_{\Sigma}=-3 J_{\Sigma}$.
We then require the existence of a metric on the elliptic three-fold  $\mathcal{T}^\tau_6$ to satisfy}
\be
    \mathfrak{R}_{\mathcal{T}^\tau_6}= 6 J_{\M_{4}}  \quad \Longleftrightarrow \quad\mathfrak{R}_{\M_{4}} +\dd Q=6 J_{\M_{4}}~.\label{M4tauRicci}
\ee
{In fact, the elliptic three-fold $\mathcal{T}^\tau_6$ is precisely that appeared in section  \ref{sec:AdS5}.} 
At fixed coordinates on $\Sigma$, the solutions can be obtained in the same way as the AdS$_5$ solutions discussed in section \ref{sec:AdS5} and will be studied in more detail in \cite{toappearAdS}. We nevertheless give a brief discussion on {global properties} of the solutions following the above.  Topologically the solution is again a $U(1)$ fibration over a K\"ahler base. Giving $\chi$ period $2 \pi \ell$ as before the first Chern class of the $U(1)$ bundle is
\be
c_{1}(U(1))=-\frac{1}{\ell}(c_{1}(\Sigma)+c_{1}( \mathcal{T}_{6}^\tau)|_{\mathcal{M}_4}) \,.
\ee
Using the same basis as previously we require
\be
\frac{1}{2\pi} \int_{\Sigma_{\alpha}}c_{1}(U(1))= \frac{1}{\ell}(c_{1}(\mathcal{T}^{\tau}_{6})\cdot \Sigma_{\alpha}) \in \Z~,~~~\frac{1}{2\pi}\int_{\Sigma} c_{1}(U(1))= \frac{\chi(\Sigma)}{\ell}\in \Z~.
\ee
Here $\chi(\Sigma)$ is the Euler number of the Riemann surface $\Sigma$. The period $\ell$ must divide both $\chi(\Sigma)$ and $c_{1}(\mathcal{T}^\tau_6)\cdot \Sigma_{\alpha}$ for all $\alpha$.


\subsubsection*{Flux Quantisation}

Recall that at fixed coordinates on the constant curvature Riemann surface $\Sigma$, the metric is no longer Einstein, though it remains Sasakian. We will refer to this space as $\M_5^\tau$ as it will be related to the $\M_5^\tau$ of section \ref{sec:AdS5}. The possible five-cycles are as before and we keep the same notation as in the previous quantisation condition. Then the quantisation condition is
\begin{align}\label{Nunitau}
N(\M_5^\tau)=\frac{1}{(2\pi \ls)^4 \gs}\int_{\M_5^\tau}F
=\frac{9}{(2 \pi m \ls)^4 \gs} \vol(\M_5^\tau) \, ,
\end{align}
which {has the same form} as for the the first class of solutions. We may rewrite the volume of $\M_5^\tau$ as
\be
\vol(\M_5^\tau)=\frac{1}{2}\int_{\M_5^\tau} \frac{1}{3} \dd \chi \wedge J_{\M_{4}}\wedge J_{\M_{4}}= \frac{\pi^3 \ell}{27} \int_{\M_{4}}\lb c_{1}(\M_{4})^2 -2 c_{1}(\M_{4}) c_{1}(\mathcal{L}_{D})+c_{1}(\mathcal{L}_{D})^2\rb \,,
\label{dfgrt}
\ee
where the integral on the right-hand side is an integer, given by the sum of three topological numbers, whose value we denote by $\widetilde{M}$. Then
\be
N(\M_5^\tau)= \frac{\ell \widetilde{M}}{2^4\cdot 3 \pi \ls^{4}m^{4}\gs}~.
\ee

The quantisation over the remaining five-cycles gives
\be
N(D_{\alpha})=\frac{ \chi(\Sigma) \ell}{2^{4}\cdot 3 \pi m^4 \ls^4 \gs} \lb \tilde{m} n_{\alpha} -\int_{\Sigma_{\alpha}}c_{1}(\mathcal{L}_{D})\rb~.
\ee
As before we impose that the fluxes are minimal integers through all integral cycles which implies the quantisation of the length scale $m$ as
\be
n= \frac{\ell h}{2^4 \cdot 3 \pi m^{4} \ls^{4} \gs}\,,\qquad 
h=\hbox{gcd}\left[ \tilde{M},\chi(\Sigma) gcd\lb\left\{\tilde{m}n_{\alpha}-\int_{\Sigma_{\alpha}}c_{1}(\mathcal{L}_{D})\right\}_{\alpha}\rb \right]~.
\ee
We have
\be\label{Quant3fold}
N(\M_5^\tau)\equiv N= \frac{\widetilde{M}n}{h}~,~~~ N(D_{\alpha})= \frac{\chi(\Sigma)n}{h}\lb \tilde{m}n_{\alpha}-\int_{\Sigma_{\alpha}}c_{1}(\mathcal{L}_{D})\rb~.
\ee


\subsection{Solutions with Calabi-Yau Factors}
\label{sec:BT}

We now consider the ansatz (\ref{M8ansatz}), with one of the factors in $\mathcal{Y}_8^\tau$ an elliptically fibered Calabi-Yau.

\subsubsection{Recovering the $(0,4)$ Solutions}
\label{sec:Recovery}

The case $s=1$, i.e. $\mathcal{Y}_8^\tau = \Y_6 \times \Sigma$, where $\Y_6$ is an elliptically fibered Calabi--Yau three-fold and $\Sigma$ is a complex curve recovers the 
 classifiaction of  $\mathcal{N}=(0,4)$ theories that we presented in \cite{Couzens:2017way}. The metric is
\be
\dd s^{2}(\mathcal{Y}_8^\tau)=  \dd s^{2}(\Y_6)+\dd s^{2}(\Sigma)~,
\ee
and as any Riemann surface is conformally flat  we may write the metric on $\M_2$ as 
\be
\dd s^{2}(\Sigma)= \me^{-2 f(x,y)}(\dd x^{2}+\dd y^{2})~.
\ee
A Riemann surface trivially satisfies $R^{2}= 2 R_{\mu\nu}R^{\mu\nu}$ and therefore \eqref{Master2s} reduces to
\be
\square_{\Sigma}R^{\Sigma}=0~.
\ee
On any smooth compact manifold any bounded harmonic function is constant and it follows that for a smooth and compact internal manifold we must have that $R^{\Sigma}$ is constant and therefore the Riemann surface is of constant curvature\footnote{Removing the smoothness assumption, there could exist further (0,2) solutions where $\Sigma$ has singularities. However, in \cite{Couzens:2017way} we did not make any global 
assumption and therefore those are indeed the most general solutions preserving (0,4) supersymmetry.}. For positive curvature, as is necessary by \eqref{warp8d}, the only possibility is a round two-sphere and it follows that the only solutions are of the form
\be
\dd s^2 = \dd s^{2}(\text{AdS}_{3})+\dd s^{2}(S^{3}/\Gamma)+\dd s^{2}(B_4)
\ee
where $B_4$ is the base of $\Y_{6}$, the elliptically fibered Calabi--Yau introduced above. This precisely reproduces the solutions discussed in \cite{Couzens:2017way}, where they were shown to be the unique $(0,4)$ solutions.


\subsubsection{Baryonic Twist Solutions}
\label{sec:BTSolution}

A new class of solutions with exactly $(0,2)$ supersymmetry can be obtained for $s=2$ in the ansatz (\ref{M8ansatz}), i.e. where the geometry consists of an elliptic K3 surface $\Y_4$ and a local K\"ahler surface $\M_4$ as factors
\be
\dd s^{2}(\mathcal{Y}_8^\tau)=  \dd s^{2}(\Y_4)+\dd s^{2}(\M_{4})~.
\ee
Any solution to the ``master equation'' (\ref{Master2s}) for the metric on $\M_{4}$ will furnish a solution with varying axio-dilaton. In fact, solutions have been found previously in the literature for $\M_{4}$,  \cite{Gauntlett:2006ns, Donos:2008ug} and in  the following section we shall discuss a particular example. We begin by writing the full local solution with varying axio-dilaton, and subsequently we will 
 investigate its global regularity, including  quantisation of the fluxes. The computations are very similar to those presented in 
\cite{Gauntlett:2006ns, Donos:2008ug} for the solutions with constant axio-dilaton.  However, we include the details below and in  appendix \ref{comoans} to be self-contained and highlight 
some subtle features, which were not emphasised before.

The solutions bear an uncanny resemblance to the five-dimensional $Y^{p,q}$  Sasaki--Einstein manifolds \cite{Gauntlett:2004yd}. 
Following the ideas in \cite{Benini:2015bwz},  this connection will be sharpened by a dual field theory discussion in section \ref{sec:FT}, where we will 
propose that the dual 2d SCFTs are obtained from a particular twisted compactification  of the $Y^{p,q}$ theories on a curve, with a varying coupling. 

The local metric in string frame is 
\be
\ba
\label{localmetric}\dd s^{2}_{IIB(SF)}&=&\frac{1}{\sqrt{a x \tau_{2}}}\left[ \dd s^{2}(\mathrm{AdS}_{3})+\left.\frac{1}{4m^{2}}\right( w[\dd \psi+g(x) D\phi]^{2}+4 a x \dd s^{2}(\Kthreebase)\right.\\
&&\left.\left.+a \lb \frac{\dd x^{2}}{x^{2}U}+\frac{U}{w}D\phi ^{2}+\dd\theta^{2} +\sin^{2}\theta \dd \chi^2\rb\rb \right]~,
\ea
\ee
with RR five-form flux
\bea
\label{localF5} F&=&-\frac{1}{m}\dd\vol(\mathrm{AdS}_{3})\wedge \lb \frac{1}{2 a x^{2}}(D \psi - g(x) D \phi)\wedge \dd x +2\dd \vol(\Kthreebase)+\frac{1}{2}\dd\vol(S^{2})\rb \nonumber\\
&&+\frac{a}{4 m^{4}}  D \psi \wedge D \phi \wedge \dd x \wedge \lb \dd \vol(\Kthreebase)+\frac{1}{4 x^{2}}\dd \vol(S^{2})\rb \nonumber\\
&&+\frac{a}{4 m^{4}}\dd \vol(\Kthreebase)\wedge \dd \vol(S^{2})\wedge \lb x D \psi -\frac{U(x)}{w(x)} D \phi\rb ~.
\eea
The axio-dilaton varies holomorphically over $B_2 = \mathbb{P}^1$, such that the total space of the elliptic fibration  $\Y_4$, 
 $
 \mathbb{E}_\tau \hookrightarrow \Y_4 \rightarrow B_2$, where the axio-dilaton parametrises the complex structure of the fiber, is a K3 surface. 
The warp factor is  $\me^{-4 \Delta (x)} =  a x$ and in the above expressions we have used the following definitions
\bea
U(x)&=&1-a(1-x)^{2}~,\nn\\
w (x) &=&1+a(2x-1)~,\nn\\
g (x) &=&-\frac{ax}{w(x)}~,\nn\\
D\phi &=& \dd \phi +\cos \theta \dd \chi~,\nn\\
D \psi &=& \dd \psi +g(x) D \phi~,
\eea
with $a$ an integration constant. {After performing the global regularity analysis, that we include in Appendix  \ref{app:Ypqstuffstuff}, one discovers that $a$ takes rational values, given in terms of two integers $\podd, \qodd$.} The resulting Type IIB solution takes the form 
\be
\AdS_3  \times \mathbb{P}^1 \times   \Yodd^{\podd,\qodd} \, , \qquad  \Yodd^{\podd,\qodd} = (S^1 \rightarrow \mathbb{F}_{0}) \,,
\ee
where $ \Yodd^{\podd,\qodd}$ is a circle fibration over $\mathbb{F}_{0}= S^2  \times S^2$, 
with Chern numbers $\podd$ and $\qodd$, respectively, that are related to the parameter $a$ as 
\be
a=\frac{\qodd^{2}}{\podd^{2}}\,.
\ee
Of course the K\"ahler metric on  this $\mathbb{F}_{0}$ is not the Einstein, direct-product metric on $S^2  \times S^2$.

Regularity of the metric requires that $a>1$, which  implies that the integers $\podd, \qodd$ obey
\be
0<\podd < \qodd ~.
\ee
This notation is closely related to the one in \cite{Gauntlett:2004yd}, and a further discussion of the relation to the standard $Y^{p,q}$ is provided in appendix \ref{app:Ypqstuffstuff}.

\subsubsection*{Flux Quantisation}

Finally, we need to check that the flux of the solution is properly quantized, i.e.
\be
N(D)=\frac{1}{(2 \pi \ell_{s})^{4}g_{s}}\int_{D}F\in \Z
\ee
for any five-cycle $D\in H_{5}(\M_{7};\Z)$. There are two independent five-cycles in $\M_7= \mathbb{P}^1 \times   \Yodd^{\podd,\qodd} $, namely  $\Yodd^{\podd,\qodd}$ at a point on the base $B_2$ of the elliptic K3,  and $E\times \Kthreebase = E\times \P^1$, where the $E$ is the unique generator of $H_{3}( \Yodd^{\podd,\qodd};\Z)$. The flux as given in \eqref{FYpq} is 
\be
m \ff=-\lb \frac{1}{ a x^{2}}(D \alphaalpha -g(x) D\phi)\wedge \dd x +2 J_{\Kthreebase}+\frac{1}{2}\sin \theta\dd \theta\wedge \dd \chi\right)~.
\ee
Due to the self-duality of the five-form flux, it is the Hodge star of the above two-form that needs to be quantised.
An explicit computation reveals that
\begin{align}
m^{4}*_{7}\ff&=  \frac{a}{4} D \alphaalpha \wedge D \phi \wedge \dd x \wedge \lb\dd \vol (\Kthreebase)+\frac{1}{4 x^{2}} \dd \vol(S^{2})\rb  \nonumber\\
&+\frac{a}{4} \dd \vol(\Kthreebase) \wedge \dd \vol(S^{2}) \wedge \lb x D \alphaalpha - \frac{U(x)}{w(x)} D \phi\rb~.
\label{banana}
\end{align}
The flux through the cycles $\Yodd^{\podd,\qodd}$ is 
\begin{align}
\int_{\Yodd^{\podd,\qodd}}*_{7}\ff &=\frac{1}{m^{4}}\int_{\Yodd^{\podd,\qodd}}\frac{a}{16 x^{2}}\sin\theta\dd x\wedge\dd \theta\wedge\dd \chi \wedge  D \alphaalpha \wedge D \phi\nonumber\\
&=-\frac{(2\pi)^{3}}{2m^{4}}\lb \frac{\qodd^{4}}{\podd(\podd^{2}-\qodd^{2})^{2}}\rb~.
\end{align}
Which implies the quantisation condition
\be
\frac{1}{(2\pi \ls)^{4}g_{s} m^{4}}=\frac{N}{4 \pi^{3}}\frac{\podd(\podd^{2}-\qodd^{2})^{2}}{\qodd^{4}} ~,
\ee
where\footnote{We chose this sign to ensure that $N>0$.} 
\be
\frac{1}{(2\pi \ls)^{4}g_{s}}\int_{\Yodd^{\podd,\qodd}}F=-N~,~~~~N\in \mathbb{N}~.
\ee
The integer $N$ is interpreted as the number of D3-branes along $\R^{1,1}\times \mathbb{P}^{1}$.

To perform the quantisation over the other five-cycle, we must first identify the correct generator for $H_{3}(\Yodd^{\podd,\qodd};\Z)$. It is not simple to identify this three-cycle in the metric as it is not a product metric, however there are four easily identifiable three-cycles at each of the degeneration surfaces, further discussion of these degeneration surfaces is provided in appendix \ref{app:sec:toric}. Let the generator of $H_{3}(\Yodd^{\podd,\qodd};\Z)$ be denoted $E$, and the three-cycles at each of the degeneration surfaces be $E^{a}$ where $a \in \{+,-, 0, \pi\}$. 
The closed three-form dual to the generator $E$ is 
\be
\omega_3= \frac{\podd^{2}-\qodd^{2}}{(4 \pi)^2} \left[ D \alphaalpha \wedge D \phi \wedge \dd x + \lb x D \alphaalpha -\frac{U(x)}{w(x)} D \phi\rb\wedge \dd \vol(S^2)\right]
\ee
and satisfies
\be
\int_{E}\omega_{3}= 1~.
\ee
One may use the above three-form to verify that the following homology relations
\be
E^{+}= (\podd +\qodd) E~,~~E^{-}=(\podd-\qodd) E~,~~ E^{0}= E^{\pi}= - \podd E~,
\ee
hold true. Then the integration of the five-form flux over the five-cycle $E\times B_{2}$  gives
\begin{align}
\int_{E\times B_{2}}\!\!\!\! *_{7}\ff &=\frac{1}{m^{4}} \int_{E\times B_{2}} \frac{a}{4}\dd \vol(B_{2}) \wedge \lb D \alphaalpha \wedge D \phi \wedge \dd x +\lb x D \alphaalpha -\frac{U(x)}{w(x)} D \phi\rb \wedge \dd \vol(S^2)\right )\nonumber\\
&=\frac{ 4a  \pi^2}{(\podd^2-\qodd^2) m^{4}} \int_{E\times B_{2}} \dd \vol(B_{2})\wedge \omega_{3}
= \frac{4 \pi^{2}}{ m^{4}}\frac{ \qodd^{2}}{\podd^{2}(\podd^{2}-\qodd^{2})} \vol(B_{2})~.
\end{align}
Flux quantisation imposes
\be
\frac{1}{(2 \pi \ls)^4 g_{s}}\int_{E\times \Kthreebase}F=-M~,~~~~~M \in \mathbb{N}~,
\ee
which may be interpreted as quantisation of the volume of $B_{2}$
\be\label{VolB1}
\vol(\Kthreebase)=\frac{M\pi}{N}\frac{\podd \qodd^{2}}{\qodd^{2}-\podd^{2}}~.
\ee

This concludes the discussion of the new $\AdS_3$ solutions in F-theory dual to $(0,2)$ SCFTs. In the following we will use these to test the duality by comparing holographic charges with the dual field theory observables.


\section{Holographic Charges}
\label{sec:HolC}

To compare physical observables with the dual SCFTs, we now turn to computing holographically the central charge as well as the R-charges and baryonic charges of baryonic 
operators, which will be compared to the dual field theories in section \ref{sec:FT}. At leading order in $N$, the results of the holographic computations presented in this section also apply, 
with minor modifications, to the holographic duals with constant axio-dilaton \cite{Benini:2015bwz}.

\subsection{General Considerations}

The leading order central charge is computed using the standard Brown-Henneaux prescription \cite{Brown:1986nw}, relating it to Newton's constant $G_N$ in 3d as 
\bea
c_{\text{sugra}}=\frac{3}{2 m G_{N}^{(3)}}~.
\eea
This can be extracted from the solution by computing the volume of the compact part of the spacetime $\M_7 $. We remark that in all the solutions presented above  the bases of the elliptic surfaces and three-folds considered above are singular, however the volumes of these spaces can be computed indirectly either by using flux quantisation or relating it to various topological quantities. Here we furthermore assume that the fibration is a smooth Weierstrass model, i.e. with only $I_1$ fibers. This will allow us to circumnavigate having to resolve any additional singularities, in passing to an M-theory picture. A similar logic was employed in \cite{Couzens:2017way}, and cross-checked against a smooth M-theory dual, field theory and anomalies. 
 Using the conventions in appendix D of \cite{Couzens:2017way} we have
\be\label{csugraeq}
c_{\text{sugra}}= \frac{3}{2m G_{N}^{(10)}}\int_{\M_{7}}\me^{\Delta}\dd\vol(\M_{7})~,
\ee
where $G_{N}^{(10)}=2^3 \pi^6 \ls^8$ is the 10d Newton's constant.

The subleading contribution to the central charge can be determined by anomaly inflow on the 7-branes as in \cite{Couzens:2017way}, which follows an argument presented in \cite{Aharony:2007dj}. Starting with a single D7-brane whose world-volume is extended along $\mathcal{W}_{8}$, the Wess--Zumino term in the effective action of the D7-brane induces a  3d  CS coupling by 
\be
S_{CS} =\frac{\mu_{7}\pi^{2}\ls^{4}}{24}\int_{\mathcal{W}_{8}}C^{(4)}\wedge  \Tr(\mathcal{R}\wedge \mathcal{R})~,\label{D7cont}
\ee
with $\mu_{7}=((2\pi)^7 \ls^{8}\gs)^{-1}$.
The results of \cite{Kraus:2005zm} allow one to extract the subleading contribution from the coefficient of the Chern-Simons term
\be\label{CLCR}
S_{CS}(\text{AdS}_{3})=\frac{c_{L}-c_{R}}{96 \pi} \int_{\text{AdS}_{3}} \omega_{CS}(\text{AdS}_{3})~.
\ee
One should then sum over all the 7-branes in the solution. 

The number (and type of) 7-branes in the background are encoded in the elliptic fibration. In the simplest case of an elliptic surface $\mathbb{E}_\tau \hookrightarrow \mathcal{S}^\tau \rightarrow \Sigma$ the number of 7-branes, assuming only $I_1$ fibers, is given by $12 \text{deg}(\mathcal{L}_{D})$.
The canonical bundle of the total space of an elliptic surface is
\be
K_{\Xtwo}= \pi^{*}\lb K_{\Sigma} +\sum_{i=1}^{|\Delta|} a_{i} P_{i}\rb \,,
\ee
where $i$ is summed over the components of the discriminant $\Delta$ of the elliptic fibration and 
$a_{i}$ are coefficients determined by the type of the singular fibers and $\pi$ is the projection to the base. For  $I_1$ fibers as considered here $a_{i}=\frac{1}{12}$. In order to satisfy 
\be
\mathfrak{R}_{\Xtwo}= -3 J_{\Sigma}= -K_{\Xtwo}~,~~\text{and}~~~\mathfrak{R}_{\Sigma}= -3 J_{\Sigma} -\dd Q=-K_{\Sigma} \,,
\ee
one obtains that the number of $I_1$ fibers is 
\be
|\Delta| = 12 \text{deg}(\mathcal{L}_{D}) \,.
\ee
Notice that for an elliptically fibered K3 surface, whose base is necessarily a $\mathbb{P}^1$,
$\text{deg}(\mathcal{L}_{D})=c_{1}(\mathbb{P}^{1})=2$ implies the well-known result of 24 7-branes. 

We will also compare R-charges and baryonic charges in the holographic duality.
Recall that in the Sasaki--Einstein setup one may compute these by evaluating the volumes of certain supersymmetric three cycles $\{\Sigma_{i}\}$. Below we present a version of this computation in the context of the AdS$_3$ solutions of interest. We assume that, similarly to their AdS$_5$ counterparts, D3-branes wrapped on $\Sigma_i$ give rise to BPS particles  moving in AdS$_3$, which we conjeture to be dual to some baryonic-type operator in the CFT$_2$.
These are BPS objects, and in 2d their conformal dimension equals their R-charge.
Denoting by $\mathcal{B}_{\Sigma_{i}}$ the operators in the dual field theory associated to the three-cycle $\Sigma_{i}$,
 the conformal dimension is 
\be
R[{\cal B}_{\Sigma_i}] = \Delta [\mathcal{B}_{\Sigma_{i}}] =\frac{M[\mathcal{B}_{\Sigma_{i}}] }{m}~,
\ee
where $M[\mathcal{B}_{\Sigma_{i}}]$ is the mass of the wrapped D3-brane. As our solutions include a warp factor for AdS$_{3}$ depending on the internal manifold, the mass of the D3-brane wrapped on the three-cycle $\Sigma_{i}$, is given by
\be
M[\mathcal{B}_{\Sigma_{i}}]  = T_{3}\int_{\Sigma_{i}}\frac{\me^{\Delta}}{m} \dd \vol(\Sigma_{i})~,
\qquad 
  T_{3}=\frac{1}{8 \pi^{3} \ls ^{4} \gs} \,,
\ee
where $T_3$ is the D3-brane tension. The factor of $\frac{\me^{\Delta}}{m}$ is precisely the warp factor due to the warping of the time coordinate. In summary
\begin{align}\label{R-chargeeq}
R[{\cal B}_{\Sigma_{i}}]=2\pi N \frac{ \int_{\Sigma_{i}} \me^{4 \Delta} \dd \widehat{\vol} (\Sigma_{i})}{\int_{\M_5} F}~.
\end{align}
The volume form with a hat is defined to be the volume form of the unwarped dimensionless metric obtained from the bracketed expression in \eqref{generalmetric}. Notice the similarity with the formulas for geometric $R$-charge in warped AdS$_4$ backgrounds \cite{Gabella:2011sg, Gabella:2012rc}.

The supersymmetric cycles are divisors in the complex cone over $\M_{7}$, which  implies that they are calibrated with respect to the four-form $\frac{\me^{4 \Delta}}{2 r^{2}} J_\text{cone}\wedge J_\text{cone}$, with $J_\text{cone}$ the K\"ahler form on the 8d metric cone $\dd s^2_\text{cone} = \dd r^2 + r^2 \dd s^2 (\M_7)$. Recall 
that unlike in the Sasaki--Einstein case, the cone is neither Ricci-flat nor K\"ahler, however as follows from \cite{Gauntlett:2007ts} we have 
\be
\dd ( r^{-4} \me^{8 \Delta} J_\text{cone} \wedge J_\text{cone} \wedge J_\text{cone})=0~.
\ee
In fact for all the solutions presented above a stronger condition holds. In each of the solutions presented above there is a distinguished Riemann surface. Define $\tilde{J}$ to be the K\"ahler form at fixed coordinate on the cone, then we have
\be
\dd (r^{-2} \me^{4 \Delta} \tilde{J}\wedge \tilde{J})=0~.
\ee
This implies that the R-charges, \eqref{R-chargeeq}, are topological quantities and independent of the coordinates chosen.

The final holographic charges that we can compute are the baryonic charges, (a summary of the related computation for the Sasaki--Einstein case is given in appendix \ref{ypqappendix}). In particular, we shall use the observation that the integral of a harmonic three-form over each of the three-cycles gives the baryonic charges of each of the baryons dual to that cycle in the field theory up to some overall normalisation which is fixed by requiring the results are integer. We note that as this result is a topological invariant we are free to multiply the metric by an arbitrary bounded and non-vanishing warp factor and perform the computation using the warped metric. We shall make use of this freedom later.

\subsection{Universal Twist Solutions: Elliptic Surface Case}
\label{sec:Holcharsol1}

Consider first the universal twist solutions,  where $\mathcal{Y}_8^\tau$ has an elliptic surface factor 
\be
\AdS_3 \times S^1\rightarrow( \M_4\times\mathcal{S}^\tau_4 )\,,\qquad \mathbb{E}_\tau\hookrightarrow \mathcal{S}^\tau_4 \rightarrow \Sigma \,.
\ee
Recall also that for a fixed coordinate on $\Sigma$ the transverse space is a Sasaki--Einstein manifold $SE_5 = (S^1 \rightarrow \M_4)$.

\subsubsection*{Central Charges}
We first consider the holographic charges of the universal twist solution with $\tau$ varying over $\Sigma$.
From \eqref{csugraeq} we have
\be\label{Csigmav1}
\ba
c_{\text{sugra}}
&=\frac{2 \cdot 3^4 \pi^2}{((2 \pi m \ls)^4 \gs)^2} \vol(\hbox{SE}_{5})\vol(\Sigma)
=\frac{2 \pi^2 N^2 \vol(\Sigma)}{\vol(SE_{5})}  \cr
&=\frac{36n^{2}M}{\tilde{m}^2 h^{2}\ell} (2(g-1)+\text{deg}(\mathcal{L_{D}}))\,,
\ea\ee
where we have used the quantisation conditions in \eqref{QuantSigma}. In the final step we have re-expressed the volume of $\Sigma$, by using the fact
that the Ricci form on $\Sigma$ satisfies \eqref{RSigma}, as
\be
\vol(\Sigma)=\int_{\Sigma}J_{\Sigma}=-\frac{1}{3}\lb \int_{\Sigma} \mathfrak{R}_{\Sigma}+\dd Q\rb= -\frac{1}{3}\lb 4 \pi (1-g) +\int_{\Sigma} \dd Q \rb~, 
\ee
and using 
\be
 -\frac{1}{2\pi}\int_{\Sigma}\dd Q= \int_{\Sigma}c_{1}(\mathcal{L}_{D})= \text{deg}(\mathcal{L}_{D})~,
 \ee
we have
\be
\vol(\Sigma)= \frac{1}{3}\lb4 \pi (g-1) + 2 \pi \text{deg}(\mathcal{L}_{D})\rb~.
\ee
 Moreover it follows that the central charge is integer for any K\"ahler--Einstein base and any surface $\Sigma$. To make contact with the field theory this can be related to the ``$a$''
central charge of the 4d quiver theory dual to the Sasaki--Einstein solution (with constant $\tau$) as
\be\label{acSE}
c_{\text{sugra}}=\frac{8 a^{4d}}{\pi} \vol(\Sigma) \,,\qquad \hbox{where} \quad a^{4d}=\frac{N^{2}\pi^{3}}{4 \vol(SE_{5})} \,.
\ee
we conclude that at leading order in $N$ the central charge is integral and given by 
\be
c_{\text{sugra}}= a^{4d}\left[ \frac{32(g-1) }{3}+\frac{16}{3}\text{deg}(\mathcal{L}_{D}) \right]~.\label{sol1c}
\ee

The first term  is precisely the result one obtains for the constant $\tau$ solution. 
Notice that even at leading order there is a correction to the central charge due to the varying axio-dilaton $\tau$, proportional to the first Chern class of the $U(1)_D$ duality bundle. 

{We note that this central charge is integer, independent of the choice of  K\"ahler-Einstein base and curve $\Sigma$. To see this one should consider the last expression in \eqref{Csigmav1}. There are three possible choices for K\"ahler-Einstein base; $\mathbb{C P}^{2}$ with $(M,\tilde{m})=(9,3)$, $S^{2}\times S^{2}$ with $(M,\tilde{m})=(8,2)$ and $\dd P_{k}$ for $k=3,..,8$ with $(M,\tilde{m})=(9-k, 1)$. Simple numerology shows that \eqref{Csigmav1} is integer for any of these choices and therefore also \eqref{sol1c}.}

By using \eqref{D7cont} and \eqref{CLCR} we find the contribution of a single 7-brane to the {difference of central charges} is 
\be
\Delta((c_{L})_\mathrm{sugra}-(c_{R} )_\mathrm{sugra})=\frac{N}{2}~.
\ee
Therefore the total contribution from the 7-branes is given by
\be\label{sol1CLCR}
(c_{L})_\mathrm{sugra}-(c_{R} )_\mathrm{sugra}=(\text{number of 7-branes})\cdot \frac{N}{2}= 6 N \text{deg}(\mathcal{L}_{D})~.
\ee

\subsubsection*{R-charges}
Recall that at fixed coordinates on $\Sigma$ the $U(1)$-fibration over the  K\"ahler--Einstein space $\M_4$ is a Sasaki--Einstein manifold, therefore the three-cycles which are dual to baryonic operators in 2d are the same as those in 4d\footnote{The Sasaki--Einstein metric (at fixed $\Sigma$ coordinates) appearing in the AdS$_3$ solution has a constant rescaling in comparison with the AdS$_5$ metric and therefore the volume form on the any three-cycle differs by a factor of $\me^{3 \Delta} \lb \frac{9}{4}\rb^{3/2}$ in comparison with the AdS$_5$ normalised metric. We shall write all volume forms with respect to the canonically normalised metric on the Sasaki--Einstein manifold with a tilde.}. From \eqref{R-chargeeq} the R-charges are
\be
R[\mathcal{B}_{\Sigma_{i}}]= 2 \pi N \frac{\int_{\Sigma_{i}}\me^{4 \Delta} \lb \frac{9}{4}\rb^{\frac{3}{2}} \dd \widetilde{\vol}(\Sigma_{i})}{9 \widetilde{\vol}(SE_{5})}= \frac{N\pi}{3} \frac{\int_{\Sigma_{i}}\dd \widetilde{\vol}(\Sigma_{i})}{\widetilde{\vol}(SE_{5})}=R^{4d}[\mathcal{B}_{\Sigma_{i}}]~,
\ee 
where we have used \cite{Berenstein:2002ke,Martelli:2004wu} to compare with the corresponding 4d R-charge.

\subsubsection*{Baryonic Charges}

During the discussion on baryonic charges we noted that the result is independent of a rescaling of the metric. Clearly this implies that the baryonic charges for these solutions will be identical to the original AdS$_5$ computation and therefore we shall not present it.

\begin{table}[h]
\begin{center}
\begin{tabular}{|c|c|}
\hline
Holographic charge& Result\\
\hline \hline
$c_{\text{sugra}}$&$\frac{32(g-1) a^{4d}}{3}+\frac{16 a^{4d}}{3} \text{deg}(\mathcal{L}_{D})$\\
\hline
$(c_{L})_\mathrm{sugra}-(c_{R} )_\mathrm{sugra}$& $6 N \text{deg}(\mathcal{L}_{D})$\\
\hline
R-charges&$ R^{(2d)}[\mathcal{B}_{\Sigma_{i}}]=R^{(4d)}[\mathcal{B}_{\Sigma_{i}}]$\\
\hline
Baryonic charges& $ B^{(2d)}[\mathcal{B}_{\Sigma_{i}}]=B^{(4d)}[\mathcal{B}_{\Sigma_{i}}]$\\
\hline
\end{tabular}
\end{center}
\caption{Holographic charges for the universal twist solution with elliptic surface $\mathcal{S}^\tau$. Here, $a^{4d}$ is the 4d central charge (\ref{acSE}) associated to the dual of the $\AdS_5 \times SE_5$ solutions.}
\end{table}

\subsection{Universal Twist Solutions: Elliptic Three-fold Case}\label{sec:uni3fold}

Consider now the universal twist solution where $\mathcal{Y}_8^\tau$ has a factor given by an elliptic three-fold. 
It will be instructive to compare these solutions to the AdS$_5$ solutions in section \ref{sec:AdS5} in an analogous manner to the way in which the discussion in the previous section referenced the Sasaki--Einstein solutions.

\subsubsection*{Central Charges}

The leading order central charge is easily found to be
\begin{align}\label{c3foldv1}
c=\frac{2\cdot 3^{4}\pi^{2}}{((2\pi m \ls)^4 \gs)^2} \vol(\M_5^\tau) \vol(\Sigma) 
=\frac{8 \pi^3(g-1) N^2}{3 \vol(\M_5^\tau)} 
=\frac{32(g-1)}{3} a^{4d}_{\tau}~,
\end{align}
where $a^{4d}_{\tau}$ is the central charge of the $\tau$ dependent 4d field theory dual to the solutions discussed in section \ref{sec:AdS5}.

As in the previous cases, the subleading contribution to the difference of central charges can be determined by anomaly inflow on the 7-branes, from the
 Wess--Zumino term  (\ref{D7cont}) in the effective action of a single  D7-brane.
 In contrast to the first case, the discriminant locus of the elliptic fibration is now a curve in $\M_{4}$. We consider only $I_1$ singular fibers and thus only single 7-branes are 
 wrapped on curves $C_x$ in the discriminant locus\footnote{With a slight abuse of notation we denote simply as $\Delta$ the locus $\{ \Delta=0 \}$.}. $\Delta$.
Imposing that the elliptic fibration satisfies \eqref{M4tauRicci} implies 
\be
[\Delta]=\sum_{x}\omega_{x}= 12 c_{1}(\mathcal{L}_{D})~,
\ee
where $\omega_{x}$ are the two-forms dual to the curves $C_{x}$, which are wrapped by the single 7-branes. Each  7-brane is extended along
 AdS$_3\times (U(1)_\chi \to \Sigma \times C_x )$, where $\chi$ is the angular coordinate with period $2\pi \ell$ along the R-symmetry direction. 
The total contribution to the WZ term is obtained by summing over all the single 7-branes, so that the effective world-volume 
 can be  written as $\mathcal{W}_{8}=  $ AdS$_3\times (\M_{3}  \to \Sigma   )$,   where $\M_{3} = U(1)_\chi \to  \Delta$  is a three-cycle in  $\M_5^\tau$.
The three-dimensional Chern-Simons term arising from the Wess-Zumino action then reads\footnote{In the following discussion the overall constant of the Wess-Zumino term in equation 
\eqref{D7cont} will cancel in the computation and therefore for simplicity we define the new constant $\tilde{\mu}_{7}=\frac{\mu_{7}\pi^2 \ls^4}{24}$.}
\begin{align}
S_{CS}&= -\tilde{\mu}_{7}\int_{\mathcal{W}_{8}} F \wedge \omega_{CS}(\text{AdS}_{3})\nonumber\\
&=-\frac{3 \tilde{\mu}_{7}}{4 m^4}  \, 2 \pi \chi(\Sigma)\, \vol(\M_{3})  \int_{\text{AdS}_{3}} \omega_{CS}(\text{AdS}_{3})~,
\end{align}
where 
\begin{align}
\vol(\M_{3}) & = \int_{\M_{3}}  \frac{1}{3} (\diff \chi + \rho)\wedge J_{\M_4} =
  \frac{ 2\pi \ell}{3}  \int_{\Delta } J_{\M_4}   \nonumber\\
&  = 8\pi \ell  \int_{\M_4} J_{\M_4} \wedge  c_{1}(\mathcal{L}_{D})= 
 \frac{8\pi^2 \ell}{3}  \int_{\M_4} \lb  c_{1}(\M_4) \wedge c_{1}(\mathcal{L}_{D}) - c_{1}(\mathcal{L}_{D})^2\rb~.
\label{newvolume3}
\end{align}
The  gravitational anomaly, by using \eqref{CLCR}, is therefore found to be 
\be
c_{L}-c_{R}= -\frac{  2^{4} 3^{2} \pi^{2} (g-1) \tilde{\mu}_{7} \vol(\M_{3})}{m^{4}}  
\label{unitauCLCR}~, 
\ee
where  notice that $\vol(\M_{3})$ is essentially an intersection number,   providing  the effective number of 7-branes,  as in  \cite{Couzens:2017way}. 

However, as for the leading order central charges, we will relate $c_{L}-c_{R}$  in the dual 2d SCFT to a corresponding holographic quantity in the parent 4d SCFT, therefore $\vol(\M_{3})$ will drop out form the equation.  Later we will show that this relationship is reproduced exactly by a field theoretic calculation, although we will not attempt to calculate the precise values of the 4d central charges in specific examples.

Concretely, we wish to identify the above result with the linear 't Hooft anomaly $k_{R}$ in the 4d theory, and therefore with 
the difference of 4d central charges $c^{4d}- a^{4d}= \tfrac{k_{R}}{16}$, where recall that
\be
a^{4d}=\frac{3}{32}(3k_{RRR}-k_R)~, \qquad c^{4d}=\frac{1}{32}(9k_{RRR}-5k_R)~.
\ee

For any 4d $\mathcal{N}=1$ SCFT with an $R$-symmetry, the $R$-symmetry current   $\mathcal{R}_\mu$   satisfies the anomalous conservation
  equation    \cite{Anselmi:1997am, Intriligator:2003jj,Cassani:2013dba}
\be
\partial_{\mu}\langle \sqrt{g} \mathcal{R}^{\mu}\rangle =\frac{k_R}{384 \pi^2} \epsilon_{\mu\nu\rho\sigma} \tensor{R}{^{\mu\nu}_{\kappa \tau}}\tensor{R}{^{\rho \sigma \kappa \tau}}+
\frac{k_{RRR}}{48 \pi^2}\epsilon^{\mu\nu\rho\sigma} F_{\mu\nu}F_{\rho\sigma} \,,
\ee
where  $F$ is the field strength of the background gauge field $A$ sourcing the R-symmetry current.

Consider the AdS$_5$ solutions of section \ref{sec:AdS5}. Recall that for the universal twist solution to be well-defined the manifold $\M_5^\tau$ is required to be quasi-regular. As such we may write the metric on $\M_5^\tau$ as a $U(1)$ fibration over a K\"ahler base $\M_{4}$ as 
\be
\dd s^{2}(\M_5^\tau)= \frac{1}{9}\lb \dd \chi +3 \sigma\rb^{2} +\dd s^{2}(\M_{4})~,
\ee  
with $\dd \sigma =2  J_{\M_4}$. As we consider only the quasi-regular cases we may fix the period of $\chi$ to be $2 \pi \ell$. By changing coordinates as $\chi=\ell \tilde{\chi}$ we define a new $2 \pi$ periodic coordinate. As the Reeb vector field is dual to the R-symmetry direction it is natural, as explained in \cite{Berenstein:2002ke}, that a shift in the coordinate $\tilde{\chi}$ induces a gauge transformation of the R-symmetry gauge field\footnote{{More precisely, here $\mathcal{A}$ is a gauge field in AdS$_5$, whose boundary value is identified with the background  R-symmetry gauge field  $A$ in the four dimensional SCFT.}} $\mathcal{A}$, that is
\be
\tilde{\chi}\rightarrow \tilde{\chi}+ \alpha \Lambda~,~~~~\mathcal{A}\rightarrow \mathcal{A}+\dd \Lambda~.
\ee
The identification of the constant $\alpha$ is fixed by using the fact that the holomorphic 3-form on the cone is associated to the superpotential and therefore has R-charge 2. The functional dependence of the holomorphic three-form on $\tilde{\chi}$ may be read off from \eqref{Omegapsi} which fixes $\alpha=\frac{2}{\ell}$. We may include $\mathcal{A}$ in the usual Kaluza-Klein ansatz by deforming the internal metric as
\be
\dd s^{2}(\M_5^\tau)\rightarrow \lb \frac{\ell}{3}\rb^{2}\lb \dd \tilde{\chi}+\frac{3}{\ell}\sigma+\frac{2}{\ell}\mathcal{A}\rb^{2}+\dd s^{2}(\M_{4})~.
\ee
Moreover, for consistency, the five-form flux must be deformed  as\footnote{The length scale associated to the AdS$_5$ will be denoted as $m_{5}$ in the following. It will be shown to be proportional to the length scale $m$ in the AdS$_3$ solutions.}
\be
F\rightarrow (1+*)\frac{2\ell}{3m_{5}^{4}}\lb \lb\dd \tilde{\chi}+\frac{3}{\ell}\sigma+\frac{2}{\ell}\mathcal{A}\rb\wedge J_{\M_4}\wedge J_{\M_4}-\frac{1}{3}\dd\mathcal{A}\wedge \lb\dd \tilde{\chi}+\frac{3}{\ell}\sigma \rb\wedge J_{\M_4}\rb \,,
\ee
which by construction is closed upon using the equation of motion for the new gauge field $\dd *\dd \mathcal{A}=0$. 
The term of the four-form potential of interest is 
\be
C_{4}\supset -\frac{2}{ 3 m_{5}^{4}} \mathcal{A}\wedge \lb \frac{\ell}{3}\lb \dd \tilde{\chi } +\frac{3}{\ell} \sigma \rb \rb \wedge J_{\M_4}\,.
\ee
In this configuration, the world-volume of each 7-brane is  AdS$_5\times (U(1)_\chi \to  C_x )$,  therefore the total contribution from all the 7-branes  is obtained by integrating
on the world-volume  $\mathcal{W}_8 = \text{AdS}_5 \times \M_{3}$ where  $\M_3= U(1)_\chi \to  \Delta $   is the same 
three-cycle in  $\M_5^\tau$ that appears in the  AdS$_3$ solution. 
 We may use this to  extract from the Wess-Zumino term  a contribution to the gravitational action in AdS$_5$ given by
\begin{align}
S_{CS}=\tilde{\mu}_{7} \int_{\mathcal{W}_{8}}C_{4} \wedge \Tr[\mathcal{R}\wedge \mathcal{R}]=-\frac{2\tilde{\mu}_{7}}{3 m_{5}^{4}}\vol(\M_{3})\int_{\text{AdS}_{5}}\mathcal{A}\wedge \Tr[\mathcal{R}\wedge \mathcal{R}]~.
\end{align}

According to the gauge/gravity duality master formula, the generating functional for (connected) current correlators in the boundary theory, $\ii W[A]= \log Z[A]$, equates 
the on-shell gravitational action, $W[A]=S_{\text{AdS}_5}[\mathcal{A}]$, and therefore as explained in  \cite{Witten:1998qj} the non-invariance under gauge transformations of the latter 
corresponds to the anomaly in the dual field theory. Specifically, a gauge transformation of the boundary gauge field $A$ induces a transformation of the Chern-Simons term
\begin{align}
\delta_\Lambda  W[A] = \delta_\Lambda S_{CS}&=- \frac{2\tilde{\mu}_{7}}{3 m_{5}^{4}}\vol(\M_{3})\int_{\text{AdS}_{5}}\dd \Lambda\wedge \Tr[\mathcal{R}\wedge \mathcal{R}]\nonumber\\
&=-\frac{2\tilde{\mu}_{7}}{3 m_{5}^{4}}\vol(\M_{3})\int_{\partial\text{AdS}_{5}}\Lambda \Tr[\mathcal{R}\wedge \mathcal{R}]~,
\end{align}
implying that  on the boundary we have
\be
\int_{\partial \text{AdS}_{5}}    \Lambda \partial_{\mu}\langle \sqrt{g} \mathcal{R}^{\mu}\rangle   \dd \vol(\partial \text{AdS}_{5})  = \frac{2\tilde{\mu}_{7}}{3 m_{5}^{4}}\vol(\M_{3})\int_{\partial\text{AdS}_{5}}\Lambda \Tr[\mathcal{R}\wedge \mathcal{R}] \,,
\ee
where 
\be
\Tr[\mathcal{R}\wedge \mathcal{R}]= -\frac{1}{4} \epsilon_{\mu\nu\rho\sigma} \tensor{R}{^{\mu\nu}_{\kappa \tau}}\tensor{R}{^{\rho \sigma \kappa \tau}} \dd \vol(\partial \text{AdS}_{5})~.
\ee 
In conclusion we find\footnote{It would be interesting to match this formula, using (\ref{dfgrt}) and   (\ref{newvolume3}),  to a purely field theoretic computation in the 4d SCFT.}
\be
k_{R}= - N\frac{\pi\vol(\M_{3})}  {12 \vol(\M_5^\tau)} =  -\frac{ 2^{6} \pi^2 \tilde{\mu}_{7}\vol(\M_{3})}{m_{5}^{4}}~,
\ee
and inserting this  into \eqref{unitauCLCR} we obtain
\be
c_{L}-c_{R}=\frac{9}{4}\frac{m_{5}^{4}}{m^{4}} (g-1) k_{R}~.
\ee
We may relate the different length scales of the two solutions by comparing the quantisation condition used to obtain the integer $N$. In both cases this gives the number of D3-branes in the solution and should therefore be fixed in flowing from the $\AdS_5$ solution to the $\AdS_3$ solution, by comparing \eqref{NAdS5} and \eqref{Nunitau} we find $9 m_{5}^4=4 m^4$ and therefore we conclude that 
\be
c_{L}-c_{R}=(g-1) k_{R}~.
\ee


\subsubsection*{R-charges}

In a similar manner to the previous section, at fixed coordinate on $\Sigma$, which is now  $\mathbb{H}^{2}/\Gamma$ with $\Gamma$ a subgroup of $SL_{2}\Z$, equipped with the constant curvature metric, one finds that the metric on $\M_5^\tau$ is the 
same (up to an overall constant factor) as the metrics discussed in section \ref{sec:AdS5}. Again we have that the three-cycles of the two solutions agree and therefore the dual baryonic operators in 2d and 4d are identified. Clearly by the same arguments as presented in section \ref{sec:Holcharsol1} the R-charges of the baryonic operators in 2d and 4d coincide.


\subsubsection*{Baryonic Charges}

As above the metrics agree up to a numerical factor. The topological nature of this computation implies that the baryonic charges of the 2d theory and the 4d theory agree. 

\begin{table}[h]
\begin{center}
\begin{tabular}{|c|c|}
\hline
Holographic charge& Result\\
\hline \hline
$c_{\text{sugra}}$&$\frac{32(g-1)}{3} a^{4d}_{\tau}$\\
\hline
$(c_{L})_\mathrm{sugra}-(c_{R} )_\mathrm{sugra}$ & $16(g-1)(c^{4d}- a^{4d})$ \\
\hline
R-charges&$ R^{(2d)}[\mathcal{B}_{\Sigma_{i}}]=R^{(4d)}[\mathcal{B}_{\Sigma_{i}}]$\\
\hline
Baryonic Charges& $ B^{(2d)}[\mathcal{B}_{\Sigma_{i}}]=B^{(4d)}[\mathcal{B}_{\Sigma_{i}}]$\\
\hline
\end{tabular}
\end{center}
\caption{Holographic charges for the universal twist with elliptic threefold $\mathcal{T}^\tau_6$. Here $a^{4d}_\tau$ is the central charge of the dual to the solutions in section \ref{sec:AdS5}.}
\end{table}


\subsection{Baryonic Twist Solution: $\mathfrak{Y}^{\podd,\qodd}$ Case}

In this final section we shall consider the baryonic twist solutions using $\mathfrak{Y}^{\podd, \qodd}$ as the example. We expect the computations to extend to other solutions with baryonic twists in a similar manner. Some such solutions will be discussed in \cite{dariochris}.

\subsubsection*{Central Charges}
From \eqref{csugraeq} we compute
\be
c_\mathrm{sugra} = \frac{6 N M \podd^{2}(\qodd^{2}-\podd^{2})}{\qodd^{2}}~.
\label{csugra}
\ee
Notice that despite the differences of our solution with respect to the constant $\tau$ version discussed in \cite{Donos:2008ug}, the value of the holographic central charge (\ref{csugra}) agrees \emph{exactly} with the value obtained in 
eq. (4.18) of \cite{Donos:2008ug}. We anticipate that this is a general property of the baryonic twist solutions, that does not depend on the details of the $\M_5^\tau$ geometry. More precisely, for any solution of the type AdS$_3\times T^2\times \M_5^\tau$ and constant axio-dilaton, 
we can construct a solution of the type AdS$_3\times \P^1 \times \M_5\tau$ with axio-dilaton varying holomorphically on $\P^1$, such that the F-theory lift has an elliptic K3 factor. These two solutions will have equal holographic central charges, at leading order in $N$.

The subleading contribution may also be simply computed from the geometry. Moreover it can be seen that the result is independent of the choice of $\M_5^\tau$, one obtains the universal contribution of $\frac{N}{2}$ for a single 7-brane. For a K3 surface the number of 7-branes for a consistent geometry is $24$ and therefore the subleading contribution is
\bea
(c_{L})_\mathrm{sugra}-(c_{R} )_\mathrm{sugra}& =  & 24 \cdot \frac{N}{2}~=~12 N~.
\label{subway}
\eea
Notice that although at leading order the central charges of the solution with constant and varying $\tau$ agree, the sub-leading contribution  (\ref{subway}) is clearly zero in the model with constant $\tau$, as there are no seven-branes. In the next section we will argue that in the dual field theory side this result is exactly reproduced combining contributions that come both from the bulk modes (3-3 strings) as well as 3-7 strings. Note that there are $O(N)$ terms in the bulk for the theory with varying coupling, that are due to the duality twisting.

\subsubsection*{R-charges}

The three-cycles in the geometry that are calibrations are the four three-cycles located at the four degeneration surfaces of the metric. Recall that at each degeneration surface a Killing vector has zero norm, which determines a codimension two locus (namely a three-manifold) in the geometry. By explicit computation one can see that the volume form on the three-manifolds will be equal to the pullback of this closed four-form and hence these cycles are calibrated.
We find
\begin{align}
R[{\cal B}_{\Sigma_+}]&=R[{\cal B}_{\Sigma_-}]=N \frac{\qodd^{2}-\podd^{2}}{\qodd^{2}}~,\nn\\
R[{\cal B}_{\Sigma_0}]&=  R[  {\cal B}_{\Sigma_\pi}]  = N \frac{\podd^2}{\qodd^2}~.
\label{susyvolumes}
\end{align}
Observe that the sum of the normalised volumes of submanifolds satisfies exactly the same relation to their Sasaki--Einstein counterparts, namely
\bea
R[{\cal B}_{\Sigma_+}] + R[{\cal B}_{\Sigma_-}] + R[{\cal B}_{\Sigma_0}] +  R[  {\cal B}_{\Sigma_\pi}]  = 2 N ~.
\eea
It would be nice to obtain a general proof of this formula, analogous to that in \cite{Martelli:2005tp}.

\subsubsection*{Baryonic Charges}

As discussed in the introduction of this section we are free to multiply the metric by an arbitrary warp factor so long as the warp factor is {bounded and} non-vanishing. We shall make use of this freedom to find such a harmonic form. 
As $\dim[H^{3}(\Yodd^{\podd,\qodd})]=1$ there is a unique closed three-form representative which may be extracted from \eqref{banana} and is given by
\be
\omega_{3}=k \lb D \alphaalpha \wedge D\phi \wedge \dd x + \dd \vol(S^{2}) \wedge \lb x D \alphaalpha -\frac{U(x)}{w(x)} D \phi\rb\rb~.
\ee
Observe that for the warped  metric on $\Yodd^{\podd,\qodd}$
\be
\dd s^{2} =\me^{-4 \Delta} \dd s^{2}(\Yodd^{\podd,\qodd})
\ee
this three-form is both closed and co-closed and therefore harmonic.  The constant $k$ is fixed by requiring that the results are integer, to be $k=-\frac{\qodd^2-\podd^2}{4}$.
Integrating this over the three-cycles we find
\begin{align}
\int_{\Sigma_{\pi}} \omega_3 =\int_{\Sigma_{0}}\omega_{3}=\podd~,~~~\int_{\Sigma_{-}} \omega_{3}= \qodd-\podd~,~~~\int_{\Sigma_{+}}\omega_{3}= -(\qodd +\podd)
\label{barionnen}
\end{align}
which gives the baryonic charges of the fields and agrees with the result in \eqref{chargematrix} for the would-be GLSM charges.

\begin{table}[h]
\begin{center}
\begin{tabular}{|c|c|}
\hline
Holographic charge& Result\\
\hline \hline
$c_{\text{sugra}}$&$\frac{6 N M \podd^{2}(\qodd^{2}-\podd^{2})}{\qodd^{2}}$\\
\hline
$(c_{L})_\mathrm{sugra}-(c_{R} )_\mathrm{sugra}$& $12N$\\
\hline
R-charges&$\begin{array}{ll} R^{(2d)}[\mathcal{B}_{\Sigma_{0}}]&=R^{(2d)}[\mathcal{B}_{\Sigma_{\pi}}]= N \frac{\qodd^2-\podd^2}{\qodd^2}\\
 R^{(2d)}[\mathcal{B}_{\Sigma_{0}}]&=R^{(2d)}[\mathcal{B}_{\Sigma_{\pi}}]= N \frac{\podd^2}{\qodd^2}
\end{array}$\\
\hline
Baryonic charges& $ \begin{array}{ll}
B^{(2d)}[\mathcal{B}_{\Sigma_{0}}]&=B^{(4d)}[\mathcal{B}_{\Sigma_{\pi}}]=\podd\\
B^{(2d)}[\mathcal{B}_{\Sigma_{0}}]&=\qodd-\podd\\
B^{(2d)}[\mathcal{B}_{\Sigma_{0}}]&=-(\qodd +\podd)
\end{array}$\\
\hline
\end{tabular}
\end{center}
\caption{Holographic charges of the Baryonic twist for $\mathfrak{Y}^{\podd,\qodd}$.}
\end{table}

\section{Dual Field Theories with Varying Coupling}
\label{sec:FT}

Field theories with spacetime varying coupling are not a new concept in itself. However the inclusion of 
S-duality monodromies specifically in 4d $\mathcal{N}=4$ SYM and, as we will see, generalizations to $\mathcal{N}=1$, have only received recent attention in \cite{Assel:2016wcr, Lawrie:2016axq}.

\subsection{Duality Twist for 4d $\mathcal{N}=4$ SYM}
\label{dualtwistedFTREV}

 For 4d $\mathcal{N}=4$ the question arose in the context of D3-branes in F-theory, which naturally implements the varying complexified coupling $\tau$ in terms of a complex structure of an elliptic curve. Field theoretically the $\tau$ variation along a curved manifold, e.g. a complex curve or surface, together with retaining some supersymmetry, implies that a particular new topological twist needs to be applied to the field theory. This topological duality twist was first introduced for abelian theories in \cite{Martucci:2014ema}, and a proposal for the non-abelian generalization was put forward based on a realization in terms of M5-branes in \cite{Assel:2016wcr}. For D3-branes wrapped on curves along which the coupling varies, the duality twist was implemented in \cite{Haghighat:2015ega, Lawrie:2016axq}. 
 
The key point about the topological duality twist is that fields and supercharges transform as sections of a duality bundle $\mathcal{L}_D$ with connection given in terms of $\tau = \tau_1 + i \tau_2$ by $Q$ in (\ref{ALMOSTTHERE}).
The transformation of the supercharges is such that they have charge  $\pm 1/2$ under this $U(1)_D$:
\be
\ba
Q_{\alpha} \ & \rightarrow & \ \me^{-i \alpha(\gamma)} Q_{\alpha} \cr 
\tilde{Q}_{\dot\alpha}  \ & \rightarrow & \  \me^{+i \alpha(\gamma)} \tilde{Q}_{\dot\alpha} 
\ea\,,
\ee
where $e^{i\alpha(\gamma)} = (c\tau + d)/|c\tau + d|$ for $\gamma ={a \ b\choose c \ d }  \in SL_2\mathbb{Z}$. The remaining fields of the $\mathcal{N}=4$ SYM theory are charged $q_{\Phi}= 0$ (scalars), 
$q_{F_\pm} =  \mp 1$ (where $F_\pm = \sqrt{\tau_2} (F\pm \star F)/2$) and $q_{\lambda} =- {1\over 2}$, $q_{\tilde\lambda} = {1\over 2}$ (fermions). 
To offset this transformation the duality twist redefines the $U(1)_D$ with an R-symmetry transformation. More generally for spacetimes of the form $M_4 = \mathbb{R}^{1,1} \times C$ the twist can involve $U(1)_C, U(1)_D$ and an R-symmetry $U(1)_R$, as discussed in \cite{Lawrie:2016axq}. 

One of the classes of solutions that we will encounter is the compactification of a 4d $\mathcal{N}=1$ theory on a curve $C= \mathbb{P}^1$, which is the base of an elliptic K3. This has many similarities to the elliptic K3 compactifications of F-theory as discussed in appendix D of \cite{Lawrie:2016axq}. Briefly, in this case the twist only requires one to combine 
\be\label{U1Twist}
U(1)_{\rm twist}:\qquad {T_{\rm twist}^{C}} = {1\over 2}  \left(T_{C} - T_D \right) \,,
\ee
without an R-symmetry twist. The resulting theory has 2d $(0,8)$ supersymmetry. The fields are counted by cohomologies  $h^{i,j} (C)$, depending on the twist charges $q_{\rm twist} = -1, 0, +1$ coresponding to $(i,j) = (1,0), (0,0), (0,1)$.

The analysis for strings arising from wrapped branes was largely performed for abelian theories. The generalisation to non-abelian is somewhat more subtle, and needs to be performed using the approach in \cite{Assel:2016wcr}, mapping the issue to M5-branes on an elliptic surface $\widehat{C}$, which geometrizes the axio-dilaton variation in terms of the elliptic fiber. Studying these theories has so far not been done, but some progress for D3-branes in CY three-folds in F-theory was put forward in \cite{Couzens:2017way} and will appear in \cite{TBA}, using anomaly arguments. 

For $\mathcal{N}=1$ theories in 4d, similar compactifications with spacetime dependent couplings can be defined. Although not every such theory has a duality group, whenever there is a holographic dual setup, and an embedding into Type IIB (or F-theory), the theory should have an induced $U(1)_D$ symmetry. One way to argue for this is presented in \cite{Intriligator:1998ig} by Intriligator, where the so-called bonus-$U(1)$, which for the abelian theories was identified with $U(1)_D$ in \cite{Assel:2016wcr}. Again there is a question as to how to generalise this to non-abelian theories, where there is no manifest way to define this duality symmetry. We should remark that this symmetry for the abelian theory is a symmetry only of the equations of motion, not of the action. From considerations in \cite{Intriligator:1998ig}, the bonus symmetry is an approximate symmetry only for certain observables in a particular limit, namely when both stringy and D-stringy corrections are suppressed, but then should also be a feature of 4d $\mathcal{N}=1$  theories.

Here we will consider well-known quiver gauge theories with Type IIB Sasaki--Einstein duals, for which we will discuss generalisations of the ``universal twist'' and ``baryonic twist'' \cite{Benini:2015bwz}. The first class of theories is characterised by having rational R-charges in four dimensions, and otherwise unspecified global symmetries; examples include ${\cal N}=4$ SYM and the Klebanov-Witten model, but more generally the theories discussed in section \ref{sec:AdS5}, which are the most general F-theory solutions with $\AdS_5$ factors dual to 4d $\mathcal{N}=1$ theories.  
The second class of theories is characterised by having a global baryonic global symmetry, and may have rational or irrational R-charges in four dimensions; our main example will be the $Y^{p,q}$ 
quivers\footnote{In order to be self-contained, we give a mini-review of the 4d $Y^{p,q}$ quiver theories in Appendix \ref{ypqappendix}.}
  \cite{Franco:2005sm}. In all cases, the R-charges of the 2d SCFTs will be rational.

In the gauge theories each node of the quiver has a complex coupling constant $\tau_i$ and the diagonal combination
\be
\tau = \sum_i \tau_i 
\ee
is identified in the dual supergravity solution with the axio-dilaton of Type IIB. 
Unlike the case of $\mathcal{N}=4$ there is no direct way to identify the charges, however we will argue that the fermions are all charged in the same way, exactly as in ${\cal N}=4$ SYM. The argument to support this uses the duality with  $\AdS_5$: although the bonus $U(1)$ is not an actual symmetry of the theory, it is a symmetry for large $N$ and for short operators. In the holographic dual these correspond to Kaluza-Klein modes on the compact part of the supergravity solution. As the latter have definite charges under $U(1)_D$, the expectation is that the dual states will also have a well-defined charge.  
The state of the art of the KK-spectrum on Sasaki--Einstein manifolds was obtained in \cite{Eager:2012hx}.

We begin with 4d $\mathcal{N}=1$ with supercharges $Q = ({\bf 2}, {\bf 1})$ and $\tilde{Q}= ({\bf 1}, {\bf 2})$ under $SO(1,3)_L$ and reduce them along the curve $C$
\be
\ba
SO(1,3)_L \quad &\rightarrow \quad SO(1,1)_L \times U(1)_C \cr 
({\bf 2}, {\bf 1})\quad  & \rightarrow \quad {\bf 1}_{++} \oplus {\bf 1}_{--} \cr 
 ({\bf 1}, {\bf 2})\quad  & \rightarrow \quad {\bf 1}_{+-} \oplus {\bf 1}_{-+} \,.
\ea
\ee
The duality charges are conjecturally $q_Q = -1$ and $q_{\tilde{Q}} = +1$. Then performing the topological twist as in (\ref{U1Twist})  results in two scalar supercharges of negative chirality (i.e. ${\bf 1}_{--}$ and ${\bf 1}_{-+}$ in the above equation). 
For abelian ${\cal N}=1$ theories the multiplets are such that the scalars are uncharged under the $U(1)_D$ and the fermions carry all the same charge, which agrees with that of  the supercharges. This is much alike the charges in the $\mathcal{N}=4$ SYM case. 
For the non-abelian theory, we propose to study the theory in a mesonic or Coulomb branch, where using anomalies we can determine the central charges \cite{TBA}.

\subsection{Twisted ${\cal N}=1$ Field Theories}
\label{ypqnotau}

Before addressing the dual field theory interpretation of the solutions we discussed in section \ref{sec:NewSol} we 
 review some aspects of  the dualities proposed in \cite{Benini:2015bwz}  for the solutions with \emph{constant} $\tau$  \cite{Gauntlett:2004zh,Gauntlett:2006qw}.
 We will follow the notation of these references, except, when we discuss the baryonic twist of the $Y^{p,q}$ theories where we will be careful in distinguishing the parameters $p,q$  in the field theories from the parameters $\podd, \qodd$ in the gravity solution  \cite{Gauntlett:2004zh,Gauntlett:2006qw}. As we 
 have already mentioned, although these parameters can be formally identified, they turn out to be  defined in disjoint domains.

A 4d ${\cal N}=1$ field theory can be compactified on a Riemann surface $C_g$ of genus $g$ by performing a topological twist that preserves ${\cal N}=(0,2)$ supersymmetry in two dimensions. 
Although the details of these two-dimensional theories may be complicated to work out, if these flow to $(0,2)$ SCFTs then many of their properties can be inferred by employing the method of $c$-extremization 
\cite{Benini:2013cda}. In particular, 
this method allows one to determine the 2d central charge $c_R$ of these theories, starting from the 't Hooft anomalies of their ``parent'' four-dimensional theories. The most reliable method to implement this is to consider 
the anomaly polynomial $I_6$ of the ${\cal N}=1$ 4d theory, that can be usually computed exactly starting from the fermionic field content of the 4d theory. This is given by
\be
I_{6}=\frac{1}{6} k_{IJK}c_{1}(\mathcal{F}_{I})\wedge c_{1}(\mathcal{F}_{J})\wedge c_{1}(\mathcal{F}_{K})-\frac{1}{24} k_{I} c_{1}(\mathcal{F}_{I})\wedge p_{1}(T_{4})~,
\ee
where the index $I$ runs over all the $U(1)$ global symmetries of the theory.   Here $c_{1} (\mathcal{F}_{I})$ are  the first Chern classes of the different $U(1)_I$ bundles and 
$p_{1}(T_{4})$ is the first Pontryagin class of the manifold the theory is placed upon.
The constants $K_{IJK}$ and $K_{I}$ are the cubic and linear `t Hooft anomalies which can be determined from the charges of the fermions in the theory, namely 
\bea
K_{IJK} = \mathrm{Tr}[U(1)_IU(1)_JU(1)_K] = \sum_{i} Q_I^i Q_J^i Q_K^i ~,  \qquad   K_{I} =  \mathrm{Tr}[U(1)_I] = \sum_{i} Q_I^i  ~,
\eea
where $Q_I^i$ denotes the charge of the $i$-th fermion under $U(1)_I$. This can be reduced to the anomaly polynomial $I_4$ of the 2d theory by integrating it over $C_g$, which in a similar notation, reads
 \be\label{AnomalyI4}
I_{4}=\frac{1}{2} k_{IJ} c_{1}(\mathcal{F}_{I})\wedge c_{1}(\mathcal{F}_{J})-\frac{k}{24}p_{1}(T_{2})\,   .
\ee
 In the $(0,2)$ SCFT we can then read off the central charges $c_R$ and the gravitational anomaly as
 \be\label{cRcL}
 c_R=3k_{RR}\,,\qquad 
 c_R-c_L=k\,.
 \ee
 In general, to compute the  $k_{IJ}$ and $k$ one requires information on the spectrum of fermions of the 2d theory, but for theories coming from a parent 4d theory with known 't Hooft anomalies, these can be extracted simply from 
 \be
 I_4  =  \int_{C_g} I_6~.
 \ee
 The 2d superconformal  $U(1)_R$ symmetry is determined by extremizing the trial $k_{RR}$.

The topological twist can be performed by switching on background gauge fields for the various global symmetries of the 4d theory, with quantised fluxes through $C_g$. 
Consider a quiver gauge theory\footnote{Twisted compatifications of various four-dimensional quiver gauge theories were studied in \cite{Amariti:2017iuz} and further examples of dual supergravity solutions will be discussed in \cite{dariochris}.} for which the global symmetries are 
\be
(U(1)_F)^{n_F}\times (U(1)_B)^{n_B} \times U(1)_{R}^{4d}~,
\ee
where $F$ stands for flavour and $B$ stands for baryonic symmetries, respectively. The superscript  on the  R-symmetry-factor emphasises that this is the exact superconformal R-symmetry of the interacting  4d SCFT, determined by $a$-maximization. 

In the notation of \cite{Benini:2015bwz}, the topological twist can be generically performed along\footnote{{$T_{\text{twist}}$ refers to the combination of symmetry generators that are used to twist the local Lorentz symmetry along the curve.}}
\bea
T_\text{twist} & = & \sum_{I}^{n_F} b_I T_I  + \sum_{I}^{n_B}B_{I} T_{B_I} + \frac{\kappa}{2} T_R^{4d}~,
\label{letstwistagain}
\eea
where $T_I,  T_{B_I}, T_R^{4d}$ are the generators of the respective global symmetries and $\kappa=1,0,-1$ for genus $g=0,1$, or $g>1$, respectively\footnote{In this equation it is  assumed that 
$C_g$ has constant curvature.}.  Here $b_I,B_I$ are suitably quantised parameters, and the factor  $\frac{\kappa}{2}$ is determined by requiring 
that the Killing spinors on $C_g$ become constants, as usual. Notice that as the Killing spinors are not charged under    the other global symmetries, this particular way of preserving supersymmetry does
not fix the  parameters $b_I,B_I$.

Note that when $\kappa \neq 0$ the twisting  (\ref{letstwistagain})
makes sense only when $U(1)_{R}^{4d}$ is a compact symmetry.  In particular,  for the $Y^{p,q}$ theories this is true iff $\bbcw\equiv \sqrt{4 p^2 -3 q^2}$  is an integer and the 4d R-charges 
(\ref{R2D2})
are rational numbers. This implies that generically the 2d R-charges will be rational numbers. When $\kappa =0$ (namely for $C_{g=1}=T^2$) there is no twist  by the 4d R-symmetry and therefore one can start from 4d field theories with irrational 
$R$-charges. In the next section we will explain a variant of this 
twisting, in which we can again start from a 4d field theory with irrational R-charges, and nevertheless compactify this on a  $C_{g=0} = \P^1$.

 The R-symmetry $U(1)_R^{2d}$  of the $(0,2)$ theory can in general mix with all the global symmetries of the 4d 
 theory\footnote{A priori, there can be global symmetries that emerge in the 2d theory. In this case $c$-extremization (like $a$-maximization) cannot be used effectively to determine the R symmetry in the IR.}, namely in terms of generators we have 
 \bea
 T^{2d}_{\text{trial}}  & = & \sum_I^{n_F}\epsilon_I T_I  + \sum_I^{n_B}\epsilon_{B_I} T_{B_I} + T_R^{4d}~,
 \label{comprare}
 \eea
 where $\epsilon_I,\epsilon_{B_I}$ are a priori real numbers that will be determined my extremizing the trial 2d central charge as a function of these parameters. This  calculation was performed in 
 \cite{Benini:2015bwz} for various examples, using the index theorem to count the fermionic zero modes in 2d \cite{Maldacena:2000mw}. However, as we discussed above, the computation using the reduction of 
 the anomaly polynomial of the 4d theory is more robust, as there is no need to assume that the theory is weakly coupled (which is not a correct assumption for most 
 ${\cal N}=1$ theories with Sasaki--Einstein duals).

 \subsubsection{Universal Twist}
 
 This twist can be applied to any theory provided the 4d R-charges are rational, and consists in taking 
  \bea
T_\text{twist} & = &  \frac{\kappa}{2} T_R^{4d}~,
\label{unitwist}
\eea
 where $\kappa=-1$.  Assuming a general parameterization as in (\ref{comprare})  the outcome of the extremization procedure is that $\epsilon_I=\epsilon_{B_I}=0$, so that the 2d  and 
 4d R-symmetries are identified\footnote{This holds if the 4d 't Hooft anomaly coefficients obey $k_{RRF}=k_F=0$ and $k_{RRB}=k_B=0$, which is true for all quiver gauge theories with toric Sasaki--Einstein duals
 \cite{Butti:2005vn}.}, namely $R^{2d}=R^{4d}$.
 
 At leading order in $N$, this yields the universal relation 
 \be
 c_R=c_L=\frac{32}{3}(g-1)a^{4d}~.
 \label{univc}
 \ee
 Recalling that (at leading order in $N$) in  4d theories 
  \be
  a^{4d} = \frac{9}{32}\sum_i(R_i^{4d}-1)^3~,
  \ee
one sees that (\ref{univc}) is indeed equivalent to $R^{2d}=R^{4d}$ and 
\be
c_R = - 3 \cdot 2 (g-1)    \sum_i \left( -\tfrac{1}{2}\right)(R^{4d}-1) (R^{4d}-1)^2 \,,
\ee
 where $-\frac{1}{2}(R^{4d}-1)$ is the net  number of 2d fermion zero modes associated to each 4d fermion.

 The results of section \ref{sec:Holcharsol1} may be used to compare with the constant $\tau$ version presented here by setting $\text{deg}(\mathcal{L}_{D})=0$. We see that, as noted in \cite{Benini:2015bwz}, the central charges match exactly. Moreover we see that the holographic computations for constant $\tau$ implies that $c_{L}-c_{R}={\cal O}(1)$ as follows from the field theory computation. Finally the results for the R-charges as presented in section \ref{sec:Holcharsol1} are in agreement with the results from the field theory computation.  
  \subsubsection{Baryonic Twist}
 \label{herearethepuzzles}
 
 Let us now consider theories that possess at least one baryonic symmetry with generator $T_B$, so that we can twist as  
 \bea
T_\text{twist} & = & B T_B + \frac{\kappa}{2} T_R^{4d}~,
\label{bariwsit}
\eea
and in particular the theories can now be compactified on a torus, $C_1=T^2$,  with $\kappa=0$. This twist is purely baryonic and 
for concreteness we focus on the $Y^{p,q}$ theories, which have $n_B=1$ and $n_F=2$.
 One finds that  extremizing $k_{RR}$ gives
\bea
\epsilon_1=0~,\quad \qquad \epsilon_2 = \frac{p+\bbcw}{3q}~, \quad \qquad \epsilon_B = \frac{p- 2\bbcw}{3q^2}~,
\label{pippi}
\eea
and 
\bea
c_R = c_L = -\frac{6Bp^2(p^2-q^2)}{q^2}N^2 ~.
\label{crcl}
\eea
Note that $B<0$. 
As remarked in  \cite{Benini:2015bwz}, from (\ref{pippi})  we see that the 2d superconformal R-symmetry involves mixing the 4d one with the baryonic symmetry. Moreover, notice that despite the mixing coefficients $\epsilon_2,\epsilon_B$
and the 't Hooft anomalies being irrational  numbers, this irrationality drops out of the final expression for $c_R$.

This result matches that of the holographic computation   \cite{Donos:2008ug}  (\emph{c.f.} (\ref{csugra})) upon making the following identifications   \cite{Benini:2015bwz}:
\be
\podd = p~, \qquad \qodd  = q~, \qquad  M=BN~.
\label{doesntwork}
\ee
However,  some comments are now in order. First of all, we note that since $\podd< \qodd$ and $p>q$, strictly speaking this identification is \emph{contradictory}. This issue was overlooked in the literature  and certainly deserves further scrutiny in the future. Here we will not attempt to resolve it, but we will make  a number of 
 checks that confirms the plausibility of these identifications.

So far the only assumptions we made on the 2d field theories are that they are $(0,2)$ SCFTs and that their global symmetries are the same as those of the 4d parent theories. 
Assuming in addition that in the 2d SCFTs there exist 2d descendants of the 4d baryonic operators, we can perform some further checks.
The 2d R-charges of the (naive) 2d reduction of the fields $Y,Z,U_\alpha,V_\alpha$ can be computed from (\ref{comprare}), namely using 
\be
R^{2d}[X_{2d}]  =  \epsilon_2 Q_{F_2}[X_{4d}]+\epsilon_B  Q_B[X_{4d}]+ R^{4d}[X_{4d}] \,.
\ee
Plugging in (\ref{pippi}) and the values of the 4d charges gathered from Table \ref{tab:YpqCharges} we obtain
\begin{align}
R^{2d}[Y_{2d}]&=R^{2d}[Z_{2d}]= \frac{q^{2}-p^{2}}{q^{2}}~,\nn \\
R^{2d}[U_{2d}]&=\frac{p^{2}}{q^{2}}~,\qquad R^{2d}[V_{2d}]=1~,
\label{pencil}
\end{align}
 in agreement  with the results (\ref{susyvolumes}) for the normalised volumes of calibrated submanifolds. However, in the field theory  the  $R$-charges associated to the fields $Y$ and $Z$ are \emph{negative}, indicating that a better understanding of the duality proposed in \cite{Benini:2015bwz} would be desirable.

\subsection{Duality Twisted  ${\cal N}=1$  Field Theories}

In this subsection we shall extend the above computations to  compactifications of the four-dimensional theories on a Riemann surface $C_g$, with  $\tau$ varying (holomorphically) over this. In particular, we shall promote the $U(1)_{D}$ symmetry obtained for varying $\tau$ to be a bundle over the Riemann surface $C_g$, with curvature two-form $\dd Q$. This implies we must introduce additional terms to the 4d anomaly polynomial for this bundle. By introducing the additional $U(1)_D$ bundle the 4d anomaly polynomial $I_6$ is modified by the inclusion of additional terms as
\begin{align}
I_{6}^\tau=& I_{6}+ \frac{1}{2}k_{D IJ}c_{1}(\mathcal{F}_D)\wedge c_{1}(\mathcal{F}_I)\wedge c_{1}(\mathcal{F}_J) + k_{DD I}c_{1}(\mathcal{F}_D)\wedge c_{1}(\mathcal{F}_D)\wedge c_{1}(\mathcal{F}_{I})\nonumber\\
&+k_{DDD}c_{1}(\mathcal{F}_D)\wedge c_{1}(\mathcal{F}_D)\wedge c_{1}(\mathcal{F}_D)- \frac{1}{24} k_{D} c_{1}(\mathcal{F}_D)\wedge p_{1}(T_4) \,,
\label{tiger}\end{align}
where $I\in \{ R, B_{I},F_{I} \}$ as before. 
The anomaly polynomial for the 2d theory, $I_4^\tau$ is again computed by integrating $I_{6}^\tau$ over $C_g$. 

With the introduction of the additional $U(1)$ bundle in the anomaly polynomial we must compute the additional `t Hooft cubic and linear anomalies involving $U(1)_D$. We shall argue that the cubic and linear anomalies involving the duality bundle will scale as $N$ and by making a   plausible  assumption we will be able to compute sub-leading contributions to the 2d anomalies, obtaining perfect agreement with the holographic computations.

 Let us consider for example the linear trace 
\bea
k_D & \equiv &   \mathrm{Tr}[U(1)_D] = \sum_{i} q_D^i  ~,
\eea
where the sum is over all the fermions (of the 4d theory) and $q_D^i$ are their charges under $U(1)_D$. However, exactly as for ${\cal N}=4$ SYM, in the non-abelian theories the bonus $U(1)_D$ is \emph{not} a 
symmetry \cite{Intriligator:1998ig} and therefore these charges are not meaningful.

To circumvent this problem, it is expedient to Higgs the ${\cal N}=1$ quiver theories with gauge group $G=SU(N)^{\chi}$ to an abelian theory, at a generic point of the (mesonic) vacuum moduli space. 
In the low energy limit this theory has gauge group $U(1)^{N-1}$ and contains $N-1$ vector multiplets and $3N$ chiral multiplets, parameterising the flat directions of the mesonic moduli space, that is 
the symmetric product of $N$ copies of the related Calabi--Yau three-fold conical singularity $\mathbb{X}=C(Y)$,  Sym$^N \mathbb{X}$. See \cite{Dymarsky:2005xt} for some  discussion in the case of the Klebanov-Witten model with $G=SU(N)^2$, and 
\cite{Berenstein:2005xa} for an explicit analysis in the $Y^{p,q}$ theories.  This is an abelian theory for which $U(1)_D$ is now  a symmetry of the equations of motion, and we can infer the charges of the fields under
$U(1)_D$ from the supergravity analysis.

As we recall in Appendix \ref{app:SL2}, in our conventions the  supergravity Killing spinors have charge $q_D=-1/2$.  In the boundary (abelian) 
field theory this translates to the fact that the scalar field $\phi$ and the fermion field $\psi$ in a chiral multiplet have $U(1)_D$  charges satisfying $q_D[\phi]-q_D[\psi]=-1/2$. The $U(1)_D$ charge of the scalar 
bifundamental fields can be fixed by an extension of the arguments in  \cite{Intriligator:1998ig}, by noting that mesonic gauge-invariants operators (closed loops in the quiver) correspond to scalar 
harmonics on the  Sasaki--Einstein manifold  $Y$ that are in 1--1 correspondence with holomorphic functions on the cone \cite{Gauntlett:2006vf}. In particular, these modes are fluctuations of a mixture of the metric and the RR four-form 
potential \cite{Kim:1985ez}. Since  these  are both inert under $SL_2\R$ transformations, it follows that an infinite tower of dual scalar operators is uncharged under $U(1)_D$. In  
 ${\cal N}=4$  SYM   these operators are Tr$X^{I_1}X^{I_2}\cdots X^{I_k}$  \cite{Witten:1998qj}
and correspond to a KK tower on $S^5$, uncharged under $U(1)_D$ \cite{Gunaydin:1984fk}. 
This clearly implies that the scalar bifundamental fields themselves must be uncharged and therefore the fermions in the chiral mutliplets have $q_D[\psi]=1/2$. The $U(1)_D$ charge of the 
gauginos is fixed by the (abelian) supersymmetry transformations to be $q_D[\lambda]=1/2$. Putting all together, we obtain
\be
k_D  =   \frac{1}{2} 3N+ \frac{1}{2} (N-1)   ~=~ 2N -\frac{1}{2}  \quad \mathrm{(at~a~generic~point~on~the~Higgs~branch)} \,,
\ee
It remains to justify the assumption that, differently from other symmetries,  for $U(1)_D$ there are no other contributions on the Higgs branch, arising from integrating out the massive off-diagonal modes 
\cite{Shimizu:2017kzs,Intriligator:2014eaa}. This is plausible, as  at the origin of the Higgs branch $U(1)_D$ ceases to be a symmetry. Moreover, this scaling with $N$ is fully consistent with the results for the $(0,4)$ 
theories that we discussed in  \cite{Couzens:2017way}.  We will return to this elsewhere \cite{TBA}.

Using this prescription it is straightforward to compute  the mixed cubic anomaly coefficients involving one $D$ index, $k_{DIJ}$. However, the result of this computation will provide us  the sub-leading term\footnote{There is no contribution to $c_R$ 
from the 37 sector, therefore this is the full contribution.} of $c_R$, which we have not attempted to compute holographically, and therefore we do not present the results here. It would be  interesting to compute this performing a KK analysis 
of the $U(1)_R$ isometry in the geometry, along the lines of \cite{Kraus:2005zm}. Below we will discuss the matching with the holographic computations of $c_R,c_L$ at leading order in $N$, and of $c_R-c_L$ at sub-leading order.


\subsubsection{Universal Duality Twist: Elliptic Surface $S_4^\tau$ Case}
\label{sec:UTFTSurface}

Let us now consider the field theory dual to the solutions discussed in section \ref{sol1deriv} and compare with the results of section \ref{sec:Holcharsol1}. Like the universal twist solutions revised above we shall compensate for the curvature of the base by coupling the 4d R-symmetry to a background field. This however is not sufficient to cancel off all of the curvature of $C_g$ and we must also twist with $U(1)_{D}$. As before we allow the 2d R-symmetry to mix with the flavour and baryonic flavour symmetries, however we do not allow it to mix with $U(1)_{D}$, as implied by the  analysis in the gravity side. 
The topological twist ensures that the Killing spinor equation on $\Sigma$ admits a constant spinor solution. To achieve this, couple to two background fields $\mathcal{A}_{i}$ (unlike the constant $\tau$ cases) as
\be\label{SigmaKSE}
(\nabla_{\Sigma} +\ii \mathcal{A}_{1} T_{D}+\ii \mathcal{A}_{2} T_{R})\epsilon=0
\ee
and tune these fields to cancel off the spin-connection on $\Sigma$. On $\Sigma$ there is a single non-trivial component of the spin connection which satisfies
\be
\dd \omega^{12}= \mathfrak{R}=-3 J -\dd Q~.
\ee
Requiring that $\tau$ is holomorphic on $\Sigma$ implies that the spinor on $\Sigma$ satisfies the projection condition $\gamma^{12}\epsilon=-\ii \epsilon$ and therefore requiring that a constant spinor satisfies \eqref{SigmaKSE} implies the topological twist 
\be
\dd \mathcal{A}_{1}= -\dd Q \,,\qquad  \dd \mathcal{A}_{2}=3 J \,,
\ee
which is precisely like the topological duality twists in \cite{Lawrie:2016axq, Couzens:2017way} and 
 results in the twisted $U(1)$
\be
T_{\text{twist}}= T_{D}-\frac{1}{2}T_{R}^{4d} \,,
\ee
whilst the trial R-charge is given by (\ref{comprare}). Concretely the twisting induces the following identifications of the curvatures of the various bundles 
\be
\begin{array}{ll}
\mathcal{F}_{R}^{4d}&\rightarrow \mathcal{F}_{R}^{2d}-\frac{3}{2}J_{\Sigma} ~,\\
\mathcal{F}_{F_{I}}^{4d}&\rightarrow \mathcal{F}_{F_{I}}^{2d}+\epsilon_{I}\mathcal{F}_{R}^{2d}~,\\
\mathcal{F}_{B_{I}}^{4d}&\rightarrow \mathcal{F}_{B_{I}}^{2d}+\epsilon_{B_{I}}\mathcal{F}_{R}^{2d}~,\\
\mathcal{F}_{D}^{4d}&\rightarrow 2\pi c_{1}(\mathcal{L}_{D})~,
\end{array}
\ee
where $F$ are the flavour symmetries and $B$ the baryonic symmetries. Upon extracting the $k_{RR}$ coefficient and extremizing with respect to the $\epsilon$'s one finds
\be
\epsilon_{I}=0=\epsilon_{B_{I}} \,,
\ee
and therefore there is no mixing in 2d of the exact R-symmetry and the flavour and baryonic symmetries. This is true at leading order but may be corrected at subleading order due to cubic `t Hooft anomalies involving $U(1)_{D}$. The central charge is 
given by $c_{R}=3 k_{RR}$ and is obtained from reducing the $I_{6}$ on the base of the elliptic surface, $\Sigma$, as
\be
c_R = \frac{16}{3}  (2(g-1)+\text{deg}(\mathcal{L}_{D}))a^{(4d)} \,,
\ee
which is in perfect agreement with \eqref{sol1c}. 
An important point to note here is that the central charge has at leading order already a $\tau$-dependence through $\mathcal{L}_D$. 

By extracting $k$ from the $I_{4}^{\tau}$ anomaly polynomial we find the subleading contribution to be
\be
(c_{L}-c_{R}  )_{\rm bulk}  = - k_{D}\int_{\Sigma} c_{1}(\mathcal{L}_{D})=-k_{D}\text{deg}(\mathcal{L}_{D})~.
\ee 
The subscript indicates that this contribution  arises from the dimensional reduction of the 4d theory, ignoring the defect modes from the 7-branes.   
Furthermore, assuming that the contributions of the 7-branes to the spectrum are again Fermi multiplets  as in \cite{Lawrie:2016axq},  we can conjecture that 
the 3-7 defect modes gives an additional contribution 
\be\label{defectmodes}
 (c_{L}-c_{R})_{\text{defect}}=8N \text{deg}(\mathcal{L}_{D}) \,.
 \ee
From the discussion at the begining of this section we have $k_{D}= 2 N -1/2$ so that at  sub-leading order we obtain the total contribution
\be
 c_{L}-c_{R}= 8 N \text{deg}(\mathcal{L}_{D}) - 2 N\text{deg}(\mathcal{L}_{D})=6 N \text{deg}(\mathcal{L}_{D}) ~,
\ee
which agrees with the result given in \eqref{sol1CLCR}. 

One may also compute the R-charges of the fields from the anomaly polynomial. As the extremization forces all the $\epsilon_{I}$ to vanish one finds that the R-charges of the 2d fields are the same as the R-charges of the 4d fields, 
in agreement with the conclusion reached in section \ref{sec:Holcharsol1}.


\subsubsection{Universal Twist: Elliptic Three-fold $\mathcal{T}^\tau_6$ Case}
\label{sec:UTFTThreeFold}

The field theory duals to the universal twists with an elliptic three-fold factor are obtained by a twisted reduction of the 4d $\mathcal{N}=1$ SCFTs in section \ref{sec:AdS5}, whose duals are F-theoretic $\AdS_5$ solutions. The field theory is reduced along a curve with constant $\tau$, so that the standard universal twist of \cite{Benini:2015bwz}  can be implemented as in \eqref{unitwist}, with the trial R-symmetry given as usual. In the following we shall assume that the 4d 't Hooft coefficients still obey $k_{RRF}=k_F=0$ and $k_{RRB}=k_B=0$ as in the toric Sasaki--Einstein case, we make no restriction on $k_{RRD}$ and $k_{D}$. This starting point implies that the 2d R-symmetry to leading order is given exactly by the 4d one, and we have $\epsilon_{I}=\epsilon_{B_{i}}=0$. In particular the central charge is
\be
c_{R}=\frac{32(g_{\Sigma}-1)}{3} a_{\tau}^{(4d)} \,,
\ee
which agrees with the holographic result. Moreover the subleading contribution is given by 
\be\label{uniclcr3fold}
c_{L}-c_{R}= (g-1) k_{R}^{\tau}~.
\ee
Since this corresponds to the twisted reduction on $\mathbb{H}^{2}/\Gamma$ above which the theory has no varying coupling this is the exact result to this order.
 In the constant $\tau$ field theory one has to subleading order $k_{R}=0$ and therefore $c_{L}=c_{R}$ at subleading order. As discussed in section \ref{sec:uni3fold}, $k_{R}^{\tau}$ is non-zero at sub-leading order in the varying $\tau$ field theory, and therefore non-trivial $\tau$ not only modifies the leading order central charge of the theory it also implies that the left and right moving central charges differ at subleading order.

As a final check of our results in section \ref{sec:uni3fold} the identification of the 2d R-symmetry with the 4d one implies that the R-charges of the fields in 2d and 4d are identical, 
which agrees with the results presented {in the holographic setup}.


\subsubsection{Baryonic Duality Twist}\label{sec:BaryonicFT}

We now discuss theories with varying coupling, which have a baryonic symmetry. We can compactify on a complex curve $C_g$ of genus $g\neq 1$ and preserve supersymmetry by twisting with $U(1)_D$, as explained in section \ref{dualtwistedFTREV}. As the supercharges are uncharged under the baryonic (and flavour) symmetries we are free to twist with these as well. In particular, we take 
$C_0=\mathbb{P}^{1}$, with  curvature given by  $-\dd Q$, which is also the connection of the duality line bundle $\mathcal{L}_{D}$. 
Concretely, the topological twist we take is 
\be
T_{\text{twist}}= B T_{B} + T_{D}~.
\ee
We again assume that  the R-symmetry does not mix with $U(1)_D$ and  therefore we take as trial R-charge 
\be
T_{\text{trial}}=\epsilon_{2}T_{2}+\epsilon_{B} T_{B}+ T_{R}^{4d}~.
\ee
Under the twisting the curvatures of the various bundles become\footnote{Note that there is a minus sign difference in the $\mathcal{F}_{B}$ term with that in equation (2.47) of \cite{Benini:2015bwz}. We fixed this by first recovering the results for constant $\tau$ on a $T^2$ via the anomaly polynomial.}
\be\ba
\mathcal{F}_{R}^{4d}&\rightarrow \mathcal{F}_{R}^{2d}~,\\
 \mathcal{F}_{F_{1}}^{4d}&\rightarrow \mathcal{F}_{F_{1}}^{2d}\\
\mathcal{F}_{F_{2}}^{4d}&\rightarrow \mathcal{F}_{F_{2}}^{2d}+\epsilon_{2}\mathcal{F}_{R}^{2d}~,\\
\mathcal{F}_{B}^{4d}&\rightarrow \mathcal{F}_{B}^{2d}+\epsilon_{B}\mathcal{F}_{R}^{2d}- B t_g~,\\
\mathcal{F}_{D}^{4d}&\rightarrow c_{1}(\mathbb{P}^{1})~.
\ea\ee
The last line is fixed as the compactification geometry is an elliptic K3-surface.
The anomaly polynomial for the 2d theory, $I_4^\tau$ is computed by integrating $I_{6}^\tau$ in (\ref{tiger}) over the base $\mathbb{P}^1$ of the elliptic $K3^\tau$\footnote{As the expression one finds for $I_{4}$ is unwieldy we present only the salient terms.} 
\begin{align}
 \int_{\mathbb{P}^1} I_6^{\tau}&=I_4^{\tau}\supset -( B(k_{R1B}+k_{12B} \epsilon_{2} +k_{1BB}\epsilon_{B})+2 (k_{R1D}+k_{12D}\epsilon_{2}+k_{1BD} \epsilon_{B}))c_{1}(\mathcal{F}_1)\wedge c_{1}(\mathcal{F}_{R})\nonumber\\
& -( B(k_{R2B} +k_{22B}\epsilon_{2}+k_{2BB}\epsilon_{B})+2 (k_{R2D}+k_{22D}\epsilon_{2}+k_{2BD}\epsilon_{B}))c_{1}(\mathcal{F}_{2})\wedge c_{1}(\mathcal{F}_{R})\nonumber\\
 &-(B(k_{RBB}+k_{2BB}\epsilon_{2}+k_{BBB}\epsilon_{B})+2(k_{RBD}+k_{2BD}\epsilon_{2}+k_{BBD}\epsilon_{B}))c_{1}(\mathcal{F}_{B})\wedge c_{1}(\mathcal{F}_{R})\nonumber\\
 &-\frac{1}{2}\left[ B \{ k_{RRB}+\epsilon_{B}(2 k_{RBB}+k_{BBB}\epsilon_{B})+\epsilon_{2}(2 k_{RRB}+k_{22B}\epsilon_{2}+2 k_{2BD}\epsilon_{B})\}\right.\nonumber\\
& \left. +2 \{k_{RRD}+\epsilon_{B}(2 k_{RBD}+k_{BBD}\epsilon_{B})+\epsilon_{2}(2 k_{R2D}+k_{22D}\epsilon_{2}+2 k_{2BD}\epsilon_{B})\}\right]c_{1}(\mathcal{F}_{R})^{2} \nonumber\\
&+\frac{1}{24}(B k_{B}+2 k_{D})p_{1}(T_2)
\end{align}
Comparing this with the general structure of the $I_4$ polynomial (\ref{AnomalyI4}) and (\ref{cRcL}) yields 
\be\ba\label{porg}
c_{R}&=3k_{RR}=-3\left[ B \{ k_{RRB}+\epsilon_{B}(2 k_{RBB}+k_{BBB}\epsilon_{B})+\epsilon_{2}(2 k_{RRB}+k_{22B}\epsilon_{2}+2 k_{2BD}\epsilon_{B})\}\right. \cr 
& \left. +2 \{k_{RRD}+\epsilon_{B}(2 k_{RBD}+k_{BBD}\epsilon_{B})+\epsilon_{2}(2 k_{R2D}+k_{22D}\epsilon_{2}+2 k_{2BD}\epsilon_{B})\}\right]~,\cr 
c_L- c_R &= - B k_{B}- 2 k_{D} \,   .
\ea\ee

The exact central charge is obtained by extremizing $c_R$ with respect to $\epsilon_{B}, \epsilon_{2}$, the expression one obtains is prohibitorily large and so we do not present it here.  
The key is to note how the various 't Hooft anomalies scale with $N$ \cite{TBA}.\footnote{We thank Craig Lawrie for discussions on this point and collaboration in \cite{TBA}.} Those not involving the duality symmetry, $U(1)_D$ will be unaffected by its inclusion and scale as $N^2$, on the other hand any term involving $U(1)_D$ will scale as $N$ and therefore it will be sub-leading.
Observe that in the universal twist solutions presented previously a non-trivial variation induces a shift in
the central charge at leading order, not just at subleading order as is the present case.

Note that so far we have not specified a theory, and therefore the conclusion that the leading order central charge is unchanged with respect to the value of the same theory, compactified on $C_{g=1}=T^2$, and twisted by 
$T_\text{twist}  =  B T_B$ is quite general. Specializing to the $Y^{p,q}$ quivers, we of course recover the result (\ref{crcl})
\be
c_R =- \frac{6 B  p^2 \left(p^2-q^2\right)}{q^2} N^2 + O(N) \,.
\ee 
This is in agreement with our observations from gravity that the corrections due to $\tau$ are sub-leading in $N$.
Using $k_B=0$ we also obtain
\be
(c_L - c_R)_{\rm bulk} = - 4 N +1 \,,
\ee
where again the subscript indicates that this contribution  arises from the dimensional reduction of the 4d theory, ignoring the defect modes from the 7-branes.   
For an elliptic K3 the 3-7 defect modes gives an additional contribution 
\be
 (c_L - c_R)_{\rm defect}  = 16 N \,,
\ee
so that  the total contribution at order $O(N)$ is precisely $12N$ as in (\ref{subway}). 


\section{Dual M-theory $\AdS_2$ Solutions}
\label{sec:M}

In all F-theory solutions, where the axio-dilaton varies non-trivially and thus there exist singular fibers, the metric on the base of the elliptic fibration, which is contained in the Type IIB spacetime, is necessarily singular. 
This is the case even if we assume a smooth Weierstrass model, which only has $I_1$ singular fibers.
In light of this, it is advisable to also consider the dual M-theory solutions, where the total space of the elliptic fibration becomes part of the full spacetime, and thus statements about the existence of smooth metrics on the total space can be used. 

For the $(0,4)$ solutions in \cite{Couzens:2017way} a dual $\AdS_3$ solution existed and the elliptic Calabi-Yau three-fold, was a factor of this solution, and various statements in F-theory could be substantiated in this way.  In the present cases of duals to $(0,2)$ SCFTs, one might therefore also wish to 
consider the dual M-theory AdS$_3$ solution obtained by T-dualising along one of the internal non R-symmetry Killing directions and uplift.
However we find that in all cases (universal twist and baryonic twist), the resulting spacetime has a key difference to the $(0,4)$ setup: the total space of the elliptic fibration does become part of the M-theory compactification space, however there is a warp-factor which only affects the base of the elliptic fibration. As it is far from clear, whether there is a smooth metric on this total space of warping plus elliptic fibration, we will here dualise to $\AdS_2$ solutions in M-theory, which do not have this problem. 
This provides M-theoretic duals to the entire class of F-theory  geometries that we studied in this paper, albeit one which admits 1d dual SCFTs. It would be very interesting to explore further the connection between these and the 2d SCFTs that are prominent in the F-theory duality frame.

\subsection{Comparison to known AdS$_2$ Solutions in M-theory}

The ``master equation'' \eqref{Master8d} is structurally the same as the equation in \cite{Kim:2006qu} governing AdS$_2$ solutions in 11d supergravity with only electric flux. As we shall see the solutions presented in this paper are dual to a subclass of those solutions when the (real) 8d K\"ahler base is taken to be elliptically fibered. To perform this duality chain we must write the AdS$_3$ metric as a foliation by AdS$_2$ \cite{Gauntlett:2006ns}, that is we use the metric
\be
\dd s^{2}(\text{AdS}_{3})= \frac{1}{4m^{2}}\lb -r^{2}\dd t^{2}+\frac{\dd r^{2}}{r^{2}}+(2 \dd \varphi +r \dd t)^{2}\rb~.
\ee
We have normalised the metric such that the Ricci-tensor satisfies $R_{\mu\nu}=-2m^{2} g_{\mu \nu}$. One may now perform a T-duality along the azimuthal coordinate $\varphi$ to obtain the metric on AdS$_2\times S^1$ with the full $SO(1,1)$ isometry group of AdS$_2$ preserved. Performing the T-duality on the general Type IIB solution given in \eqref{generalmetric} and \eqref{ffdef} results in the string frame Type IIA solution
\begin{align}
m^{2} \dd s^{2}(\M_{IIA})&= \frac{\me^{2 \Delta}}{\sqrt{\tau_{2}}}\lb \frac{1}{4}\lb -r^{2}\dd t^{2}+\frac{\dd r^{2}}{r^{2}}\rb+\frac{1}{4}(\dd \chi+\rho)^{2}+\me^{-4 \Delta}\dd s^{2}(\M_{6})\rb+\sqrt{\tau_{2}}\me^{-2 \Delta}\dd \varphi^{2}~, \nonumber\\
F^A_{4}&=\frac{1}{4 m^{2}}\dd\vol(\text{AdS}_{2})\wedge \ff~,\nonumber\\
F^A_{2}&= \frac{1}{m} \dd \tau_{1}\wedge \dd \varphi~,\\
H&=\frac{1}{2} \dd \varphi \wedge \dd \vol(\text{AdS}_2)~,\nonumber\\
\me^{-2 \Phi_{IIA}}&=\tau_{2}^{\frac{3}{2}} \me^{2 \Delta} \,, \nonumber
\end{align}
which uplifts to 11d supergravity as an $\AdS_2\times \mathcal{M}_9$ solution
\begin{align}
m^2 \dd s^{2}(\M_{11})&= \me^{\frac{8 \Delta}{3}}\lb \frac{m^2}{4}\dd s^{2}(\text{AdS}_{2})+\frac{1}{4}(\dd \chi+\rho)^{2}+\me^{-4 \Delta}\lb \dd s^{2}(\M_{6}) + \tau_{2}\dd \varphi^{2}+\frac{1}{\tau_{2}}(\dd \psi+\tau_{1}\dd \varphi)^{2}\rb\rb~,\nonumber\\
G_{4}&= \frac{1}{4m}\dd\vol(\text{AdS}_{2})\wedge \left[ -2 J_{8}-\frac{1}{2} \dd (\me^{4 \Delta}(\dd \chi+\rho))\right]~, 
\end{align}
which agrees with the general form presented in \cite{Kim:2006qu} upon making the identifications 
\be
A_{KP}=\frac{4\Delta}{3}~,~~~~B_{KP}= \frac{1}{2}\rho~,~~~~\psi_{KP}=\frac{\chi}{2}~.
\ee
Observe that the elliptically fibered space $\mathcal{Y}_{8}^\tau$, that underlies the F-theory solutions of section \ref{F-theory reformulation}, now appears in the solution explicitly as part of the geometry. In the following we shall analyse this map for the previously discussed solutions.


\subsection{Flux Quantisation and Central Charges}

As in the Type IIB solutions presented above it is necessary to quantise the flux through all compact integral cycles in the geometry.
To quantise the flux we impose that over all integral seven-cycles in the geometry, $\{ A_{i}\}\in H_{7}( \M_{9};\Z)$ and integral four-cycles $D_i\in H_{4}(\M_{9};\Z)$ \cite{Witten:1996md}\footnote{One should also quantise $G_4$ through all four-cycles however as the legs always lie along $\AdS_2$ no quantisation is necessary.}
\be\label{Mtheoryquant}
\frac{1}{(2 \pi \lp)^6} \int_{A_{i}} * G_{4} \in \Z~ \,,\qquad 
\int_{D_{i}} \frac{p_{1}}{4}\in \mathbb{Z}~.
\ee
The leading order holographic central charge of the dual 1d SCFTs may be extracted from  \cite{Castro:2014ima} 
\be
{c^{1d}} =\frac{3}{4 \pi G_{N}^{(2)}} \,,\label{2dcentral}
\qquad 
\frac{1}{G_{N}^{(2)}}=\frac{1}{G_{N}^{(11)}}\int_{\M_{9}} \dd \vol(\M_{9})~,
\ee
where $G_N^{(2)}$ is the 2d Newton's constant\footnote{Notice that in the evaluation of the 3d Newton's constant, \eqref{csugraeq}, there is a factor of $\me^{\Delta}$ coming from the warping of AdS$_3$, this is not the case for $G_{N}^{(2)}$. Recall that reduced Newton's constant is computed by extracting the coefficient of the reduction of $\int_{\M_{10(11)}}\sqrt{g}R$. Under a conformal transformation $g\rightarrow \me^{2 \Delta}g$ we have $\sqrt{g}R\rightarrow \me^{(d-2)\Delta} \sqrt{g}R$. The factors of the warp factor appearing (or not appearing) in the formula for the dimensional reduction of Newton's constant is therefore clear.}.

\subsection{$\AdS_2$ Solutions with Elliptic Surface Factor}

From the general form of the AdS$_2$ solutions we have
\begin{align}
\dd s^{2}(\M_{11})&=\frac{\me^{\frac{8 \Delta}{3}}}{4}\left[ \dd s^{2}(\text{AdS}_{2}) +\frac{9}{m^2}\lb \frac{1}{9}(\dd \chi +\rho)^2 + \dd s^{2} (\mathcal{M}_{4}) +\dd s^{2}(\mathcal{S}^\tau_{4})\rb \right]~,\\
G_{4}&= -\frac{1}{2m} \dd \vol(\text{AdS}_{2}) \wedge \left[ J_{\mathcal{S}^\tau_4} +\frac{1}{3}(J_{\M_{4}}+ J_{\Sigma})\right]~,\\
\me^{-4 \Delta}&= \frac{9}{4}~,\\
\rho&=- 6 \mathcal{A}_{\M_{4}}+3 \mathcal{A}_{\Sigma}~,
\end{align}
where as before $\dd \mathcal{A}_{i}=J_{i}$. Note that the factors in the metric have conspired such that the metric on $\Sigma$ and the elliptic fibration combine into the smooth metric on the elliptic surface $\mathcal{S}^\tau_{4}$.

\subsubsection*{Flux Quantisation}

First consider the condition on the Pontryagin class following (\ref{Mtheoryquant}). There are two four-cycles to consider, $\M_{4}$ and $\mathcal{S}^\tau_4$. First note that
\be
p_1(\M_{9})=p_{1}(\M_{4})+p_{1}(\mathcal{S}^\tau_4)~.
\ee
We see that both the manifold $\mathcal{S}^\tau_4$ and $\M_{4}$ must have first Pontryagin class divisible by four. For the base of $Y^{p,q}$ for example this has vanishing first Pontryagin class. 

There are three distinct types of seven-cycle in the geometry; the seven cycle, $D_{\Sigma}$ given by the five-cycle $\M_5^\tau$ fibered over $\Sigma$, the five-cycle $\M_5^\tau$ with the elliptic fiber which we call $D_{E}$ and finally the seven-cycles $D_{\alpha}$ consisting of a three-cycle in $\M_5^\tau$ fibered over $\Sigma$ along with the elliptic fibration. From the rules of T-duality and uplift one finds the relations
\be
\lp^3=\frac{m \ls^4}{\beta}~,~~~\vol(\mathbb{E}_{\tau})=\frac{(2\pi)^2 \lp^6 m^2}{\ls^4 \gs}~,
\ee
where $\beta$ is the period of the $U(1)$ isometry dualised along.
We find
\be
\ba
n(D_{E})&= \frac{1}{(2 \pi \lp)^6}\int_{\M_5^\tau\times \mathbb{E}_{\tau}}*G_{4}= \frac{1}{(2 \pi \lp)^6}\int_{\M_5^\tau\times \mathbb{E}_{\tau}}\lb -\frac{3}{2 m^6}\rb D\chi\wedge J_{4}\wedge J_{4}\wedge J_{\mathbb{E}_{\tau}}\cr 
&=\frac{9 \vol(SE_{5})}{(2 \pi m \ls)^4 \gs}\equiv N \,,
\ea
\ee
which is the same quantisation condition as in Type IIB/F-theory. Consider next the second type of seven-cycle containing the elliptic fibration, then
\be
\ba
n(D_{\alpha})=\frac{1}{(2 \pi\lp)^6}\int_{\Sigma_{\alpha}\times \mathcal{T}^\tau_6}-\frac{3}{ 4 m^6} D\chi \wedge J_{4} \wedge J_{\Sigma} \wedge J_{\mathbb{E}_{\tau}}=\frac{(2 \pi)^3 \ell \tilde{m} n_{\alpha}}{(2 \pi m \ls)^4 \gs }(2(g-1) +\text{deg}(\mathcal{L}_{D}))~.
\ea
\ee
Over the final type of seven-cycle we have 
\begin{align}
\frac{1}{(2\pi \lp)^6}\int_{D_{\Sigma}}*G_{4}= \frac{2 \beta^{2} a^{4d}}{9}(2(g-1)+\text{deg}(\mathcal{L}_{D}))~.
\end{align}
Flux quantization implies that $\beta$ needs to take particular values, depending on $a^{4d}$ and the geometric data, and only in these cases do we expect the M-theory solution to be consistent\footnote{This expression is in fact proportional to the central charge in the  $\AdS_3$ dual, and it would be interesting to understand what the physical interpretation of this quantization is.}.

\subsubsection*{Central Charge}
Using the formula for the central charge in equation \eqref{2dcentral} we find
\begin{align}
{c^{1d}}&=\frac{3}{4\pi}\frac{1}{2^4 \pi^7 \lp^9}\lb \frac{3}{2m}\rb^9 \me^{12\Delta} \vol(SE_{5})\vol(\Sigma) \vol(\mathbb{E}_{\tau})\nonumber\\
&=\frac{16 a^{4d}}{3}(2(g-1)+\text{deg}(\mathcal{L}_{D})~.
\end{align}
We find agreement with the central charge obtained from the Type IIB solution as expected.

\subsection{AdS$_2$ Solutions with Elliptic Three-fold Factor}

From the general form of the AdS$_2$ solutions we have
\begin{align}
\dd s^{2}(\M_{11})&=\frac{\me^{\frac{8 \Delta}{3}}}{4}\left[ \dd s^{2}(\text{AdS}_{2}) +\frac{9}{m^2}\lb \frac{1}{9}(\dd \chi +\rho)^2 + \dd s^{2} (\mathbb{H}^2) +\dd s^{2}(\mathcal{T}^\tau_6)\rb \right]~,\\
G_{4}&= -\frac{1}{2m} \dd \vol(\text{AdS}_{2}) \wedge \left[ J_{\mathbb{E}_{\tau}} +\frac{1}{3}(J_{\M_{4}}+4 J_{\Sigma})\right]~,\\
\me^{-4 \Delta}&= \frac{9}{4}~,\\
\rho&=- 6 \mathcal{A}_{\M_{4}}+3 \mathcal{A}_{\Sigma}~,
\end{align}
where as before $\dd \mathcal{A}_{i}=J_{i}$. By assumption the metric on $\mathcal{T}^\tau_6$ is smooth. 

%
%
\subsubsection*{Flux Quantisation}
As before we begin by considering the condition on the first Pontryagin class. There are two types of four-cycles to consider; the base $\M_{4}$ and a two-cycle in $\M_{4}$ with the elliptic fibration over it and 
\be
p_1(\M_{9})=p_{1}(\mathcal{T}^\tau_6)~.
\ee
We must impose the topological restriction that the first Pontryagin class of $\mathcal{T}^\tau_6$ is divisible by 4.

As in the previous class of solution there are three distinct types of seven-cycle in the geometry; the seven cycle, $D_{\Sigma}$ given by the five-cycle $\M_5^\tau$ fibered over $\Sigma$, the $U(1)$ fibration over $\mathcal{T}^\tau_6$ which we call $D_{E}$ and finally the seven-cycles $D_{\alpha}$ consisting of the elliptic fibration over a three-cycle in $\M_5^\tau$ fibered over $\Sigma$. The same relations between $\ls, \lp$ and $\vol(\mathbb{E}_{\tau})$ hold as in the previous case.
The first quantization condition is 
\be
n(D_{E})= \frac{1}{(2 \pi \lp)^6}\int_{D_{E}}*G_{4}
=\frac{9 \vol(\M_5^\tau)}{(2 \pi m \ls)^4 \gs}\equiv N \,,
\ee
which is the same as in Type IIB. Consider next the second type of seven-cycle containing the circle fibration over the two-cycle in $\M_{4}$ with the elliptic fibration over it all fibered over {the curve, which we assume to be simply} $\mathbb{H}^{2}$ then  
\be
n(D_{\alpha})=\frac{1}{(2 \pi\lp)^6}\int_{D_{\alpha}}-\frac{3}{ 4 m^6} D\chi \wedge J_{4} \wedge J_{\Sigma} \wedge J_{\mathbb{E}_{\tau}}=\frac{\pi^3 \chi(\Sigma) \ell}{3(2 \pi m\ls)^4 \gs} \lb \tilde{m} n_{\alpha} -\int_{\Sigma_{\alpha}}c_{1}(\mathcal{L}_{D})\rb~.
\ee
Over the final type of seven-cycle we have 
\begin{align}
\frac{1}{(2\pi \lp)^6}\int_{D_{\Sigma}}*G_{4}= \frac{2 \beta^{2} a^{4d}_{\tau}}{9}2(g-1) \,.
\end{align}
As in the previous case we find that this is proportional to the central charge in the $\AdS_3$ case as can be seen by using \eqref{c3foldv1}. We may again tune the period of the $U(1)$ dualised to give an integer result as necessary. 

\subsubsection*{Central Charge}

Using the formula for the central charge in equation \eqref{2dcentral} we find
\begin{align}
{c^{1d}}&=\frac{3}{4\pi}\frac{1}{2^4 \pi^7 \lp^9}\lb \frac{3}{2m}\rb^9 \me^{12\Delta} \vol(\M_5^\tau)\vol(\mathbb{H}^{2}) \vol(\mathbb{E}_{\tau})\nonumber\\
&=\frac{32(g-1) a^{4d}_{\tau}}{3}={c^{2d}}~.
\end{align}
As in the previous case we find agreement with the central charge obtained from the Type IIB solution as expected.


\subsection{AdS$_2\times \Yodd^{\podd,\qodd}\times K3$ Solutions}

Following the discussion in the previous subsection there exists an AdS$_2$ solution of 11d supergravity
which is M/F dual to the baryonic twist solutions in F-theory, of the form  
\begin{align}
\dd s^2=& \frac{\me^{\frac{8 \Delta}{3}}}{4}\lb \dd s^{2}(\text{AdS}_{2})+ w[\dd \psi+g(x) D\phi]^{2}+a \lb \frac{\dd x^{2}}{x^{2}U}+\frac{U}{w}D\phi ^{2}+\dd\theta^{2} +\sin^{2}\theta \dd \chi^2\rb\right. \nonumber\\
&\left. \frac{}{}+4 \me^{-4 \Delta} \dd s^{2}(K3)\rb~,\\
G_4=&- \frac{1}{4m} \dd \vol(\mathrm{AdS}_2) \wedge \lb  \frac{1}{ 2a x^{2}}(D \psi -g(x) D\phi)\wedge \dd x +2 J_{K3}+\frac{1}{2}\sin \theta\dd \theta\wedge \dd \chi\right)~,
\end{align}
where K3 denotes an elliptically fibered K3 surface, $\dd s^{2}(K3)$ is the smooth Ricci-flat K\"ahler metric and $J_{K3}$ is the K\"ahler two-form on the K3. 
By assumption we only have $I_1$ singular fibers, and so the K3 surface is smooth -- recall that the metric induced on the base $B_2= \mathbb{P}^1$ of the fibration, which is part of the Type IIB spacetime,  has singularities.
If we allowed for general singular fibers, then in this M-theory setup, this would allow us to resolve these singularities. In either case we are considering a smooth K3 surface.
Furthermore, the previous regularity analysis for the metric on $\Yodd^{\podd,\qodd}$ shows that the full 11d solution is perfectly smooth.

\subsubsection*{Flux Quantisation}

Let us first consider the Pontryagin class in \eqref{Mtheoryquant}. There are two four-cycles in the geometry the first is the base $\newbase$ of the $U(1)$ fibration of $\Yodd^{\podd,\qodd}$ and the second is the K3 surface. As the metric is topologically 
\be
\text{AdS}_2 \times \Yodd^{\podd,\qodd}\times K3
\ee
it is trivial to check that the first Pontryagin class of the 11d metric splits as $p_1(\M_{11})=p_1(\newbase)+p_{1}(K3)$. As was shown earlier in section \ref{newbase} the signature of $\newbase$ is zero which is a third of the first Pontryagin class for a four-manifold. Similarly $\sigma(K3)=-16$ and therefore no additional constraint is imposed. Instead consider the quatisation of the flux over compact integral seven-cycles. There are two distinguished classes of seven-cycles in the geometry, the K3 surface with the unique three-cycle generator $E\in H_{3}(\Yodd^{\podd,\qodd})$ which we call $\Sigma_{E}$ and the seven-cycles arising from a two-cycle, $\{C_{i}\} \in H_{2}(K3;\Z)$ and $\Yodd^{\podd,\qodd}$.
The flux may be written as
\be
*_{11} G_{4}= \dd \vol(\mathbb{E}_{\tau})\wedge \hat{*}_{7} \ff+ 2m \me^{8 \Delta} \dd \vol(\M_{7}) \,,
\ee
where $\hat{*}_{7}$ is the Hodge star on the unwarped internal manifold in the Type IIB solution.
Consider first the seven-cycle $\Sigma_{E}$
\begin{align}
\frac{1}{(2 \pi \lp )^{6}}\int_{\Sigma_{E}} *G_{4}&=\frac{1}{(2 \pi \lp m)^{6}}\vol(\Kthreebase)\vol(\mathbb{E}_{\tau})\frac{4 \pi^{2}\qodd^{2}}{\podd^{2}(\podd^{2}-\qodd^{2})}\nonumber\\
&=\frac{\vol(\Kthreebase)}{(2 \pi \ls m)^{4}\gs} \frac{4 \qodd^{2}\pi^{2}}{\podd^{2}(\podd^{2}-\qodd^{2})}\equiv M \,,
\end{align}
where we have used
\be
\vol(\mathbb{E}_{\tau})= \frac{(2 \pi)^2\lp^{6} m^{2}}{\ls ^{4} \gs}
\ee
as follows from correctly identifying the periods in the duality chain. Consider instead the seven-cycle $\Yodd^{\podd,\qodd}\times \mathbb{E}_{\tau}$, we find
\begin{align}
\frac{1}{(2 \pi \lp)^{6}}\int _{\Yodd^{\podd,\qodd}\times \mathbb{E}_{\tau}} * G_{4}&= \frac{ 4 \pi^{3}\qodd^{4}}{ (2 \pi \ls m)^4 \podd (\podd^2-\qodd^2)^2}\equiv N~.
\end{align}
Observe that the integers that we have introduced are the same as those appearing in the Type IIB analysis.

\subsubsection*{Central Charge}
For the central charge we obtain
\begin{align} 
{c^{1d}}=\frac{3. 2^3 \pi}{( 2 \pi \lp m)^9}\int _{\M_9} \me^{4 \Delta} \dd \vol(\Yodd^{\podd,\qodd})\wedge \dd \vol(\Kthreebase)\wedge \dd \vol(\mathbb{E}_{\tau})={c^{2d}} \,,
\end{align}
where use has been made of the relation $\lp^3 =\ls^4 m$ as follows from the T-duality and uplift.


\section{Discussion and Outlook}

\label{sec:DO}

The aim of this paper was to extend the $\AdS_3$/CFT$_2$ dictionary in the F-theory context to 2d superconformal field theories with $(0,2)$ supersymmetry.  
There are at least two motivations for pursuing this program: on the one hand, it is interesting 
to explore the relatively uncharted territory of holography in the context of F-theory. 
On the other hand, we believe that it is worthwhile broadening the set of field theories, which have varying coupling, such as was done in the case of 4d ${\cal N}=4$ SYM with topological duality twist \cite{Martucci:2014ema, Assel:2016wcr}.   

The starting point of our analysis is the  derivation of the general constraints for $\AdS_3$ solutions in Type IIB, with varying axio-dilaton, that follow 
 from the Killing spinor equations on the internal geometry (\ref{Master}). 
We then investigated various solutions to these equations.  A  summary of all solutions obtained in this paper can be found in Table \ref{starwars}.

The first class of solutions have enhanced supersymmetry to $(2,2)$, and in this case the axio-dilaton was shown to always be constant. If  $\AdS_3$ is allowed to become  a slicing of an $\AdS_5$ solution, $\tau$ can vary, and the solutions are the most general F-theory solutions dual to 4d $\mathcal{N}=1$ theories. 
The brane-setup is given in terms of D3-branes probing F-theory geometries that are elliptic Calabi-Yau four-folds. A further analysis of these will appear in \cite{CDJS}.

For duals to 2d $(0,2)$ SCFTs there are two classes of solutions discussed in this paper, which are all based on the general form of the F-theory solution (i.e. including the axio-dilaton into the geometric description in terms of the elliptic fibrations)   given by 
\be
\AdS_3\times (S^1\rightarrow \mathcal{Y}_8^\tau) \,.
\ee
Here $\mathcal{Y}_8^\tau$ is elliptically fibered. The base of this elliptic fibration $\widetilde{\mathcal{M}}_6$ is a K\"ahler three-fold. 
The first class of solutions are of the type $\widetilde{\mathcal{M}}_6= \Sigma \times \mathcal{M}_4$, i.e. a product of a curve and a surface. 
This gives rise to the universal twist solutions, where the elliptic fibration is non-trivial only over one of the two factors. The key characteristic of these universal twist solutions in F-theory is that they do not have any Calabi-Yau factors, i.e. the elliptic fibration restricted to $\Sigma$ and $\mathcal{M}_4$, respectively, cannot be Ricci flat! 
The second class of solutions is obtained by imposing that there is explicitly a Ricci-flat {factor in the direct product} $\mathcal{Y}_8^\tau = \mathcal{M}_4 \times K3^\tau $. The resulting solutions are of the type $\AdS_3\times K3^\tau \times \mathfrak{Y}^{\podd, \qodd}$, or as Type IIB solution $\AdS_3 \times \mathbb{P}^1 \times \mathfrak{Y}^{\podd, \qodd}$, where $\mathfrak{Y}^{\podd, \qodd}$ are circle-fibrations over $\mathbb{F}_0$. 
These are the baryonic twist solutions. 
In each case we determined the holographic central charges and matched them to dual field theory, where the central charge is obtained using c-extremization applied in the context of 4d $\mathcal{N}=1$ field theories with varying coupling. {Key to our analysis are various topological twists of the 4d theories that involve the
 $U(1)_D$ ``bonus'' symmetry inherited from Type IIB supergravity. In particular, we have demonstrated in several examples how this twisting affects the F-theory geometry as well as the dual field theories, through an analysis based on an $U(1)_D$-augmented anomaly polynomial of these theories.}

{For the baryonic twist solutions, based on the $\Yodd^{\podd,\qodd}$ geometries, we have uncovered some puzzling aspects 
(see Section (\ref{herearethepuzzles})) of the proposed duality  with the $Y^{p,q}$ quiver gauge theories \cite{Benini:2015bwz}, already present in the solutions with constant $\tau$.  It is clearly an interesting question to resolve these puzzles, and we hope  to return to this in the near future.}

Let us mention some other directions following from this work. Enhancement to $(2,2)$ supersymmetry comes at the cost of choosing $\tau$ constant. 
Clearly one of the extensions of this work is to find more general solutions to (\ref{Master}) in tandem with the dual field theories, both for $(0,2)$ theories with varying coupling, as well as the $(2,2)$ theories, with constant $\tau$. Extensions to the holographic dictionary in F-theory to higher dimensions could build also on the work \cite{DHoker:2016ujz, DHoker:2017mds, DHoker:2017zwj}, which could be put into a more F-theoretic setting. 

{In Section \ref{sec:M} we have discussed  M-theory duals to the entire class of F-theory AdS$_3$ solutions, and argued that these are more naturally represented as AdS$_2$ solutions in eleven dimensions. In particular, in the baryonic twist solutions the $K3^\tau$ factor is geometrized as in the standard M/F duality. However, the universal twist solutions result in M-theory geometries, where the elliptically fibered part of the space is \emph{not} a Calabi--Yau. This hints at a universal relation between 2d and 1d SCFTs. We have shown that for all these solutions the leading order holographic central charge agrees precisely with the F-theory result. How to extract the sub-leading contributions, remains an interesting open question.}

Finally,  in this paper we have shown that a simple  extension of the anomaly polynomial to the ``bonus'' $U(1)_D$ symmetry provides a powerful tool for studying field theories with varying couplings.  It will be interesting to  make 
more rigorous the arguments that we employed in Section \ref{sec:FT} to deduce the contribution of the seven-brane modes to the central charges of the two-dimensional theories. We anticipate that doing this will improve our understanding
of the still  elusive field theories with varying couplings, including the case of non-abelian ${\cal N}=4$ SYM and $\mathcal{N}=1, 2$ theories. Work in this direction is under way  \cite{TBA}.

\subsection*{Acknowledgments}

We especially thank Craig Lawrie for discussions and collaboration on a related work, and  Jenny Wong for collaboration at an earlier stage of this paper.
 We also thank Benjamin Assel, Francesco Benini, Nikolay Bobev, Heeyeon Kim, Yolanda Lozano, Dave Morrison, Sameer Murthy,  Itamar Shamir, and James Sparks for discussions. 
CC is supported by STFC studentships under the STFC rolling grant ST/N504361/1.
DM is supported by the ERC Starting Grant 304806 ``The gauge/gravity duality and
geometry in string theory''. SSN is supported by the ERC Consolidator Grant 682608 ``Higgs bundles: Supersymmetric Gauge Theories and Geometry (HIGGSBNDL)''.



\appendix



\section{S-duality and $U(1)_D$ in Type IIB}
\label{app:SL2}

In this appendix we collect some basic facts about the transformation of Killing spinors in Type IIB supergravity under the $SL_2\R$ duality, as well as the relevance of the $U(1)_{D}$ symmetry with connection $Q$.

\subsection{Duality $U(1)_D$}

Let $\gamma$ be an element of the Type IIB self-duality group $SL_2\R$, which acts on the axio-dilaton $\tau$ as
\be
\gamma. \tau=\frac{a \tau+b}{c \tau +d}~,~~~~~ad-bc=1\,.
\ee
Due to quantum corrections the true duality group is $SL_2\Z$. 
Define $\alpha(\gamma)$ in terms of the phase 
\be
\me^{\ii \alpha(\gamma)}=\frac{c \tau+d}{|c \tau+d|}~.
\ee
We define the $U(1)_D$ gauge field $Q$ by
\be
Q=-\frac{1}{2 \tau_{2}}\dd \tau_{1} \,,
\ee
which transforms under $SL_2\R $ as 
\be
Q \rightarrow Q -\dd \alpha(\gamma) \,,\qquad 
\dd \alpha(\gamma)=c\frac{(d +c \tau_{1})\dd \tau_{2}-c \tau_{2} \dd \tau_{1}}{|c \tau+d|^{2}} \,.
\ee
Furthermore, the Killing spinors transform under the duality transformation as
\be
\epsilon \rightarrow \me^{\ii q \alpha(\gamma)} \,,
\epsilon
\ee
where $q$ is the charge under the $U(1)_{D}$. 
We see that the $\tau$ dependent part of the Killing spinor equation transforms as
\be
\ba
\mathcal{D}\epsilon=\lb\nabla -\frac{\ii}{2} Q\rb\epsilon \ \rightarrow\   & \lb\nabla -\frac{\ii}{2} Q'\rb\me^{\ii q \alpha(\gamma)}\epsilon 
=\me^{\ii q \alpha(\gamma)}\lb \ii q ((\nabla \alpha(\gamma))\epsilon+\frac{\ii}{2}(\dd \alpha(\gamma)) \epsilon+ \mathcal{D}\epsilon\rb\cr 
=\me^{\ii q \alpha(\gamma)}\mathcal{D}\epsilon
\ea\ee
which holds with $q=-\frac{1}{2}$. 

To determine how the Killing spinor equations with a general background transform under the $U(1)_D$, recall that the following combination of supergravity fields transform under the duality group
\be\ba
G =\frac{\ii}{\sqrt{\tau_{2}}}\lb \tau \dd B -\dd C^{(2)}\rb \,,\qquad 
P=\frac{\ii}{\tau_{2} }\dd \tau
\ea
\ee
as 
\be
\ba
G\ &\rightarrow \ \frac{|c \tau+d|}{c \tau +d} G = \me^{- \ii \alpha(\gamma)} G\cr 
P\ &\rightarrow \ \frac{c \bar{\tau}+d}{c \tau+d} P =\me^{-2 \ii \alpha(\gamma)}P \,.
\ea
\ee
The five-form field $F$ is of course invariant. 
The gravitini  and diliatini  Killing spinor equations then transform as follows
\be
\ba
\delta \Psi_{M}\ &\rightarrow \me^{-\frac{\ii}{2} \alpha(\gamma)} \delta \Psi_{M} \cr 
\delta \lambda \ & \rightarrow \me^{-\frac{3 \ii}{2} \alpha(\gamma)}\delta \lambda \,,
\ea
\ee
where we have used that $\epsilon^{c}$ has charge $-q$ where $q$ is the charge of the Killing spinor $\epsilon$.

Let us now consider the action. Of course for Type IIB supergravity there is no action that imposes the correct equations of motion and the self-duality of the five-form. One may however take as action
\begin{align}
S=\frac{1}{2 \kappa_{10}^{2}}\int \lb R -\frac{1}{2 \tau_{2}^{2}}\partial_{\mu}\tau \partial^{\mu} \bar{\tau}-\frac{1}{2}|G|^{2}-\frac{1}{4} |F|^{2}\rb *1+\frac{\ii}{4 \kappa_{10}^{2}}\int C_{(4)}\wedge \bar{G}\wedge G
\end{align}
to derive the equations of motion and impose the self-duality of $F$ afterwards. From the above transformations it follows that the action is invariant under $SL_2\R$.

\subsection{Gauge theory couplings from Supergravity}

In this section we recall how the gauge couplings in the field theory are identified in the holographic dual.

 We shall begin with describing how this works in the Klebanov-Witten theory \cite{Klebanov:1998hh}. Recall that the Klebanov-Witten theory is a quiver theory with two nodes. At each of these nodes there is an associated complexified coupling constant which we shall denote by $\tilde{\tau}_{i}$. The complex coupling constant is
\be
\tilde{\tau}_{i}=\frac{\theta_{i}}{2 \pi}+\frac{4 \pi\ii}{g_{i}^{2}}~.
\ee
From \cite{Klebanov:1998hh} we know that there are two complex moduli in the theory and these correspond to the sum and difference of the two gauge coupling constants. The sum of the two is identified on the gravity side with the $\tau$, see \cite{Morrison:1998cs} for an early example and \cite{Benini:2006hh} for a later use,
\be
\tau= \tilde{\tau}_{1}+\tilde{\tau}_{2}~.
\ee
The second marginal coupling corresponds to the integration over the unique two-cycle of the two-form
\be
 \tau B-C^{(2)} \,,
\ee
where $C^{(2)}$ and $B$ are the RR and NS-NS two form potentials respectively. One has \cite{Klebanov:1999rd} (see also \cite{Martelli:2008cm})
\be
\tilde{\tau}_{1}-\tilde{\tau}_{2}= \frac{1}{2\pi}\left[\int_{S^{2}}\lb  \tau B-C^{(2)}\rb -\pi \right] \text{mod}2 \pi~.
\ee

With these expressions we can compute how the two combinations of couplings transform under $SL_2\R$. We have
\be
\tilde{\tau}_{1}+\tilde{\tau}_{2}=\tau \rightarrow \frac{a (\tilde{\tau}_{1}+\tilde{\tau}_{2})+b}{c (\tilde{\tau}_{1}+\tilde{\tau}_{2})+d}
\ee
whilst
\be
\tilde{\tau_{1}}-\tilde{\tau}_{2}\rightarrow \frac{1}{c (\tilde{\tau}_{1}+\tilde{\tau}_{2})+d} (\tilde{\tau}_{1}-\tilde{\tau}_{2})~.
\ee
As remarked in \cite{Benini:2006hh} these formulae are derived for the $\mathcal{N}=2$ orbifold theory and in the literature are assumed to hold for the conifold also. The difference between the two couplings has the interpretation as the distance between two NS5 branes in the T-dualised theory which is associated to the non-anomalous baryonic symmetry.


\section{Details for the Derivations in Section \ref{sec:AdS302}}

\subsection{Torsion conditions}\label{Torsion-conditions}
\label{sec:torsion}

This appendix summarises the torsion conditions relevant for section \ref{sec:AdS302}. They are the same as computed in \cite{Couzens:2017way}, and we refer the reader there for further details. 

\noindent {\bf Scalar differential equations}
\begin{align}
\dd S_{ij}&= \frac{\ii m}{2}(\alphai_{i}-\alphai_{j})K_{ij}~,\label{dsij}\\
\me^{-2 \Delta}\D(\me^{2 \Delta}A_{ij})&= -\frac{\ii m}{2}(\alphai_{i}-\alphai_{j})B_{ij}\label{daij}~.
\end{align}
{\bf One-form differential equations}
\begin{align}
\me^{-4 \Delta}\dd \lb \me^{4 \Delta}K_{ij}\rb&= -\ii m (\alphai_{i}+\alphai_{j})U_{ij}-S_{ij}\me^{-4 \Delta}\ff  \label{dKij}\\
\D(\me^{2 \Delta}B_{ij})&=0\label{dbij}
\end{align}
{\bf Two-form differential equations}
\begin{align}
\me^{-4 \Delta}\dd (\me^{4 \Delta}U_{ij})&=-\frac{\ii m}{2}(\alphai_{i}-\alphai_{j})X_{ij}~,\label{dUij}\\
\me^{-6 \Delta}\D \lb \me^{6 \Delta}V_{ij}\rb &=-\frac{3 \ii m}{2}(\alphai_{i}-\alphai_{j})Y_{ij}+\me^{-4 \Delta}\ff \wedge B_{ij} \label{DVij}
\end{align}
{\bf Three-form differential equations}
\begin{align}
\me^{-8 \Delta}\dd \lb \me^{8 \Delta}X_{ij}\rb &= 2 m (\alphai_{i}+\alphai_{j})*X_{ij}-\me^{-4 \Delta}\ff\wedge U_{ij}~,\label{dXij}\\
\me^{-6 \Delta}\D\lb \me^{6 \Delta}Y_{ij}\rb &= m (\alphai_{i}+\alphai_{j})*Y_{ij} \label{DYij}
\end{align}
{\bf Four-form differential equations}
\begin{align}
\me^{-8 \Delta}\dd \lb \me^{8 \Delta}*X_{ij}\rb &=-\frac{3\ii m}{2}  (\alphai_{i}-\alphai_{j})*U_{ij}\label{d*Xij}\\
\me^{-10 \Delta}\D\lb \me^{10 \Delta}*Y_{ij}\rb&=-\frac{5\ii m}{2}(\alphai_{i}-\alphai_{j})*V_{ij}-\ii \me^{-4 \Delta}\ff\wedge Y_{ij}~,\label{D*Yij1}\\
\me^{-6 \Delta}\D\lb \me^{6 \Delta}*Y_{ij}\rb &=-\frac{\ii m}{2}(\alphai_{i}-\alphai_{j})*V_{ij}- \me^{-4 \Delta}A_{ij}*\ff\label{D*Yij2}
\end{align}
{\bf Five-form differential equations}
\begin{align}
\me^{-8 \Delta}\dd \lb \me^{8 \Delta}*U_{ij}\rb&= \ii m (\alphai_{i}+\alphai_{j})*K_{ij}~,\label{d*Uij}\\
\me^{-10 \Delta}\D\lb \me^{10 \Delta}*V_{ij}\rb &= \ii m (\alphai_{i}+\alphai_{j})*B_{ij}\label{D*Vij}
\end{align}
{\bf Six-form differential equations}
\begin{align}
\me^{-12 \Delta}\dd \lb \me^{12 \Delta}*K_{ij}\rb &=\ii m (\alphai_{i}-\alphai_{j})S_{ij}\text{Vol}(\M_{7})~,\label{d*Kij}\\
\me^{-10 \Delta}\D\lb \me^{10\Delta}*B_{ij}\rb &=-\frac{3 \ii m}{2}(\alphai_{i}-\alphai_{j})A_{ij}\text{Vol}(\M_{7})~.\label{D*Bij}
\end{align}

\subsection{Derivation of the ``Master Equation''}
\label{app:Mast}

In this appendix we provide an extensive discussion on the derivation of the ``master equation'' (\ref{Master}). 
Supersymmetry implies that a solution satisfies the Einstein equation and the Bianchi identity for $\ff$ but not the equation of motion for $\ff$. In this appendix we show that the equation of motion for $\ff$ is equivalent to \eqref{Master}. In \cite{Kim:2005ez} the $\ff$ equation of motion is shown to be equivalent to the differential equation
\be
\square_{6}R -\frac{1}{2}R^{2}+R_{\mu\nu}R^{\mu\nu}=0
\ee
on the K\"ahler base. We shall find that a similar equation governs the existence of a solution when $\tau$ becomes non-trivial.

In the main text it was shown that the internal space is a $U(1)$-fibration over a warped six-dimensional K\"ahler base. In the following it will be necessary to reduce along the Killing direction and to express everything in terms of the K\"ahler metric rather than the warped one, as such it is necessary to first clarify the notation we shall be using. We denote by $*_{7}$ the Hodge dual operator on the internal space, $*_{6}$ is the Hodge dual operator on the base of the $U(1)$ fibration and $\hat{*}_{6}$ the Hodge dual operator on the K\"ahler metric. The Ricci tensor, Ricci scalar and Ricci-form appearing are that of the K\"ahler metric and the K\"ahler two form is denoted by $J$.

Supersymmetry implies that the flux satisfies
\bea
m*_{7}\ff&=&*_{7}\lb- 2  J-4 m\me^{4 \Delta}\dd \Delta\wedge K-\frac{1}{2}\me^{4 \Delta}\dd\rho \rb\nonumber\\
&=&  \frac{\me^{-4 \Delta}}{m^{2}}K\wedge J\wedge J-\frac{1}{m^{3}}\hat{*}_{6}\dd \me^{-4 \Delta}-\frac{1}{2m^2}\hat{*}_{6}(\mathfrak{R}+\dd Q)\wedge K~.
\eea
Making use of the identities (which are easily proven)
\bea
\hat{*}_{6}\mathfrak{R}&=&\frac{R}{4}J\wedge J-\mathfrak{R}\wedge J~,\\
\hat{*}_{6}P\wedge P^{*}&=&-\frac{\ii |P|^{2}}{2}J\wedge J-P\wedge P^{*}\wedge J
\eea
we have
\bea
m*_{7}\ff=-\frac{1}{8m^{3}}\hat{*}_{6}\dd (R-2 |P|^{2})+\frac{1}{2m^2}\lb\mathfrak{R}\wedge J-\ii P\wedge P^{*}\wedge J\rb \wedge K
\eea
Imposing \eqref{d*F} is then equivalent to
\bea
0&=&\dd \hat{*}_{6}\dd(R-2|P|^{2})+2 \mathfrak{R}\wedge\mathfrak{R}\wedge J+4\ii \mathfrak{R}\wedge P\wedge P^{*}\wedge J~.
\eea
Taking the Hodge dual of the above and using the identities
\bea
\hat{*}_{6}\mathfrak{R}\wedge \mathfrak{R}\wedge J&=&\frac{1}{4}R^{2}-\frac{1}{2}R_{\mu\nu}R^{\mu\nu}~,\\
\hat{*}_{6}\mathfrak{R}\wedge P\wedge P^{*} \wedge J&=&-\ii \lb\frac{1}{2}R |P|^{2}-R_{\mu\nu}P^{\mu}P^{*\nu}\rb
\eea 
one obtains
\bea
\hat{\square}_{6}(R-2|P|^{2})=\frac{1}{2}R^{2}-R_{\mu\nu}R^{\mu\nu}-2 |P|^{2}R+4 R_{\mu\nu}P^{\mu}P^{*\nu}~,
\eea
where
\be
\hat{\square}_{6}=\hat{*}_{6}\dd \hat{*}_{6}\dd~.
\ee
Equation (\ref{Master}) determines the K\"ahler metric from which the remaining geometry may be recovered. Notice that for constant axio-dilaton one recovers the equation of \cite{Kim:2005ez} as expected.

\section{Derivation of the $\mathcal{N}=(2,2)$ Solutions}
\label{app:22}

In this appendix we derive the results of section \ref{sec:AdS322} and show that the classic AdS$_{3} \times S^{3}\times T^4$ or $K3$ obtained from purely D3-branes and the solutions discussed in \cite{Kim:2012ek} fit into this classification. All these solutions are contained within a co-homogeneity one ansatz that was successfully used in \cite{Gauntlett:2004zh} in the context of AdS$_5$ M-theory solutions. We conclude with a short discussion about a more general ansatz. 

Recall from section \ref{sec:AdS322} that the scalar bilinear $A_{12}$ was non-trivial and therefore $\tau$ was constant. The analysis may be split into two further subcases, the first when the scalar bilinear $S_{12}$ is non-constant and the second being the constant case. In the first case of non-constant $S_{12}$ one can show that it is not possible to satisfy the torsion conditions and therefore we restrict to the case of constant $S_{12}$ in the following. In fact integrability of the torsion conditions forces $S_{12}=0$ and the two spinors are orthogonal.

The $SU(2)$ structure specifies a basis of three vectors which may be chosen to be the bilinears\footnote{For ease of notation we shall drop the `$12$' subscript on $A_{12}$ and $B_{12}$ in this appendix.}
\be
\{ K_{11},~K_{22},~ \Im[ A^{*} B] \}~.\label{one-form basis}
\ee
Both $K_{11}$ and $K_{22}$ are Killing vectors and are dual to the right and left moving $U(1)$ R-symmetries in the field theory. This defines a foliation of the 7d metric and the metric on the space defined by the three vectors is found to be
\be
\dd s^{2}(\M_{3})= \frac{1}{4 |\A|^{2}(1-|\A|^{2})} \lb |\A|^{2}(K_{11}+K_{22})^{2}+(1-|\A|)^{2}(K_{11}-K_{22})^{2}+ 4 \imab^{2}\rb~.
\ee
The canonical $SU(2)$ structure two-forms may be expressed in terms of the bilinears as
\begin{align}
U_{11}&=-\ii \lb j_{4} -\frac{1}{2 |\A|^{2}(1-|\A|^{2})}(K_{22}+(2 |\A|^{2}-1)K_{11})\wedge \imab\rb ~,\\
U_{12}&=-\ii \sqrt{1-|A_{12}|^{2}}\omega_{4}~,
\end{align}
Here $j_{4}$ and $\omega_{4}$ are the two $SU(2)$ two-forms and after putting a vielbein on $\M_{4}$ take the canonical form
\be
j_{4}=e^{12}+e^{34}~,~~~\omega_{4}=(e^{1}+\ii e^{2})\wedge (e^{3}+\ii e^{4})~.
\ee
All other forms may be expressed in terms of these two two-forms, the three one-forms in \eqref{one-form basis} and $A$. We use this to reduce the torsion conditions to the minimal set acting on this basis of bilinears. The torsion conditions on the bilinears, that are not in the basis, must all be either automatically satisfied or they impose additional algebraic relations. We find that integrability will constrain the warp factor and flux $\ff$ to take specific forms.

\subsection{Torsion Conditions}

From \eqref{daij} we find that \eqref{dbij} is automatically satisfied. Moreover one finds that $\imab$ is conformally closed,
\begin{align}
\dd (\me^{4 \Delta}|\A|^{2})&= 2 m \me^{4 \Delta} \imab~,\label{dIMAB}
\end{align}
and this allows us to introduce a coordinate for $\imab$. Using the differential equations for $U_{11},~U_{12},~K_{11},~K_{22}$ and $A$ one finds that all the other torsion conditions are satisfied. From \eqref{D*Yij1} and \eqref{D*Yij2} one may find an expression for $*F$ which reads
\be
*\ff=\frac{\me^{4 \Delta}}{\A}\lb 4 \dd \Delta \wedge *Y_{12}+4 \ii m *V_{12}+\ii \me^{-4 \Delta} \ff \wedge Y_{12}\rb \,,
\ee
and it follows that $\dd *\ff=0$ is satisfied. Contrast this with the $(0,2)$ case here and in \cite{Kim:2005ez} where it is necessary to impose the ``master equation'' for the metric in order to satisfy \eqref{d*F}. A similar expression for the flux $\ff$ as in the $(0,2)$ analysis is possible and will be given later.

The torsion conditions for the $SU(2)$ structure two-forms imply that $\M_{4}$ is complex and that the flux $\ff$ and warp factor give an obstruction to $\M_{4}$ being K\"ahler,
\begin{align}
\me^{-4 \Delta}\dd (\me^{4 \Delta} j_{4})&= -\frac{1}{1-|\A|^{2}}\left[ \frac{2}{1-|\A|^{2}}\dd \Delta \wedge (K_{11}+K_{22}) +2m j_{4}+\me^{-4 \Delta}\ff\right]\wedge \imab~,\\
\me^{-4 \Delta}\dd (\me^{4 \Delta}\omega_{4})&=-\frac{1}{2(1-|\A|^{2})}\left[ 4 |\A|^{2}\dd \Delta+ \ii m (K_{11}+K_{22})\right] \wedge \omega_{4}~.
\end{align}
In light of the above equations it is natural to make the following redefinitions\footnote{The first line of the redefinitions extracts out a conformal factor from the metric on $\M_{4}$. In the second we mix the two Killing vectors for later simplicity. This implies that the Killing vectors $S$ and $T$ are now dual to a combination of the left and right moving R-currents.}
\begin{align}
m^{2}\me^{4 \Delta}j_{4}\rightarrow J_{4}~,~~~m^{2}\me^{4\Delta} \omega_{4}\rightarrow \Omega_{4}~,\\
S=K_{11}+K_{22} ~,~~~T=K_{11}-K_{22}~.
\end{align}
The conditions that we must impose to preserve $(2,2)$ supersymmetry become
\begin{align}
\me^{-4 \Delta}\dd (\me^{4 \Delta} S)&= \frac{2m}{1-|\A|^{2}}S \wedge \imab -2 \me^{-4 \Delta}\ff -\frac{4 \me^{-4 \Delta}}{m} J_{4}~,\label{dSeq}\\
\me^{-4 \Delta} \dd (\me^{4 \Delta} T)&=-\frac{2m}{|\A|^{2}} T\wedge \imab~,\label{dTeq}\\
\dd (\me^{4 \Delta} |\A|^{2})&= 2m \me^{4 \Delta} \imab~,\\
\dd J_{4}&=-\frac{1}{1-|\A|^{2}}\left[ \frac{2 m^2 \me^{4 \Delta}}{1-|\A|^{2}}\dd \Delta \wedge S+2 m J_{4}+m^2 \ff\right] \wedge \imab~,\label{dJ4initial}\\
\dd \Omega_{4}&=-\frac{1}{1-|\A|^{2}}\left[\frac{\ii m}{2} S+ 2 |\A|^{2} \dd \Delta \right]\wedge \Omega_{4}~.\label{dOmega4initial}
\end{align}
In the next section we introduce local coordinates which leads to a simplification of these equations. The 7d metric takes the form
\be
\dd s^{2}(\M_{7})=\frac{1}{4 |\A|^{2}(1-|\A|^{2})}\lb |\A|^{2} S^{2}+(1-|\A|^{2})T^{2}+ 4 \imab^{2}\rb +\frac{\me^{-4 \Delta}}{m^{2}}\dd s^{2}(\widetilde{\M}_{4})~,
\ee
where $\widetilde{\M}_{4}$ is a 4d space with $SU(2)$ structure defined by the two two-forms $J_{4}$ and $\Omega_{4}$.


\subsection{Reducing the Conditions}

To proceed we introduce coordinates for each of the three vectors. We may introduce local coordinates adapted to each of the Killing vectors $S^{\#}$ and $T^{\#}$. Moreover we have seen that $\imab$ is conformally closed and we may therefore introduce a further local coordinate for this direction. In light of equation \eqref{dIMAB} we introduce the coordinate $y$, via the equation 
\be
y \me^{-4 \Delta}\equiv|\A|^{2}\label{ydef}
\ee
such that
\be
\frac{\me^{-4 \Delta}}{2m} \dd y = \imab~.
\ee
It can be shown that this defines an integrable almost product structure.\footnote{If one defines the unit norm form
\be
\Pi= \frac{1}{|\A|\sqrt{1-|\A|^{2}}} \imab
\ee 
then the almost product structure defined by
\be
\tensor{J}{_{\mu}^{\nu}}=\Pi_{\mu}\Pi^{\nu}-\delta_{\mu}^{\nu}
\ee
is integrable. This implies that the remaining metric may have $y$ dependence however there are no $\dd y$ terms appearing in the metric other than the one in $\imab$, this will become pertinent soon.} 

Explicit computation by Fierz identities (or equivalently via an orthonormal frame computation) gives the conditions
\be
S_{\mu}T^{\mu}=0~,~~ S_{\mu}S^{\mu}=4(1-y\me^{-4 \Delta})~,~~ T_{\mu}T^{\mu}=4 y \me^{-4 \Delta}~,
\ee
the first signifies that the two Killing vectors are orthogonal. Introducing local coordinates, $\psi_{1}$ and $\psi_{2}$ for these Killing directions we have
\begin{align}
S^{\#}&=2m \frac{\partial}{\partial \psi_{1}}~,~~~S=\frac{2(1-y \me^{-4 \Delta})}{m}(\dd \psi_{1}+\sigma_{1})~,\\
T^{\#}&=2m \frac{\partial}{\partial \psi_{2}}~,~~~T=\frac{2 y \me^{-4 \Delta}}{m}(\dd \psi_{2}+\sigma_{2})~.
\end{align}
The integrable almost product structure implies that the $\sigma_{i}$ have no $\dd y$ term but may otherwise depend on $y$ non-trivially.

Using the local coordinates defined above the one-form equations \eqref{dSeq} and \eqref{dTeq} become
\begin{align}
m \ff &=-\lb \frac{}{}  (\me^{4 \Delta}-y) \dd \sigma_{1}+2 J_{4} +4 \me^{4 \Delta} \dd \Delta \wedge (\dd \psi_{1}+\sigma_{1})\rb~,\label{dsigma1}\\
\dd \sigma_{2}&=0~.
\end{align}
Equation \eqref{dsigma1} will be used as the defining equation for the flux $\ff$. As $\sigma_{2}$ is closed and therefore locally exact, it may, through a local change of coordinates,  be set to zero. The Killing vector $T$ is then unfibered. The metric on $\M_{3}$ in local coordinates is
\be
\dd s^{2}(\M_{3})= \frac{1}{m^{2}}\lb (1- \me^{-4 \Delta}y)(\dd \psi_{1}+\sigma_{1})^{2}+ y \me^{-4 \Delta} \dd \psi_{2}^{2} +\frac{1}{4 y (\me^{4 \Delta}-y)} \dd y^{2}\rb~.
\ee
The two two-form equations \eqref{dJ4initial} and \eqref{dOmega4initial} become
\begin{align}
\dd J_{4}&= \frac{1}{2}\dd \sigma_{1}\wedge \dd y~,\label{dJ4}\\
\dd \Omega_{4}&=-\lb \ii (\dd \psi_{1}+\sigma_{1})+\frac{2 y \me^{-4 \Delta}}{1-y \me^{-4 \Delta}} \dd \Delta\rb \wedge \Omega_{4}~.\label{dOmega4}
\end{align}
Having introduced three local coordinates the exterior derivative splits as 
\be
\dd = \dd_{4}+ \dd y \wedge \frac{\partial}{\partial y}+\dd \psi_{1}\wedge \frac{\partial}{\partial \psi_{1}}+\dd \psi_{2} \wedge \frac{\partial}{\partial \psi_{2}} \,,
\ee
and decompose the remaining torsion conditions. Let the coordinates  on $\widetilde{\M}_{4}$ be denoted by $x_{i}$ and let the metric on $\widetilde{\M}_{4}$ be $g^{(4)}_{ij}(x,y)$ where non-trivial $y$ dependence of the metric is permitted. 
The decomposition of \eqref{dOmega4} gives
\begin{align}
\partial_{\psi_{1}}\Omega_{4}&=- \ii \Omega_{4}~,\label{psi1Omega}\\
\partial_{\psi_{2}}\Omega_{4}&=0~,\label{psi2Omega}\\
\partial_{y}\Omega_{4}&= -\frac{2 y \me^{-4 \Delta}}{1- y \me^{-4 \Delta}} \partial_{y}\Delta \Omega_{4}~,\label{yOmega}\\
\dd_{4}\Omega_{4}&=-\lb  \ii \sigma_{1}+\frac{2 y \me^{-4 \Delta}}{1- y \me^{-4 \Delta}}\dd_{4}\Delta\rb \wedge \Omega_{4}\label{d4Omega}~.
\end{align}
From the decomposition of \eqref{dJ4} we find
\begin{align}
\partial_{\psi_{1}}J_{4}&=\partial_{\psi_{2}}J_{4}=0~,\\
\partial_{y}J_{4}&=\frac{1}{2} \dd_{4}\sigma_{1}~,\label{yJ4}\\
\dd_{4}J_{4}&=0~.\label{d4J4} \,.
\end{align}
A well known fact of complex geometry is that if the maximal holomorphic form $\Omega$ satisfies the differential equation $\dd \Omega=\ii \widehat{P} \wedge \Omega$ for some one-form $\widehat{P}$ then the almost complex structure defined by $\Omega$ is integrable and the manifold is complex of real dimension $2n$, in fact it follows that the one form $\widehat{P}$ satisfies $\mathfrak{R}=\dd \widehat{P}$ as has been used previously in the paper.\footnote{$\widehat{P}$ should not be confused with the one-form $P$ which depends on the axio-dilaton and is vanishing in this case.} Equation \eqref{d4Omega} shows that $\widetilde{\M}_{4}$ is a complex manifold, furthermore as discussed in \cite{Gauntlett:2004zh} this implies that the complex structure $\tensor{J}{_{i}^{j}}$ is independent of $\psi_{1},\psi_{2}$ and $y$. One may solve \eqref{psi1Omega} by extracting out a suitable $\psi_{1}$ dependent phase from $\Omega_{4}$. Equation \eqref{yOmega} fixes the $y$ variation of the volume of $g^{(4)}$. From $\Omega_{4}\wedge \bar{\Omega}_{4}=4 \vol(\widetilde{\M}_{4})$ we find
\be
\frac{\partial}{\partial y} \log \sqrt{g}= -\frac{4 y \me^{-4 \Delta}}{1- y \me^{-4 \Delta}}\partial_{y}\Delta~.\label{detcon}
\ee
Finally \eqref{d4Omega} implies
\be
\sigma_{1}= - \widehat{P}_{4} +\frac{2y \me^{-4 \Delta}}{1- y \me^{-4 \Delta}} \dd_{4}^{c}\Delta\,,
\label{sigma1def}
\ee
where $\dd_{4}^{c}=  \ii (\bar{\partial}_{4}-\partial_{4})$ with $\partial_{4}, ~\bar{\partial}_{4}$ the Dolbeault operators on $\widetilde{\M}_{4}$. From \eqref{d4J4} we see that $J_{4}$ is closed on $\widetilde{\M}_{4}$ and therefore $g^{(4)}$ locally defines a family of K\"ahler metrics on $\widetilde{\M}_{4}$ parametrised by $y$. 

For a supersymmetric solution we must solve the two differential equations
\begin{align}
\partial_{y}J_{4}&=\frac{1}{2}\dd_{4}\sigma_{1}~,\label{final12app}\\
\partial_{y}\log \sqrt{g} &=-\frac{4 y \me^{-4 \Delta}}{1- y \me^{-4 \Delta}} \partial_{y}\Delta~,
\end{align}
where $\sigma_{1}$ and the flux $\ff$ are given by
\begin{align}
\sigma_{1}&= -\widehat{P}_{4} +\frac{2y \me^{-4 \Delta}}{1- y \me^{-4 \Delta}} \dd_{4}^{c}\Delta~,\label{sigma1def}\\
m \ff &=-\lb \frac{}{}  (\me^{4 \Delta}-y) \dd \sigma_{1}+2 J_{4} +4 \me^{4 \Delta} \dd \Delta \wedge (\dd \psi_{1}+\sigma_{1})\rb~.\label{Fluxdef}
\end{align}
The seven-dimensional metric is
\be
m^{2}\dd s^{2}( \M_{7})=  (1- y\me^{-4 \Delta}) (\dd \psi_{1}+\sigma_{1})^{2}+ y \me^{-4 \Delta} \dd \psi_{2}^{2}+\frac{\me^{-4 \Delta}}{4y(1- y \me^{-4 \Delta})}\dd y^{2}+ \me^{-4 \Delta} g^{(4)}(y,x)_{ij}\dd x^{i}\dd x^{j}\ee
with $g^{(4)}$ K\"ahler at fixed $y$. Following the arguments in \cite{Gauntlett:2004zh}, we define the self-dual and anti-self-dual combinations of an arbitrary two-form, $\zeta$ on $\widetilde{\M}_{4}$ to be $\zeta^{\pm}=\frac{1}{2}\zeta\pm *\zeta$, then we have the identity
\be
(\partial_{y} J_{4})^{+}=\frac{1}{2}\partial_{y}\log \sqrt{g} J_{4}
\ee
which is valid when the complex structure $\tensor{J}{_{4 i}^{j}}$ is independent of $y$. We may use this to rewrite the equation for the volume as
\be
(\dd_{4}\sigma_{1})^{+}=-\frac{4 y \me^{-4 \Delta}}{1- y \me^{-4 \Delta}}\partial_{y}\Delta J_{4}\label{d4sigma}~.
\ee
The necessary conditions to solve are \eqref{final12app} and \eqref{d4sigma} along with the definitions \eqref{sigma1def} and \eqref{Fluxdef}.

\subsection{Recovering known $(2,2)$ Solutions}
\label{app:Recover22}

In this subsection we show that the classic AdS$_{3} \times S^{3}\times T^{4}$ or $K3$, obtained from purely D3-branes and the solutions discussed in \cite{Kim:2012ek} fit into this classification. We shall use a co-homogeneity one ansatz inspired by \cite{Gauntlett:2004zh}. The distinguished coordinate in this co-homogeneity one ansatz is $y$. We shall take the warp factor to satisfy $\Delta=\Delta(y)$ and moreover we impose $\partial_{y}\sigma_{1}=0$.
With these assumptions the Ricci-form on $\widetilde{\M}_{4}$ is
\begin{align}
\mathfrak{R}=- \dd_{4}\sigma_{1}~,
\end{align}
and therefore
\begin{align}
\mathfrak{R}&=-2\partial_{y}J~,\label{co1yJ}\\
\mathfrak{R}^{+}&=\frac{4 y \me^{-4 \Delta}}{1- y \me^{-4 \Delta}} \partial_{y}\Delta J~.\label{co1Deltaeq}
\end{align}
We see that $\mathfrak{R}^{+}$ is pointwise proportional to $J$ and that the proportionality factor is independent of the coordinates on $\widetilde{\M}_{4}$. As the metric is K\"ahler and the proportionality factor constant on $\widetilde{\M}_{4}$ the Ricci scalar must also be constant on $\widetilde{\M}_{4}$, that is
\be
\dd_{4}R=0~.\label{d4R}
\ee
From the assumption $\partial_{y}\sigma_{1}=0$ we see that $\partial_{y}\mathfrak{R}_{ij}=0$ and therefore as the complex structure is independent of $y$ we have 
\be
\partial_{y}R_{ij}=0~.
\ee
As the complex structure is $y$ independent, equation \eqref{co1yJ} is equivalent to
\be
R_{ij}=-2 \partial_{y}g^{(4)}_{ij} \quad \Rightarrow \quad 
R_{ij}R^{ij}=2 \partial_{y}R\,,
\ee 
and therefore 
\be
\dd_{4}R_{ij}R^{ij}=0~.\label{d4R2}
\ee
On a K\"ahler manifold the eigenvalues of the Ricci tensor come in pairs and therefore for a 4d metric there are at most two distinct eigenvalues with degeneracy at least two. Equations \eqref{d4R} and \eqref{d4R2} show that both the sum of the eigenvalues and sum of the squares of the eigenvalues are constant over the base. This implies that the two pairs of eigenvalues of the Ricci tensor are constant on $\widetilde{\M}_{4}$. Using \cite{2000math......7122A},
 which assumes that the Goldberg conjecture is true, we find that the K\"ahler surface is the sum of two complex curves each with constant curvature. In the case where the two pairs of eigenvalues are the same, by definition the manifold is K\"ahler--Einstein. There are then two natural classes to consider, either $\widetilde{\M}_{4}$ is K\"ahler--Einstein or it is the product of two Riemann surfaces.

\subsection{Case 1: K\"ahler--Einstein} 

When $\widetilde{\M}_4$ is K\"ahler--Einstein, the Ricci-form is
\be
\mathfrak{R}=\frac{\kappa}{F(y)}J \,,
\ee
where $\kappa \in\{0,\pm1\}$ and $F(y)>0$, the last inequality follows from requiring the metric to be positive definite. For $\kappa=0$ $\widetilde{\M}_{4}$ is Ricci-flat and K\"ahler and therefore is Calabi--Yau. The warp factor in this case is found to be constant and we recover the classic AdS$_3\times S^{3}\times T^4 (K3)$ solution with just five-form flux and quotients preserving the Calabi--Yau condition.

We consider the case of $\kappa\neq0$ in the remainder of this section. From \eqref{co1yJ}
\be
F(y)=c_{1}- \frac{\kappa}{2} y
\ee
where $c_{1}$ is an integration constant. Note that in \cite{Gauntlett:2004zh} the analogous function was found to be quadratic rather than linear. We may then solve \eqref{co1Deltaeq} for the warp factor to obtain
\be
\me^{-4 \Delta}= \frac{ c_{2} +2 y -4 c_{1} \kappa \log y}{(y \kappa -2 c_{1})^{2}}~.
\ee
We must fix the range of $y$. To do so we look at the values of $y$ at which the metric degenerates. For smoothness $y$ must be prevented from attaining the value 0 due to the logarithmic term and therefore we require that $1-y \me^{-4 \Delta}$ has two roots. We were unable to find two such roots. One may be tempted to set the constant $c_{1}=0$, this implies $\kappa=-1$. One then finds that there is only one solution for the root of $1-y \me^{-4 \Delta}$. Note that as the warp factor is singular at $y=0$ we must avoid this value of $y$ here and it follows that the metric will be non-compact. Instead let us consider the case of a product of Riemann surfaces.

\subsection{Case 2: Product of Riemann Surfaces}

The second case to consider is when for fixed $y$, $\tM_4 = \Sigma_1 \times \Sigma_2$ is a product of two Riemann surfaces of constant curvature, i.e. 
\be
\dd s^{2}(\widetilde{\M}_{4})=\dd s^{2}(\Sigma_{1})+\dd s^{2}(\Sigma_{2})~.
\ee
Denoting K\"ahler forms for each curve as $J_{i}$ the Ricci-forms are given by
\be
\mathfrak{R}_{1}=\frac{\kappa_{1}}{G_{1}(y)} J_{1} ~,~~~\mathfrak{R}_{2}= \frac{\kappa_{2}}{G_{2}(y)} J_{2} \,,
\ee 
where as before $\kappa_{i}\in \{ 0, \pm1\}$ and $G_{i}(y)>0$ for all values of $y$ within its domain. On $\widetilde{\M}_{4}$ The K\"ahler form and Ricci-form factor as
  \be
  J=J_{1}+J_{2}~,~~~~\mathfrak{R}=\mathfrak{R}_{1}+\mathfrak{R}_{2}~.
  \ee
We may solve for the functions $G_{i}(y)$ by using equation \eqref{co1yJ}
\be
G_{i}(y)= c_{i}-\frac{\kappa_{i}}{2} y~.
\ee
Using \eqref{co1Deltaeq} the warp factor is found to be
\be
\me^{-4 \Delta}= \frac{2 \kappa_{1}\kappa_{2} y + k -2 (\kappa_{1} c_{2}+\kappa_{2}c_{1})\log y}{(2 c_{1}-\kappa_{1} y)(2 c_{2}-\kappa_{2}y)}~,\label{warpcase2}
\ee
where $k$ is an integration factor. We analyse the regularity of these solutions in the remainder of this section.

\subsubsection*{\underline{$\Sigma_{1}=T^{2}$}}
First consider the case where one of the Riemann surfaces is a $T^2$, without loss of generality let us set $\kappa_{1}=0$. Observe that for $\kappa_{2}=0$ we return to the Calabi--Yau case which was discussed previously, therefore restrict to  $\kappa_{2}\neq1$. The warp factor becomes
\be
\me^{-4 \Delta}= \frac{k-2 \kappa_{2} c_{1}\log y}{2 c_{1} (2 c_{2}-\kappa_{2} y)}
\ee
As $y$ appears in the metric we require that it is strictly positive. We wish to find the range of $y$ such that the metric is both compact and smooth. For both $\kappa_{2}=\pm 1$ the solution is found to be singular or to not close.

\subsubsection*{\underline{$S^{2}\times H^{2}$}}

The metric in this case is 
\begin{align}
m^{2}\dd s^{2}( \M_{7})=& 4 (1- y\me^{-4 \Delta}) (\dd \psi_{1}+\sigma_{1})^{2}+4 y \me^{-4 \Delta} \dd \psi_{2}^{2}+\frac{\me^{-4 \Delta}}{4y(1- y \me^{-4 \Delta})}\dd y^{2}\nonumber\\
&+\me^{-4 \Delta}\lb G_{1}(y)\dd s^{2}(\Sigma_{1})+G_{2}(y)\dd s^{2}(\Sigma_{2})\rb~,\label{7dmetricproduct}\\
\sigma_{1}=&-\frac{1}{2}(\hat{P}_{1}+\hat{P}_{2})
\end{align}
where $\widehat{P}_{i}$ satisfy $\dd \widehat{P}_{i}=\mathfrak{R}_{i}$. 
Due to the Logarithmic term appearing in the warp factor \eqref{warpcase2} we shall choose the constants such that the term $c_{2}\kappa_{1}+c_{1}\kappa_{2}$ vanishes. Clearly for this case we must take $c_{1}=c_{2}=c$ and we have
\begin{align}
G_{1}(y)&=c-\frac{y}{2}~,~~G_{2}(y)=c+\frac{y}{2}~,\\
\me^{-4 \Delta}&=\frac{k- 2 y}{(2 c-y)(2 c +y)}= \frac{k- 2 y}{4 G_{1}(y) G_{2}(y)}~.
\end{align}
There are two constants however we may remove one of these constants by a rescaling of the $y$ coordinate. Observe that under the rescaling $y\rightarrow q y$ we have
\begin{align}
G_{i}(y)&\rightarrow q\lb \frac{c}{q}\pm \frac{y}{2}\rb=q \tilde{G}_{i}(y)~,\\
\me^{-4 \Delta}&\rightarrow \frac{\frac{k}{q}-2 y}{4 q \tilde{G}_{1}(y)\tilde{G}_{2}(y)}=\me^{-4 \tilde{\Delta}}~,
\end{align}
and the full metric is seen to be rescaled by an  irrelevant constant factor. We may then use this coordinate transformation to set without loss of generality $k=4$. 

For a smooth compact metric we require that both $\me^{-4 \Delta}$ and $G_{i}(y)$ are strictly positive for all values of $y$. Moreover as is clear from the explicit metric in \eqref{22metric} we require $y\geq0$. In order to satisfy these conditions we take $c>0$ and positivity of the metric then implies $0\leq y\leq \text{min}\{2, 2c\}$. Computing the scalar invariants of the metric one finds that the Ricci scalar vanishes as expected from the equation of motion, whilst the contraction of the Ricci tensor into itself has poles when $G_{i}(y)=0$ or $y=2$. This latter singularity may be rephrased in terms of the divergence of the warp factor at these points. We must therefore require that the warp factor does not vanish or degenerate for all $y$. This implies that the manifold is topologically the direct product of AdS$_3$ with $\M_{7}$ and regularity of the solution is equivalent to regularity of $\M_{7}$.

The $y$ coordinate is fixed by finding points where the metric degenerates and making a choice of coordinate period such that it is smooth. The metric on $\M_{7}$ degenerates at $y=0$ and a real positive root of $1-y \me^{-4 \Delta}$. As $y$ is strictly positive $G_{2}(y)$ is strictly positive for all values of positive $y$. On the other hand $G_{1}(2 c)=0$ and unless $c=1$\footnote{This value of $c$ will be studied separately later} the metric is singular. We are then restricted to have $0\leq y\leq y^{*}$ where $y^{*}$ is a positive root of $1-y \me^{-4 \Delta}$ which is smaller than $2$. The roots are
\be
y_{\pm}=2 (1\pm \sqrt{1-c^{2}})
\ee
and therefore to allow real roots we fix $0\leq c\leq1$ and the range of $y$ to be
\be
0\leq y \leq y_{1}= 2(1- \sqrt{1-c^{2}})~.
\ee
We first analyse the metric at the two endpoints for $0\leq c <1$, returning to the special case of $c=1$ after. Expanding the metric around $y=0$ we find
\begin{align}
m^{2}\dd s^{2}&=\frac{1}{c^{2}}\lb 4 y \dd \psi_{2}^{2}+ \frac{\dd y^{2}}{4 y}\rb+ 4 (\dd \psi_{1}+\sigma_{1})^{2}+\frac{1}{c}(\dd s^{2}(S^{2})+\dd s^{2}(H^{2}))~.
\end{align}
The bracketed term is $\mathbb{R}^{2}$ if $\psi_{2}$ has period $ \frac{\pi}{2}$. The remaining part of the metric is a $U(1)$ fibration over $S^{2}\times H^{2}$ and is regular. We choose to redefine the coordinate $\psi_{2}$ so that it has the canonical $2\pi$ period. Performing the expansion around $y^{*}$ the $\psi_{1}$ coordinate is fixed to have period $\frac{\pi}{2}$, and again we perform a redefintion of the coordinate to give the canonical $2 \pi$ period. The metric is then smooth for all $0\leq c <1$ and is topologically $S^{3}\times S^{2}\times H^{2}$.

We return to the special case of $c=1$. The analysis around $y=0$ is identical and so we shall not repeat it here. Around $y=y^{*}=2$ the metric becomes
\begin{align}
m^{2}\dd s^{2}&= \dd \tilde{\psi}_{2}^{2}+\dd s^{2}(H^{2})+\frac{2-y}{4} \lb (\dd \psi_{1}+\cos \theta \dd \phi)^{2}+\dd \theta^{2}+\sin^{2}\theta \dd \phi^{2}\rb+\frac{1}{4(2-y)}\dd y^{2} \,,
\end{align}
which after a coordinate transformation can be seen to be $S^{1}\times H^{2}\times \mathbb{R}^{4}$ if $\psi_{1}$ has period $4 \pi$. Again the metric is smooth for $c=1$ but is topologically different to the $0\leq c <1$ case.

It is instructive to put the solution into the form of the classification of \cite{Kim:2005ez}. The change of coordinates one needs to perform is
\be
\psi_{1}=\chi+z~,~~\psi_{2}= \chi-z
\ee
which puts the metric into the form
\begin{align}
m^{2}\dd s^{2}(\M_{7})&=\lb \dd \chi+2(1- y \me^{-4 \Delta})\sigma_{1}+(1-2 y \me^{-4 \Delta})\dd z \rb^{2}+\me^{-4 \Delta}\dd s^{2}(\widetilde{\M}_{6})\\
\dd s^{2}(\widetilde{\M}_{6})&= \frac{\dd y^{2}}{4 y (1-y \me^{-4 \Delta})}+ 4 y (1- y \me^{-4 \Delta})(\dd z +\sigma_{1})^{2}+ G_{1}(y)\dd s^{2}(S^{2})+G_{2}\dd s^{2}(H^{2})\nonumber
\end{align}
where the metric on $\widetilde{\M}_{6}$ is K\"ahler and satisfies the defining equation 
\be
\square_{6} R= \frac{1}{2}R^{2}-R_{\mu\nu}R^{\mu\nu}~.
\ee
We have checked that all the remaining conditions in \cite{Kim:2005ez} are satisfied and the flux given in \eqref{Fluxdef} can be repackaged into the form given in \cite{Kim:2005ez}.

We observe that for $c=1$ this is the solution explicitly presented in \cite{Kim:2012ek}, (11)-(12), after a change of coordinates. The form of the solution presented in \cite{Kim:2012ek}\footnote{We correct some terms that are missing from both (11) and (12) in \cite{Kim:2012ek}.} is
\begin{align}
\dd s^{2}&=\lb \dd z-\frac{\cos^{2} \theta}{1+\sin^{2}\theta}(\dd \varphi +\cos \theta_{1} \dd \phi_{1} -\frac{1}{x_{1}}\dd x_{2})\rb^{2}\nonumber\\
&+\me^{-4 \Delta}\lb 2 (1+\sin^{2}\theta)\dd \theta^{2}+ \frac{2 \sin^{2}\theta \cos^{2}\theta}{1+\sin^{2}\theta}\lb\dd \varphi +\cos \theta_{1} \dd \phi_{1} -\frac{1}{x_{1}}\dd x_{2}\rb^{2}\right.\nonumber\\
&\left.+\cos^{2}\theta (\dd \theta_{1}+\sin^{2}\theta_{1}\dd \phi_{1}^{2})+(1+\sin^{2}\theta)\lb \frac{\dd x_{1}^{2}+\dd x_{2}^{2}}{x_{1}^{2}}\rb\rb~,\label{Kimsol}\\
\me^{-4 \Delta}&=\frac{1}{1+\sin^{2}\theta}~.
\end{align}
Performing the change of coordinates
\be
y=2 \sin^{2}\theta~,~~z=-\psi_{2}~,~~\varphi=-(\psi_{1}+\psi_{2})
\ee
puts the metric \eqref{Kimsol}  in the form \eqref{7dmetricproduct}. In later pages of \cite{Kim:2012ek} a solution in different dimensions is presented with this additional parameter included.

\subsubsection*{\underline{$S^2 \times S^2$}}

Again the strategy to obtain a smooth solution is to eliminate the logarithmic term appearing in \eqref{warpcase2}. We impose that $c_{1}=-c_{2}$. By definition the coordinate $y$ satisfies $y\geq 0$ and it is trivial to see that it is impossible for $G_{1}(y)$ and $G_{2}(y)$ to be simultaneously positive. Therefore if we tune the parameters such that the logarithmic term vanishes then no solution exists.

\subsubsection*{\underline{$H^{2}\times H^{2}$}}

Eliminating the logarithmic term appearing in \eqref{warpcase2} we impose $c_{1}=-c_{2}\equiv c$. Due to the symmetry in $c$ we may take $c\geq 0$ without loss of generality. Contrary to the $S^{2}\times S^{2}$ case both $G_{1}(y)$ and $G_{2}(y)$ can be made positive if $y\geq 2 c$. However if we compute the roots of $1-y \me^{-4 \Delta}$ we find that they are both negative and therefore there is no way to make the space close and have a positive definite metric. We conclude that there are no smooth solutions of this form.

\subsection{Further Generalisations}

In the preceding sections we have recovered known $(2,2)$ solutions in the literature. A natural ansatz to use to find new solutions is that presented in \cite{Bah:2013qya}. The ansatz imposes a $U(1)$ isometry for the base $\widetilde{\M}_{4}$ and is the most general complex metric in 4d with a $U(1)$ isometry. The 7d internal metric will admit three commuting $U(1)$ Killing vectors. A natural interpretation of this ansatz is that this is the near horizon of $D3$ branes in the background $\R^{1,1}\times CY_{4}$ where the $CY_{4}$ is non-compact and is decomposed as the sum of three line bundles over a Riemann surface.


\section{$\AdS_3$ to $\AdS_5$}
\label{app:AdS5}

In this appendix we provide some of the computational derivations for section \ref{sec:AdS5}. We look at the AdS$_3$ solutions with $\mathcal{N}=(2,2)$ and varying $\tau$ by relaxing the compactness condition of the internal space. We find the only solutions of this problem decompactify to an $\AdS_5$ solution. In fact the resulting $\AdS_5$ varying $\tau$ solutions of IIB supergravity are the most general of this kind, which we show in section \ref{app:AdS5General}. In \cite{Kehagias:1998gn} AdS$_5$ solutions with five-form flux and varying axio-dilaton were considered. We recover the analysis presented there and give an F-theoretic interpretation in terms of an elliptically fibered Calabi--Yau four-fold.

\subsection{$\AdS_3$ Solutions with $(2,2)$ and Varying $\tau$}
\label{app:AdS3to5}

\subsubsection{Torsion Conditions}

The starting point for this analysis is (\ref{APeq}) where for $P$ to be non-zero and thus $\tau$ varying, we need  $A_{12}=0$. It is easy to see that by setting $S_{12}$ to be constant it must in fact vanish. Moreover it is trivial to see that it is impossible to satisfy the torsion conditions if both of these scalars simultaneously vanish. We shall therefore restrict to the case when $S_{12}$ is non-constant in the remainder of this subsection. As before we find that both $S_{11}$ and $S_{22}$ are constant and therefore we may normalise the spinors such that they are both unity.

Recall that $\M_7$ admits an $SU(2)$ structure which implies there is a $3+4$ splitting, such that the '$3$' part, $\widetilde{M}_{3}$ has a vielbein given by the three vectors of the SU(2) structure. One may take as a basis for the three independent vectors 
\be
\{ K_{11},~K_{22}, ~ \Im[S_{12}^{*}K_{12}]\}\label{tau1formbasis}
\ee
in terms of which we may write the metric on $\M_3$ as 
\be
\dd s^{2}(\M_{3})=\frac{1}{4 |S_{12}|^{2}(1-|S_{12}|^{2})}\lb K_{11}^{2}+K_{22}^{2}+2( 1- 2 |S_{12}|^{2}) K_{11}\otimes K_{22} +4 \Im[S_{12}^{*}K_{12}]^{2}\rb~.
\ee
The canonical $SU(2)$ structure two-forms are written in terms of the bilinears as
\begin{align}
j&=\ii U_{11}-\frac{1}{2|S_{12}|^{2}(1-|S_{12}|^{2})}\lb K_{22}+(1-2|S_{12}|^{2}K_{11}\rb \wedge \Im[S_{12}^{*}K_{12}]~,\label{taujdef}\\
\omega&=\frac{1}{(1- |S_{12}|^{2})^{\frac{1}{2}}}V_{12}^{*}~.\label{tauomegadef}
\end{align}
We may construct a basis of independent bilinears consisting of the scalar $S_{12}$, the three one-forms in \eqref{tau1formbasis} and the two canonical $SU(2)$ two-forms in \eqref{taujdef} and \eqref{tauomegadef}. All other bilinears may be obtained from wedge products of these bilinears. The torsion conditions of the non-basis elements should  then be either automatically satisfied by imposing the equations for the basis forms or impose additional algebraic constraints. 

Integrability of the torsion conditions (\ref{iKijdDelta}) imply the warp factor satisfies 
\be
\Delta=-\frac{1}{2}\log [1- |S_{12}|^{2}]~.
\ee
We may use it as a coordinate for $\Im[S_{12}^{*}K_{12}]$. 
Moreover integrability of the torsion conditions implies that the flux $\ff$ is fixed to be 
\be
\ff = -\frac{1}{|S_{12}|^{2}}\dd \Delta \wedge (K_{11}+K_{22})~,
\ee
which is easily shown to be both closed and co-closed and therefore $\ff$ satisfies both its Bianchi identity and its equation of motion.
The torsion conditions for $K_{11}$ and $K_{22}$ imply for the K\"ahler form on $\M_4$ that
\be
j= -\frac{\me^{-2 \Delta}}{4m} \dd \lb \me^{2 \Delta}(K_{11}-K_{22})\rb \,,
\ee
which is conformally closed. In light of this we define the rescaled real and complex two forms
\be
J=m^2 \me^{2\Delta} j \,,\qquad 
\Omega= m^{2}\me^{3 \Delta}V_{12}^{*}\,,
\ee
for the resulting four-fold as $\tM_4$, which satisfy 
\be\label{DOmega7d}
\dd J =0\,,\qquad 
\bar{\mathcal{D}}\Omega=-\frac{3 \ii m }{2}\me^{2 \Delta} (K_{11}-K_{22})\wedge\Omega \,.
\ee
From \eqref{dilalg} we see that $P$ is holomorphic with respect to the induced complex structure defined by $J$. The metric after this redefinition takes the form
\be
\dd s^{2}(\M_{7})= \frac{\me^{2 \Delta}}{4}(K_{11}-K_{22})^{2}+\frac{1}{4(1-\me^{-2 \Delta})}(K_{11}+K_{22})^{2}+\frac{\me^{-2\Delta}}{m^2 (1-\me^{-2 \Delta})}\dd \Delta^{2}+\frac{\me^{-2 \Delta}}{m^{2}} \dd s^{2} (\widetilde{\M}_{4})
\ee
where $\dd s^{2}(\widetilde{\M}_{4})$ is K\"ahler. As $K_{11}$ and $K_{22}$ are Killing vectors so are the linear combinations
\begin{align}
K&= K_{11}-K_{22}~,~~~
L=K_{11}+K_{22}~,
\end{align}
and they satisfy the algebraic conditions
\be
||K||^{2}=4 \me^{-2 \Delta}~,~~~~
||L||^{2}= 4 (1- \me^{-2 \Delta})~,~~~~
K_{\mu}L^{\mu}=0
\ee
and the differential equations
\be\label{dL}
\dd(\me^{2 \Delta} K)=-\frac{4}{m}  J~,\qquad 
\dd \lb \frac{1}{1-\me^{-2 \Delta}}L\rb =0~.
\ee

\subsubsection{Decompactification to $\AdS_5$}

We may introduce local coordinates adapted to these two Killing directions as
\begin{align}
K^{\#}&=m\frac{\partial}{\partial \psi}~,~~~~
L^{\#}= m\frac{\partial}{\partial \varphi}~,
\end{align}
with dual one-forms
\begin{align}
K&=\frac{4}{m} \me^{-2 \Delta}\left(\dd \psi+\frac{1}{2}\rho\right)~,~~~~
L= \frac{4}{m} (1- \me^{-2 \Delta})(\dd \varphi+\sigma)~.
\end{align}
The one-forms $\rho$ and $\sigma$ are both independent of $\psi$ and $\varphi$. From \eqref{dL} we see that $\sigma$ is closed and therefore locally exact and may be set to zero by a local change of coordinates. The metric takes the form
\be
m^{2}\dd s^{2}(\M_{7})=\frac{\me^{-2 \Delta}}{ 1-\me^{-2 \Delta}}\dd \Delta^{2}+4(1-\me^{-2 \Delta})\dd \varphi^{2}+  \me^{-2 \Delta} \lb 4 \left(\dd \psi+\frac{1}{2}\rho\right)^{2}+\dd s^{2}(\widetilde{\M}_{4})\rb~.
\ee
These explicit coordinates induce a splitting of the exterior derivative as
\be
\dd\rightarrow \dd \varphi \frac{\partial}{\partial \varphi}+\dd \Delta \frac{\partial}{\partial \Delta}+\dd \psi \frac{\partial}{\partial \psi} +\dd_{4}~.
\ee
With this splitting equation \eqref{DOmega7d} decomposes as
\begin{align}
\partial_{\varphi}\Omega&=\partial_{\Delta}\Omega=0~,\\
\partial_{\psi}\Omega&=- 6 \ii \Omega~,\label{dpsiomega}\\
\bar{\mathcal{D}}_{4}\Omega&=-\frac{3 \ii}{2} \rho \wedge \Omega~.\label{Domega}
\end{align}
Equation \eqref{dpsiomega} may be solved by extracting a phase from $\Omega$. Equation \eqref{Domega} implies that the Ricci form on $\widetilde{\M}_{4}$ is
\be
\mathfrak{R}=6  J-\dd Q~.
\ee
Combining these terms, the full $10d$ metric is
\be\ba
\dd s^{2}=&\me^{2 \Delta}\lb \dd s^{2}(\text{AdS}_{3})+\frac{\me^{-2 \Delta}}{m^{2}(1-\me^{-2\Delta})}\dd \Delta^{2}+\frac{ 4(1-\me^{-2 \Delta})}{m^{2}}\dd \varphi^{2}\rb +\frac{1}{m^{2}}\left [ \frac{}{} \lb2 \dd \psi+ \rho\rb^{2}+\dd s^{2}(\widetilde{\M}_{4})\right]\cr 
=& \dd s^{2}(\text{AdS}_{5}) +\frac{1}{m^{2}}\left [ \frac{}{} \lb2 \dd \psi+ \rho\rb^{2}+\dd s^{2}(\widetilde{\M}_{4})\right] \,.
\ea\ee
The first term in the brackets with the warp factor included is in fact the metric on AdS$_{5}$ with Ricci-tensor satisfying $R_{\mu\nu}=-4 m^{2}g_{\mu\nu}$.

\subsection{Generality of the $\AdS_5$ Solution with varying $\tau$}
\label{app:AdS5General}

In this section we show briefly how the $\AdS_5$ solution obtained in the last section, are in fact the most general solutions with varying axio-dilaton, and dual $\mathcal{N}=1$ supersymmetry. We perform the general analysis of these solutions directly from an AdS$_5$ ansatz. These AdS$_5$ solutions with varying $\tau$ were originally studied in \cite{Kehagias:1998gn}, though the F-theoretic Calabi--Yau four-fold interpretation given in this paper was not noticed there. A more in depth analysis of the solutions and their holographic duals will be relegated to \cite{toappearAdS}.

\subsubsection{Conditions for Supersymmetry}

We make an ansatz for an $\AdS_5$-solution
\bea
\dd s^{2}&=&\me^{2 \Delta}\lb \dd s^{2}(\text{AdS}_{5})+\dd s^{2}(\M_5^\tau)\rb\nonumber\\
F&=&f (\vol(\text{AdS}_{5})+\vol(\M_5^\tau))\,,
\eea
with $\Delta\in \Omega_{0}(\M_5^\tau,\R)$, $\tau\in \Omega_{0}(\M_5^\tau,\C)$ and $f$ a constant. The metric on AdS$_{5}$ is such that the Ricci tensor satisfies $R_{\mu\nu}=-4 m^{2}g_{\mu\nu}$ and $\dd s^{2}(\M_5^\tau)$ is an arbitrary five-dimensional metric on the internal space $\M_5^\tau$. This allows us to use the supersymmetry equations in $d=5$ directly from \cite{Gauntlett:2005ww} by setting $G=0$ and we obtain
\bea
0&=&D_{m}\xi_{1}+\frac{\ii}{4}(f\me^{-4\Delta}-2m)\gamma_{m}\xi_{1} ~,\label{d1}\\
0&=&\bar{D}_{m}\xi_{2}-\frac{\ii}{4}(f\me^{-4\Delta}+2m)\gamma_{m}\xi_{2}~,\label{d2}\\
0&=&\gamma^{m}\partial_{m}\Delta \xi_{1}-\frac{\ii}{4}(f \me^{-4\Delta}-4m)\xi_{1}~,\label{a1}\\
0&=&\gamma^{m}\partial_{m}\Delta \xi_{2}+\frac{\ii}{4}(f \me^{-4\Delta}+4m)\xi_{2}~,\label{a2}\\
0&=&P_{m}\gamma^{m}\xi_{2}\label{a3}~,\\
0&=&P_{m}^{*}\gamma^{m}\xi_{1}\label{a4}~.
\eea
The first implication from  (\ref{a1}) and (\ref{a2}) is that  $\partial_{m}\Delta=0$. Substituting this back into (\ref{a1}) and (\ref{a2}) it is necessary to set one of $\xi_{1}$ or $\xi_{2}$ to be zero and to set $f=\pm 4 m \me^{4 \Delta}$ depending on which Killing spinor we keep. Without loss of generality we set $\xi_{2}=0$ and set $\Delta=0$ so that $f=4 m$. The SUSY equations reduce to
\bea
0&=&D_{m}\xi_{1}+\frac{\ii m}{2}\gamma_{m}\xi_{1} \label{d12}~,\\
0&=&P_{m}^{*}\gamma^{m}\xi_{1}\label{a32}~.
\eea
Computing the integrability conditions for the Killing spinor equation implies that both the Einstein equation and $P$ equation of motion are satisfied. 

\subsubsection{Torsion Conditions}

To compute the necessary and sufficient conditions for the existence of bosonic supersymmetric solutions in this class we shall use again $G$-structure techniques. For a single Killing spinor in five dimensions the solution must admit an $SU(2)$ structure, which in 5d is specified by the existence of a real one-form, which defines a foliation of the space with a transverse 4d space admitting an $SU(2)$-structure. The latter consists of a real two-form of maximal rank and a holomorphic two-from. First define the spinor bilinears
\be\label{vectors}
A =\bar{\xi}_{1}\xi_{1}~,\quad 
K= \bar{\xi}_{1}\gamma_{(1)}\xi_{1}\,,\quad 
j= \ii \bar{\xi}_{1}\gamma_{(2)}\xi_{1}\, ,\qquad 
\omega= \bar{\xi}_{1}^{c}\gamma_{(2)}\xi_{1} \,  .
\ee
From Fierz identities it follows that the one-form satisfies 
\be
K\lrcorner j=0~,~~K\lrcorner \omega=0~,
\ee
and therefore the space transverse to the one-from admits indeed an $SU(2)$-structure.
From \eqref{d12} and \eqref{a32} follow the torsion conditions
\be\label{Kjeq}
\dd A=0~,\quad 
\dd K= 2m~ j~,\quad 
\dd j=0~, \quad 
D\omega=-3 \ii m *\omega=-3 \ii m ~K\wedge \omega \,,
\ee
and the algebraic conditions
\be\label{Phol}
\tensor{J}{^{\mu}_{\nu}}P^{\nu}=\ii P^{\mu}~,\quad 
i_{K}P=0~. 
\ee
Equation \eqref{Phol} implies that the complex field $P$ is holomorphic with respect to the complex structure. From the torsion conditions it follows that the vector $K$ is not only a  Killing vector but a symmetry of the full solution, corresponding in the dual field theory to the $U(1)$ R-symmetry. We shall use this as customary to be the \emph{Reeb} vector field, or simply Reeb, in the following. To proceed we introduce coordinates adapted to the Killing direction. Define
\be
K^{\#}= m \frac{\partial}{\partial \psi}\,,\quad 
K=\frac{1}{m}\lb \dd \psi+\sigma\rb~,
\ee
where the factor of $m$ has been introduced for later convenience and we have used the fact that the Reeb has unit norm. The one-from $\sigma$ is independent of $\psi$. It is also convenient at this point to extract out a dimensionful parameter from the metric on the transverse space, and we extract out a factor of $\frac{1}{m^{2}}$. The metric on $\M_5^\tau$ is
\be
m^{2}\dd s^{2}(\M_5^\tau)=\lb \dd\psi+\sigma \rb^{2}+ \dd s^{2}(\widetilde{\M}_4)\,,
\ee
where $\widetilde{M}_4$ is a K\"ahler surface, as follows from \eqref{Kjeq}, and we shall refer to it as the `transverse' space. Introducing a vielbein on the transverse space yields
\begin{align}
j&=\frac{1}{m^{2}} J=\frac{1}{m^{2}}(e^{23}+e^{45})~,\\
\omega&=\frac{1}{m^{2}}\bar{\Omega}=\frac{1}{m^{2}}(e^{2}-\ii e^{3})\wedge (e^{4}-\ii e^{5})~.
\end{align}
Here $J$ and $\Omega$ are the canonical K\"ahler-form and holomorphic two-form respectively.
Having introduced a coordinate along the Killing direction we may perform a splitting of the exterior derivative as
\be
\dd = \dd \psi \wedge \frac{\partial}{\partial \psi}+\dd _{4}~.
\ee
 With this decomposition \eqref{Kjeq} becomes
\be\label{dsigma}
\dd_{4} \sigma=2 J~,\qquad 
\dd J=0~,
\ee
and \be \label{Omegapsi}
\partial_{\psi}\Omega=3\ii \Omega~,\quad 
\dd_{4}\Omega=\ii(3  \sigma-Q) \wedge \Omega~.
\ee
The first equation is easily solved by extracting out a $\psi$ dependent phase, which we shall implicitly do in the following and by an abuse of notation keep the notation $\Omega$. The second equation determines the Ricci form on $\widetilde{\M}_4$ to be
\be
\mathfrak{R}_{4}=6 J- \dd Q~.\label{Ricciformeq}
\ee
To find a solution we should solve this final equation (\ref{Ricciformeq}). Notice that for constant $\tau$ this reduces to the case of Sasaki--Einstein. This may be written as a fourth order equation for the K\"ahler potential of {$\M_4$}, however there is a nice interpretation of this geometry arising from the base of an elliptically fibered Calabi-Yau four-fold as is explained in section \ref{sec:AdS5}. 
This classifies all possible $\mathcal{N}=1$ AdS$_{5}$ solutions with $G=0$ and varying $\tau$.

\section{Details for the Baryonic Twist Solution }
\label{app:Ypqstuffstuff}

In this appendix we provide some more details on the baryonic twist solution of section \ref{sec:BTSolution}.
The  solution to \eqref{Master2s} that we shall use was found in \cite{Gauntlett:2006ns} for general values of $s$, here we are interested in the $s=2$ case\footnote{In the notation of  \cite{Gauntlett:2006ns} $s=n+1$.
 The authors of \cite{Gauntlett:2006ns}  were mainly interested in the cases $s=3$ and $s=4$.}.
This  was later discussed in \cite{Donos:2008ug}, where it was interpreted as a Type IIB solution of  the form AdS$_3\times T^2\times \M_5^\tau$, with the regularity analysis performed therein. As we show here, the same solution to \eqref{Master2s} yields an F-theory geometry of the form  AdS$_3 \times K3 \times \M_5^\tau$, with the same manifold (in particular the same metric) 
$\M_5^\tau$. After reviewing the derivation of the local form of the solution, for completeness, we shall perform a similar analysis of the regularity and global properties, with some minor changes from \cite{Donos:2008ug}.

The starting point is a cohomegenity one  ansatz for the K\"ahler metric on $\M_4$,
\be
\dd s^{2}(\M_4) =\frac{\dd r^{2}}{U(r)}+U(r) r^{2} \lb\dd \varphi+\frac{1}{2}\cos\theta\dd \chi\rb^{2}+\frac{r^{2}}{4}(\dd \theta^{2} +\sin^{2} \theta \dd \chi^{2})~,
\ee
for which, after changing variable to $x=1/r^2$,  one can find the explicit solution to \eqref{Master} as 
\bea
U(x)&= &1-a (x-1)^{2}
\eea
depending on one integration constant $a$. This is reviewed below. 

\subsection{Polynomial Solution to the ``Master Equation''}
\label{comoans}

Let us denote $C=\frac{1}{2}\cos\theta\dd \chi$  and $KE_2=S^2$  the round two-sphere. 
The associated K\"ahler form reads
\bea
J=-\lb r \dd r\wedge (\dd \varphi+C)+r^{2} J_{KE}\rb
\eea
where $J_{KE}$ is the K\"ahler form on $KE_{2}$. Then
\bea
\dd J= r \dd r\wedge \dd C-2 r \dd r \wedge J_{KE}
\eea
and $\dd J=0$ implies  
\be
\dd C =2 J_{KE}~.
\ee
The holomorphic 2-form is given by
\bea
\Omega=\me^{2\ii  \varphi} r \lb \frac{1}{\sqrt{U(r)}}\dd r +\ii \sqrt{U(x)} r (\dd \varphi+C)\rb \wedge \Omega_{KE}~,
\eea
where $\Omega_{KE}$ is the holomorphic one-form on  $S^2$ and satisfies
\be
\dd_{2}\Omega_{KE}=2 \ii C\wedge \Omega_{KE}~.
\ee 
Then
\bea
\dd \Omega=\ii \lb 2(1-U(r))-\frac{r}{2}\frac{\dd U(r)}{\dd r}\rb (\dd \varphi +C)\wedge \Omega~,
\eea
and defining 
\be
f(r)\equiv 2(1-U(r))-\frac{r}{2}\frac{\dd U(r)}{\dd r}
\ee
we have that the Ricci form is
\bea
\mathfrak{R}&=&\dd (f (\dd \varphi+C))\nonumber\\
&=&\frac{\dd f}{\dd r} \dd r\wedge (\dd \varphi+C)+2 f J_{KE}~.
\eea
The Ricci scalar is 
\be
R=\frac{4 f}{r^{2}}+\frac{2}{r}\frac{\dd f}{\dd r}~.
\ee
It is convenient to make the change of coordinates $x=1/r^{2}$, under which the metric becomes
\bea
\dd s^{2}&=&\frac{1}{x}\lb \frac{\dd x^{2}}{4 x^{2}U}+U (\dd \varphi+C)^{2}+\dd s^{2}(KE_2) \rb~,
\eea
and we have
\bea
f(x)&=&2(1-U(x))+x \frac{\dd U(x)}{\dd x}~,\label{deffx}\\
R&=& 4 f x-4 x^{2}\frac{\dd f}{\dd x}~,\label{Rfx}\\
\mathfrak{R}&=& \frac{\dd f}{\dd x}\dd x \wedge (\dd \varphi+C)+ 2 f J_{KE}~.
\eea
Using 
\bea
\Box R &=& 2 x^{3} \partial_{x}\lb 2 U \partial_{x}\rb R~,
\eea
 (\ref{Master2s}) reduces to the following ODE
\bea
 \frac{\dd }{\dd x} \lb U \frac{\dd R}{\dd x}+2 f^{2}\rb  &=& 0 ~, 
\eea
which is immediately integrated once to 
\be
 U \frac{\dd R}{\dd x}+2 f^{2}=8c_1~,\label{eqtosolve}
\ee
where $c_1$ is an arbitrary constant. Inserting (\ref{deffx}) and (\ref{Rfx}) we obtain the third order non-linear equation
\be
-8 U(x)+8 x U'(x)+x^{2}(2 U'(x)^{2}-4 U(x)U''(x))-4 x^{3}U(x)U'''(x)=8c_1~,
\ee
where $a$ is an arbitrary constant. This may be rewritten as
\be
\frac{\dd }{\dd x} g(x)-\frac{2}{x}g(x)=8c_1~,\label{firstordergeq}
\ee
where
\be
g(x)\equiv 8 U(x)x +x^{3}(2 U'(x)^{2}-4 U(x)U''(x))~.
\ee
It is simple to find the most general solution to (\ref{firstordergeq})
\be
g(x)= x^{2}8 c_2 -8 c_1 x~,
\ee
where $c_2$ is another integration constant. Thus finally we obtain a non-linear second order equation for $U(x)$ that reads
\be
 8 U(x) +x^{2}(2 U'(x)^{2}-4 U(x)U''(x)) -  x 8 c_2  + 8 c_1  = 0 ~.
\ee
 The most general solution to this equation remains unknown, but we one may find two simple solutions. Setting $c_1=1$, and only for $c_1=1$, we find the solution
 \be
 U(x)= \frac{4 \sqrt{c_{2}}}{\sqrt{3}} \sqrt{x}-8~,
 \ee
 however this solution is problematic for furnishing a smooth metric as it possesses only one root. Instead we find the quadratic solution
 \be
 U(x)= -c_{1} +c_{2} x +\frac{c_{2}^2}{4 (c_{1} -1)} x^{2}~,
 \ee
 which has in principle two arbitrary parameters. By making the redefinition $c_{1}=a-1$ and a coordinate rescaling $x \rightarrow \frac{2 a}{c_{2}}x$ we see that in fact only one of the parameters is physical and that $U(x)$ takes the form
 \be
 U(x)=1-a(x-1)^2~.
 \ee

\subsection{The local F-theory geometry}

The full  geometry can then be reconstructed as follows. The Ricci form of the  K\"ahler base $\widetilde{\M}_{6} = \P^1 \times \M_4$ is 
\be
\mathfrak{R}_6  =-( \dd \rho+\dd Q) =-\dd \rho+\mathfrak{R}_{\Kthreebase}
=\mathfrak{R}_{4}+\mathfrak{R}_{\Kthreebase}~,
\ee
with 
\bea 
\mathfrak{R}_{4}&=& \dd \lb f(x) \lb \dd \varphi+\frac{1}{2}\cos \theta \dd \chi\rb \rb \cr 
f(x)&=&2(1-U(x))+x \frac{\dd U(x)}{\dd x}=2a(1-x)~.
\eea 
Thus we take 
\be
\rho=-f(x) \lb\dd \varphi +\frac{1}{2}\cos \theta \dd \chi\rb~.
\ee
Furthermore the Ricci scalar of  $\widetilde{\M}_{6} = \P^1 \times \M_4$ is 
\be
R_{6}=2|P|^{2}+8 \me^{-4 \Delta}
=R_{\Kthreebase}+R_{4}~.
\ee
From which we may identify
the warp factor as 
\be
\me^{-4 \Delta}= \frac{R_4}{8 } = \frac{1}{8}\lb 4 x f - 4 x^{2} \frac{\dd f}{\dd x}\rb = ax~.
\ee
The 5d part of the metric takes the form 
\bea
 m^{2}\dd s^{2} (\M_5)&=&\frac{1}{4}\lb \dd \psi- 2a (x-1)  D\phi\rb ^{2}\nn\\
&+&a \left[  \frac{\dd x^{2}}{4 x^{2}U}+U D\varphi^{2}+\frac{1}{4}\lb\dd \theta^2+\sin^{2}\theta \dd  \chi^{2})\rb +x \dd s^{2}(\Kthreebase)\right]~,
\label{localpub}
\eea
where $D \varphi= \dd \varphi +\frac{1}{2} \cos \theta \dd \chi$. Using (\ref{ffdef}) we can read off  the expression for the flux 
\be
m \ff= -\frac{1}{2 a x^{2}} \dd \psi \wedge \dd x -2 \dd \vol(\Kthreebase) -\frac{1}{2} \dd \vol(S^{2})~. \label{FYpq}
\ee
The regularity analysis performed in the next subsection shows that the base of this local $U(1)$ fibration is itself not a manifold\footnote{A similar situation occurs with the Sasaki--Einstein $Y^{p,q}$ manifolds. 
The K\"ahler base of $Y^{p,q}$ in the canonical Sasaki--Einstein coordinates is not in general a manifold.}, instead a change of coordinates is useful  to describe the global geometry and results in the solution in the form 
presented in \eqref{localmetric} and \eqref{localF5}.  The profile of the axio-dilaton is determined (implicitly) by the condition that the metric on $\Y_4$ is a Ricci-flat metric, and thus $\Y_4$ is an elliptically fibered 
$K3$,  with base $B_2=\P^1$. Note that 
we will not determine explicitly the metric on $\P^1$ and in particular this cannot be the Einstein metric, but the stringy-cosmic string metric of \cite{Greene:1989ya}, induced by the elliptic fibration. 
In particular, the metric will have singularities at the discriminant loci. We thus continue to distinguish the two two-spheres in the geometry by referring to them as $S^2$ and $\P^1$, 
respectively.

At this stage the background depends on two arbitrary constants $m$, $a$  and we now determine which values of these allow for a globally defined solution.

\subsection{Regularity}

We first consider regularity of the metric and later address the quantisation of the flux (similar discussions have appeared in \cite{Gauntlett:2004yd, Donos:2008ug}). 
The metric on $\M_{7}$ is 
\begin{align}
\dd s^{2}(\M_{7})=&\frac{1}{4}\lb\dd \psi-2 a(1-x)\lb\dd \varphi+\frac{1}{2}\cos\theta \dd \chi\rb\rb^{2} \\
&+a \lb \frac{\dd x^{2}}{4 x^{2}U}+U \lb \dd \varphi+\frac{1}{2}\cos \theta \dd \chi\rb^{2}+\frac{1}{4}(\dd \theta^{2}+\sin^{2}\theta \dd \chi^{2})+x \dd s^{2}(B_2)\rb\nonumber~.
\end{align}
We require that the warp factor does not vanish and therefore the range of the coordinate $x$ cannot include $x=0$. This implies that the 7d geometry  is topologically $\M_5^\tau\times \P^1$, with $\M_5^\tau$ the five-dimensional space defined by $x=$ constant, and therefore we need only analyse the regularity of $\M_5^\tau$, subject to $x$ avoiding $x=0$.
The range of $x$ is fixed to lie between the two roots of $U(x)$
\be
x_{\pm}=1\pm \frac{1}{\sqrt{a}}~.
\ee
Clearly to avoid $x=0$ it is necessary to have $x_->0$, so that  $a>1$, and  it follows that $U(x)$ is positive between the two roots for all values $a>1$.

\subsubsection*{The Base $\newbase$}\label{newbase}

Let us first consider the four-dimensional part of the metric, namely the K\"ahler  base $\M_{4}$. 
The round $S^{2}$ appearing in $\M_{4}$ has  coordinates $\theta$ and $\chi$ with the canonical coordinate periodicities $\theta\in [0,\pi]$ and $\chi \in [0,2\pi]$. 
Near to the zeroes of $U$ at $x=x_{\pm}$, the degenerating part of the metric is 
\be
\left.\left.\frac{1}{x_{\pm}^{2}U'(x_{\pm})}\right( \dd \rho^{2}+(x_{\pm} U'(x_{\pm}))^{2}\rho^{2}\dd \varphi^{2}\rb~,
\ee
where $\rho= 2 \sqrt{x -x_{\pm}}$, respectively.
For this to be locally $\R^2$ at both end-points it is necessary that $x_{\pm}U'(x_{\pm})$ has the same value at both roots; a trivial calculation shows this is not the case and therefore there is no choice of  periodicity 
 of $\varphi$ that gives a smooth metric.

As in \cite{Gauntlett:2004yd}, the way to proceed is to show that one can still view the five-dimensional space as a circle fibration over a base $\newbase$, albeit one with metric different from the local K\"ahler metric on $\M_4$. 
Changing coordinates from $(\psi,\varphi)$ to $(\alpha,\phi)$ by 
$\alphaalpha = \psi$, $\phi=2\varphi+\psi$, the 5d metric takes the form
\begin{align}
\dd s^{2} (\M_5) =&\frac{w(x)}{4}\lb \dd \alphaalpha \ +g(x) (\dd \phi +\cos \theta \dd \chi)\rb^{2}\nn\\
&+\frac{a}{4} \left[ \frac{\dd x^{2}}{x^{2} U}+\frac{U}{w}(\dd \phi+\cos\theta \dd \chi)^{2}+\dd \theta^{2}+\sin^{2}\theta \dd \chi^{2}\right]~.
\label{U1Z4}
\end{align}
With this change of coordinates we may avoid potential conical singularities at the endpoints of $x$ if $\phi$ has period $2 \pi$ due to the remarkable fact that
\be
\frac{(U'(x_{\pm})x_{\pm})^{2}}{w(x_{\pm})}=1~.
\ee
As in \cite{Gauntlett:2004yd} we may introduce a new angular coordinate defined by
\begin{align}
\cos\zeta=-\frac{1+a(x-1)}{\sqrt{w}}~,~~~~\sin\zeta=\frac{\sqrt{a U}}{\sqrt{w}}~,
\end{align}
with $\zeta\in [0,\pi]$. However, performing this change of coordinates  is not particularly useful and so we shall keep the $x$ coordinates in the following.

At fixed $x$ between the two roots the base $\newbase$ with metric given in the second line of  (\ref{U1Z4}), is a circle bundle over the round two-sphere, where the $U(1)$ fiber coordinate is $\phi$. The Chern number of this bundle is obtained by computing the integral of the curvature two-form of the connection on the $U(1)$ and gives
\be
\frac{1}{2\pi}\int_{S^{2}}\dd (-\cos\theta \dd \chi)=2~.
\ee
This identifies the three-dimensional space at fixed $x$ to be $S^{3}/\mathbb{Z}_{2}$. Furthermore it follows that the four-dimensional base $\newbase$ has topology $S^{2}\times S^{2}$. For the following it is useful to have an explicit basis for the homology group $H_{2}(\newbase;\Z)=\Z\oplus \Z$. The two natural two-cycles are the two $S^{2}$'s, whose cycles we denote by $C_{1},C_{2}$ in keeping with the notation in \cite{Gauntlett:2004yd}. Since the metric on $\newbase$  is not a product metric the location of the two $S^{2}$'s is not clear, however we may take $C_{1}$ to be the fiber $S^{2}$ at fixed $\theta,\chi$ on the round two-sphere. There are two two-cycles which are visible in the geometry; namely the two $S^{2}$'s at the south and north poles of the fiber $S^{2}$ (at $\zeta=0,\pi$ respectively or equivalently $x=x_{-},x_{+}$), let us call them $S_{1}$ and $S_{2}$. Then the two-cycles $C_{i}$ are
\be
2 C_{1}=S_{1}-S_{2}~,~~~2 C_{2}=S_{1}+S_{2}~,
\ee
with dual cohomology elements
\begin{align}
\omega_{1}&=-\frac{1}{4\pi}\lb \sin \zeta \dd \zeta \wedge (\dd \phi+\cos\theta \dd \chi)- \cos \zeta \sin \theta \dd \theta \wedge \dd \chi\rb~,\nn\\
\omega_{2}&=\frac{1}{4\pi}\sin \theta \dd \theta\wedge \dd \chi~,
\end{align}
satisfying
\be 
\int_{C_{i}}\omega_{j}=\delta_{ij}~.
\ee
As we wish to be precise in comparing the geometry here with that of the $Y^{p,q}$ manifolds, we will perform some additional 
checks on the base $\newbase$.

The Euler characteristic of a four-manifold $\M_4$ may be computed by using the Chern-Gauss-Bonnet theorem
\be
\chi(\M_4)= \frac{1}{32 \pi^{2}}\int_{\M_{4}}\sqrt{g}\lb |W|^{2}-2 \left|\text{Ric}-\frac{R}{4} g\right|^{2} +\frac{1}{6}R^{2}\rb~,
\ee
where norms are computed using the metric, $W$ denotes  the Weyl tensor and $\text{Ric}$  the Ricci tensor. Computing this for our metric we find\footnote{In general, for the product of two Riemann surfaces $\Sigma_1\times \Sigma_2$ of genus $g_1$, $g_2$, respectively, we have $\chi (\Sigma_1\times \Sigma_2 )  = 4 (1-g_1)(1-g_2)$.}
\be
\chi(\newbase)=4~.
\ee
We may compute the signature using the Hirzebruch signature theorem
\be
\sigma (\M_4) =\frac{1}{48 \pi^{2}}\int_{\M_{4}} \sqrt{g}(| W^{+}|^{2}-|W^{-}|^{2})
\ee
and indeed we find\footnote{The signature of the product of two Riemann surfaces $\Sigma_1\times \Sigma_2$ vanishes by Rohlin's theorem, as this is the boundary of its handlebody.} 
\be 
\sigma (\newbase) =0~.
\ee

Let us also check that $\newbase$ is a complex manifold. To do so we compute the exterior derivative of the associated $(0,2)$ two-form. As is well-known   the exterior derivative of the holomorphic $n$-form on a complex manifold of complex dimension $n$ satisfies 
\be
\dd \Omega=\ii \widehat{P}\wedge \Omega~,
\ee
where $\widehat{P}$ is a one form potential for the Ricci-form $\mathfrak{R}$, that is, $\dd \widehat{P}=\mathfrak{R}$. For the metric on $\newbase$ we have  
\be
\Omega=\frac{a}{x}\lb \frac{1}{x \sqrt{U}}\dd x +\ii \frac{\sqrt{U}}{\sqrt{w}}(\dd \phi+\cos \theta \dd \chi)\rb \wedge (\dd \theta +\ii \sin \theta \dd \chi)
\ee
and upon taking the exterior derivative one finds that the manifold is complex, with the one-form Ricci potential given by
\be
\widehat{P}=\lb 1+\frac{1+a(2(2x-1)-a(x-1)(1+x(x-3)))}{w^{3/2}}\rb (\dd \phi+\cos\theta \dd\chi)~.
\ee
The corresponding Ricci-form $\mathfrak{R}=  \dd \widehat{P}$ is proportional to the first Chern class of the tangent bundle of the manifold, and may be integrated over the two two-cycles discussed above. We find
\begin{align}
\frac{1}{2\pi}\int_{S_{1}} \mathfrak{R}=0~,\qquad \frac{1}{2 \pi}\int_{S_{2}}\mathfrak{R}=4~,
\end{align}
which implies 
\be
c_{1}(C_{1})=c_{2}(C_{2})=2~.
\ee
This discussion establishes that in fact $\newbase $ is complex-diffeomorphic to the Hirzebruch surface $\mathbb{F}_0 = S^2\times S^2$, exactly as for the 4d base that appeared in the $Y^{p,q}$ construction in \cite{Gauntlett:2004yd}\footnote{Recall that for a Hirzebruch surface $\F_n$, in a basis of $H_2(\F_n;\Z)$ with intersection matrix 
$$ 
\begin{pmatrix}
- n & 1 \\
1 & 0 
\end{pmatrix}
$$ 
we have the following invariants $\chi(\F_n)=4, \sigma (\F_n) = 0, c_1 (C_1) = -n+2, c_1(C_2)=2$. We have checked that these invariants identify the base manifold 
$B_4$ in   \cite{Gauntlett:2004yd} as $B_4\simeq \F_0$. We have also checked that computing explicitly these for 
the metric on $\F_1$ found in \cite{Martelli:2004wu}, gives correctly $\chi(\F_1)=4, \sigma (\F_1) = 0, c_1 (C_1) = 1, c_1(C_2)=2$, where $C_1=H-E, C_2=E$.}.

\subsubsection*{The Circle Fibration}

We now turn to the circle fibration. The norm of the Killing vector $\partial/\partial\alphaalpha$ is $w(x)/4$ and this is nowhere vanishing in the range between the zeroes of $U(x)$. In order to get a compact five-dimensional manifold we need  the coordinate $\alphaalpha$ to describe an $S^{1}$ bundle over $\newbase$. We then take it to have period
\be
0\leq \alphaalpha \leq 2\pi \ell~,
\ee
where $\ell$ parametrises the arbitrariness of the period of $\alphaalpha$. We may then rescale $\alphaalpha$ by $\ell^{-1}$ which implies that
\be
\ell^{-1}A=\ell^{-1}g (\dd \phi+\cos \theta \dd \chi)
\ee
should be a connection on a $U(1)$ bundle over $\newbase \simeq S^{2}\times S^{2}$. In general such $U(1)$ bundles are completely specified topologically by the gluing on the equators of the two $S^{2}$ cycles $C_{1}$ and $C_{2}$. These are measured by the corresponding Chern numbers in $H_{2}(S^{2};\mathbb{Z})=\Z$ which we label $\podd$ and $\qodd$. These are given by the integrals of the $U(1)$-curvature two-form $\dd A/2\pi$ over the two two-cycles which form the basis of $H_{2}(\newbase;\Z)=\Z\oplus\Z$. We may choose $\ell$ such that $\podd$ and $\qodd$ are coprime, $(\podd,\qodd)=1$. We first check that $\dd A$ is a globally defined two-form. 
At fixed $x$ between the two roots $x_-,x_+$ we see that $\dd A$ is proportional to the ``global angular form'' on the $U(1)$ bundle with fibre $\phi$ and is a globally well-defined one-form, therefore so is $\dd A$ on a fixed $x$ slice of $\newbase$. We must also check how the curvature two-form behaves near to the zeroes of $U$. We find that the only piece that may be troublesome is the term proportional to $\dd x \wedge \dd \alphaalpha$ near the poles, however the true radial coordinate is $r=(x-x_{i})^{1/2}$ and so this term is proportional to the volume form near the fibre poles and thus is well-defined.
Consequently $\dd A$ is a globally well-defined smooth two-form on $\newbase$~.

Let us now calculate the periods
\be
P_{i}\equiv \frac{1}{2\pi}\int_{C_{i}}\dd A~.\label{chernclasscondition}
\ee
The corresponding integrals of $\ell^{-1}\dd A/2\pi$ give the Chern numbers $\podd,\qodd$, so that we have $P_{1}=\ell \podd$ and $P_{2}=\ell \qodd$. These are most easily found by first computing the integrals of $\dd A$ over the two cycles $S_{i}$, namely
\bea
\frac{1}{2\pi}\int_{S_{i}}\dd A= 2 g(x_{i})~,
\eea
from which we find
\be
P_{1}=\frac{2\sqrt{a}}{1-a}~, \qquad P_{2}~=~\frac{2a}{1-a}\qquad \Rightarrow \qquad 
\frac{P_{1}}{P_{2}}=   \frac{1}{\sqrt{a}} = \frac{\podd}{\qodd}
\ee
which implies that 
\be\label{aqp}
a=\frac{\qodd^{2}}{\podd^{2}}\,,\qquad 
\ell =\frac{2\qodd}{\qodd^{2}-\podd^{2}}\,,\qquad x_{\pm}=1\pm \frac{\podd}{\qodd}\,.
\ee
Recall that the regularity of the metric required that $a>1$, which  implies that the integers $\podd, \qodd$ obey
\be
0<\podd < \qodd ~,\label{oddrange}
\ee
for which there is clearly an infinite number of solutions. We have deliberately used a notation as close as possible to \cite{Gauntlett:2004yd}, and found that topologically the base $\newbase$ and the circle fibration are formally identical. More precisely, $\newbase$ here and the base $B_4$ in \cite{Gauntlett:2004yd} are diffeomorphic as complex manifolds, and the  two corresponding circle bundles are characterised by a pair of coprime integers. However, the regularity of the metric here, implies that the Chern numbers
$\podd$, $\qodd$  characterising the fibration obey an inequality that is \emph{opposite} to those obeyed by the integers $p$,  $q$ in the $Y^{p,q}$ Sasaki--Einstein manifolds, which was $p>q>0$!  We denote the corresponding five-dimensional manifolds as $\M_5=\Yodd^{\podd,\qodd}$.

To  summarise,  the geometry of the full Type IIB solution is 
\be
\AdS_3  \times \mathbb{P}^1 \times   \Yodd^{\podd,\qodd} \, , \qquad  \Yodd^{\podd,\qodd} = S^1 \rightarrow \mathbb{F}_{0} \,,
\ee
where $ \Yodd^{\podd,\qodd}$ is a circle fibration over $\mathbb{F}_{0}= S^2  \times S^2$. Of course the K\"ahler metric on  this $\mathbb{F}_{0}$ is not the Einstein, direct-product metric on $S^2  \times S^2$.

 As already mentioned, the same  $\M_5=\Yodd^{\podd,\qodd}$ geometry enters in the solutions with constant $\tau$ presented in \cite{Donos:2008ug}.
 Indeed,   one can show that the global  analysis conducted in \cite{Donos:2008ug} matches that presented above\footnote{Denoting the integers $p,q$ in \cite{Donos:2008ug} as 
 $p_{DGK}, q_{DGK}$, one has the following identifications
$\podd=q_{DGK}$ and $ \qodd=p_{DGK}+q_{DGK}$.}.

\subsection{Toric Geometry of $\Yodd^{\podd,\qodd}$}\label{app:sec:toric}

The fact that the manifolds  $\Yodd^{\podd,\qodd}$ are not Sasaki--Einstein\footnote{In fact, they are neither Einstein nor Sasakian. They are not even contact manifolds \cite{CDJJ}.}  leads to the cones constructed over these, 
$C(\Yodd^{\podd,\qodd})$, not being Calabi--Yau. In fact, the cone over these $\M_5^\tau$ geometries  admit an integrable complex structure, but not a symplectic structure. In particular they are not K\"ahler 
\cite{Gauntlett:2007ts}.  However, both the five-dimensional manifolds $\Yodd^{\podd,\qodd}$ and their cones $C(\Yodd^{\podd,\qodd})$ admit an isometric (and holomorhic) $\T^{3} \simeq U(1)^{3}$ action. Therefore, on the one hand, the methods from 
\emph{toric symplectic geometry} employed in \cite{Martelli:2004wu} cannot be applied here. In particular, we do not have moment maps whose images would determine the convex polyhedral cones underlying several  properties of the toric 
Sasaki--Einstein geometries  \cite{Martelli:2004wu,Martelli:2005tp}.  On the other hand, we still have a $\mathbb{T}^{3}$ action and one may  attempt to understand these geometries from a complex toric geometry viewpoint \cite{CDJJ}. Below we will use the example of  the $\Yodd^{\podd,\qodd}$ solution to illustrate some features of these geometries, that we expect to persist more generally.

A key property of toric Calabi-Yau singularities is that the image of the moment map associated to the $\T^{3}$ action is a convex polyhedral cone. The primitive normals to the facets of this cone can be projected to a plane, where they provide the  toric diagram of the singularity. Equivalently, these normals correspond to the vanishing of different (Killing) vectors in  $\T^{3}$, and thus define co-dimension two loci that are toric divisors in the Calabi--Yau cone, or equivalently calibrated three-manifolds in the Sasaki--Einstein base. These vectors may be extracted from an analysis of the explicit metric, and written in a basis for $\T^{3}$ they yield the toric diagram 
\cite{Cvetic:2005ft,Martelli:2005wy}. Following these references, below we will employ this method for obtaining a toric diagram assocated to the $\Yodd^{\podd,\qodd}$
geometries, albeit one that will not be convex. As we will explain, this diagram is formally in 1--1 correspondence with that of the $Y^{p,q}$ geometries.

The analysis  below will follow closely the discussion in \cite{Martelli:2005wy,Cvetic:2005ft} for the regularity of the five-dimensional $L^{a,b,c}$ toric Sasaki--Einstein metrics. 
This gives an alternative method to performing the regularity analysis of the metric, and in particular to determine the constraint $\podd<\qodd$. The starting point is the local five-dimensional metric (\ref{localpub}) depending on the parameter $a$.
There are  four  codimension two fixed point sets,
where the metric degenerates; these are at $x=x^+,x=x^-, \theta=0$ and $\theta=\pi$.  At each of these points a Killing vector has vanishing norm. 
We may introduce a $2\pi$ periodic coordinate for each of these angular directions at the degeneration loci by normalising the associated Killing vector such that its surface gravity, defined as
 \be
 \kappa = \frac{\partial_{\mu} |V|^{2} \partial^{\mu}|V|^{2}}{4 |V|^{2}}~,
 \ee
  is unity on the degeneration surface. With this choice of periodicity the Killing vector degenerates smoothly onto the degeneration surface. 
  
  The most general Killing vector one can construct is
\be
V= S \partial_{\alphaalpha}+T \partial_{\phi}+W \partial_{\chi}\,,
\ee
where $S,T,W$ are three constants. This has norm
\be
|V|^{2}= \frac{w(x)}{4}( S + g(x)(T +\cos \theta W))^{2}+\frac{a}{4} \lb \frac{U(x)}{w(x)} (T+\cos\theta W)^{2}+\sin^{2}\theta W^{2}\rb~.
\ee
The norm is a sum of three positive terms and therefore for it to vanish each of these terms must independently be zero. We find that the Killing vectors after being suitably normalised are
 \begin{align}
 &&k^{+}&= \frac{1}{x^{+}}\lb \partial_{\alphaalpha}+x^{+} \partial_{\phi}\rb~,&  k^{-}&= \frac{1}{x^{-}}\lb \partial_{\alphaalpha}+x^{-} \partial_{\phi}\rb~,\nn&&\\
&& k^{0}&=\partial_{\phi}-\partial_{\chi}~,  &k^{\pi}&= \partial_{\phi}+\partial_{\chi}~,&&
 \end{align}
 where the superscript denotes the associated degeneration point. Clearly these four Killing vectors are not linearly independent as they span a three-dimensional space and therefore they must satisfy
 \be 
 H k^{+}+J k^{-}+K k^{0}+L k^{\pi}=0~,
 \ee
 for some constant coefficients. As explained in \cite{Cvetic:2005ft} the constant coefficients must be integers. This follows because each of the Killing vectors generate $2\pi$ periodic translations, and therefore the coefficients must be rational. Then by taking integer combinations of translations around these circles one generates a translation which would identify arbitrarily close points. To prevent this from occurring one must take the coefficients to be integers which may be assumed to be coprime. One finds that the integers satisfy
 \begin{align}\label{HJKLrelation}
H+J +K+L=0~,~~~ K =L~,
 \end{align} 
 and
 \begin{align}
 \frac{H}{x^{+}}+\frac{J}{x^{-}}=0~\Rightarrow ~~~ \sqrt{a}= \frac{H-J}{H+J}~.
 \end{align}
 Taking into account the constraints above, 
 we may redefine the integers $H$ and $J$ as 

 \be
 H -J= 2\qodd~,~~~H +J=2 \podd
 \ee
for consistency with the previous section's notation, i.e. (\ref{aqp}).
Then from the constraint that $a>1$ it follows again that $\podd<\qodd$. Moreover, rewriting the linear relation between the vectors in terms of these two integers we find
 \be\label{pqxpm}
 (\podd+\qodd)k^{+}+(\podd-\qodd)k^{-} -\podd k^{0}-\podd k^{\pi}=0~.
 \ee
From this we can read off what in the GLSM language is called the ``charge matrix'' (up to an overall sign) to be
\be
(\podd,\podd,-\podd+\qodd, -\podd -\qodd)~.\label{chargematrix}
\ee
Notice that this is formally identical to the charge matrix of  $Y^{p,q}$ singularities, in particular the sum of all these charges vanishes. 
However, due to the different sign of $\podd-\qodd$, here there are three positive charges
 and one negative for $\Yodd^{\podd,\qodd}$, in contrast to the two positive and two negative for $Y^{p,q}$. In the Calabi--Yau context, these charges can be used to reconstruct the 
 singularity (and all its resolutions) from the K\"ahler quotient $\C^4 /\!\!/U(1)$. Then two charges of the same sign give rise to toric non-orbifold singularities, whereas three charges with the same 
 sign produce a $\C^3/\Z_n$ orbifold. However the cone over $\Yodd^{\podd,\qodd}$ is \emph{not} an orbifold singularity, as it follows from the preceding analysis, this is not in  contradiction because the 
 cone is not K\"ahler.

To extract a toric diagram from the previous analysis\footnote{For a similar analysis in $L^{a,b,c}$ and conventions see \cite{Franco:2005sm,Butti:2005sw}.} we need to write the four vectors above in an effectively acting basis of $\T^{3}$.   Locally the $\mathbb{T}^{3}$ action is generated by the vector fields $\partial_{\alphaalpha}$, $\partial_{\phi}$ and $\partial_{\chi}$, however they do not give an effectively acting basis. 
Let an effectively acting basis of these Killing vectors be the set $\{e_{1},e_2, e_{3}\}$, which are linear combinations of $\partial_{\alphaalpha},\partial_{\phi}, \partial_{\chi}$ and are taken to be suitably normalised such that all have period $2 \pi$. Any $SL_3\Z$ transformation of this basis will also generate the effective $\T^3$ action.
Writing the degenerating Killing vectors as a linear combination of the $e_{i}$ and applying $SL_3\Z$ transformations to bring the first row and column to a canonical form, this becomes
\be
\begin{pmatrix}
k^{+}\\
k^{-}\\
k^{0}\\
k^{\pi}
\end{pmatrix}=\begin{pmatrix}
1 &0&0\\
1& A&B\\
1&C&D\\
1&E &F
\end{pmatrix}
\begin{pmatrix}
e_{1}\\
e_{2}\\
e_{3}
\end{pmatrix}~.
\ee
Consider the degeneration surface defined by $x=x^{+}$ with degenerating Killing vector $k^{+}= e_{1}$. The $\T^{3}$ fibration reduces smoothly to a $\T^{2}$ fibration over this surface which is spanned by $\{ e_{2}, e_{3}\}$. At the intersection of this degeneration surface with the degeneration surfaces located at $\theta=0$ and $\theta=\pi$ we have an additional degenerating Killing vector. Recall from previous arguments that this vector must be $2\pi$ periodic for the degeneration to be smooth. At $\theta=0$ we have the Killing vector $C e_2+ D e_{3}$ degenerating on the surface. For this to be $2\pi$ periodic it is necessary that $C$ and $D$ are relatively prime, $\hbox{gcd}(C,D)=1$. A similar argument follows for the degeneration surface at $\theta=\pi$ and so we also have $\hbox{gcd}(E,F)=1$. 
Notice that there is no condition on $A,B$ as the degeneration surface at $x=x^{-}$ does not intersect with the one at $x=x^{+}$.
As $\hbox{gcd}(C,D)=1$ there exist integer solutions to $ R C + S D=1$ and therefore by an $SL_2\Z\subset SL_3\Z$ transformation we may set $C=1, D=0$. We find 
\be
\begin{pmatrix}
k^{+}\\
k^{-}\\
k^{0}\\
k^{\pi}
\end{pmatrix}=\begin{pmatrix}
1 &0&0\\
1& A&B\\
1&1&0\\
1&E &F
\end{pmatrix}
\begin{pmatrix}
e_{1}\\
e_{2}\\
e_{3}
\end{pmatrix}~.
\ee
Next, using the linear relation between the four Killing vectors we find
\be
(\podd-\qodd) B - \podd F=0~,~~~~~(\podd -\qodd) A -\podd -\podd E =0~.
\ee
These can be solved by 
\be
B= \podd\,,\qquad G= \podd -\qodd\,,\qquad A=0\,,\qquad E=-1\,,
\ee
and we obtain
\be
\begin{pmatrix}
k^{+}\\
k^{-}\\
k^{0}\\
k^{\pi}
\end{pmatrix}=\begin{pmatrix}
1 &0&0\\
1& 0&\podd\\
1&1&0\\
1&-1 &\podd-\qodd
\end{pmatrix}
\begin{pmatrix}
e_{1}\\
e_{2}\\
e_{3}
\end{pmatrix}~.
\ee
We may now introduce three  $2\pi$ periodic coordinates $\psi_{i}$ for each of the three $e_{i}$, the change of coordinates from the original set is
\begin{align}
\alphaalpha= \frac{1}{x^{+}}(\psi_{1}-\psi_{2}) +\lb \frac{\mu -\nu}{x^{+}} -\frac{\nu}{x^{-}}\rb \psi_{3}~,~~~ \phi=\psi_{1}~,~~~ \chi= -\psi_{2}+\mu \psi_{3}~,
\end{align}
where the integers $\mu$ and $\nu$  satisfy $ \mu (\podd-\qodd) -\podd \nu=1$ and are guaranteed to exist by the fact $\hbox{gcd}(\podd,\podd-\qodd)=1$. With these coordinates the $\T^3$ action acts effectively.


\begin{figure}
\begin{center}
\begin{minipage}[t]{0.45\linewidth}
  \centering
  \begin{tikzpicture}
    \coordinate (Origin)   at (0,0);

    \clip (-2.9,-1.5) rectangle (4.1cm,4.2cm);

  \foreach \x in {-2,-1,...,3,4}{
      \foreach \y in {4}{
        \node[draw,circle,inner sep=1.5pt,fill][blue] at (\x,  \y) {};
          }
    }

    \foreach \x in {-1,...,3,4}{
      \foreach \y in {-1}{
        \node[draw,circle,inner sep=1.5pt,fill][blue] at (\x,  \y) {};
           }
    }
  \foreach \x in {-2,-1,2,3,4}{
      \foreach \y in {0}{
        \node[draw,circle,inner sep=1.5pt,fill][blue] at (\x,  \y) {};
                 }
    }
  \foreach \x in {-2,-1,0,2,3,4}{
      \foreach \y in {1}{
        \node[draw,circle,inner sep=1.5pt,fill][blue] at (\x,  \y) {};
       }
    }

  \foreach \x in {-2,-1,0,1,3,4}{
      \foreach \y in {2}{ 
        \node[draw,circle,inner sep=1.5pt,fill][blue] at (\x,  \y) {};
                }
    }

  \foreach \x in {-2,-1,0,1,2,4}{
      \foreach \y in {3}{ 
        \node[draw,circle,inner sep=1.5pt,fill][blue] at (\x,  \y) {};
               }
    }

\node[draw,circle,inner sep=1.5pt,fill][black] at (-2,-1){};
\node[draw,circle,inner sep=1.5pt,fill][black] at (3,3){};
\node[draw,circle,inner sep=1.5pt,fill][black] at (0,0){};
\node[draw,circle,inner sep=1.5pt,fill][black] at (1,0){};

\node[draw,circle,inner sep=1.5pt,fill][blue] at (1,1){};
\node[draw,circle,inner sep=1.5pt,fill][blue] at (2,2){};

\draw[thick, black] (0,0)-- (1,0) node [below] {\small{(1,0)}}-- (3,3) node [above] {\small{$(\podd,\podd)$}}-- (-2,-1) node [below] {\small{($\podd$-$\qodd$ -1,$\podd$-$\qodd)$}}--(0,0) node[below] {\small{(0,0)}};
 \draw[thick,black]   ; 
  \end{tikzpicture}
  \caption{Toric diagram for $\Yodd^{\podd,\qodd}$. Notice that this is not convex, however the external lines do not intersect. The figure  is representative of the choice $\podd=3$, $\qodd=4$.}
  \label{figure:toricdata1}

\end{minipage}
\begin{minipage}{0.5\linewidth}
\end{minipage}
\begin{minipage}{0.5\linewidth}
\end{minipage}
\begin{minipage}{0.5\linewidth}
\end{minipage}
\begin{minipage}{0.5\linewidth}
\end{minipage}
\begin{minipage}{0.5\linewidth}
\end{minipage}
\begin{minipage}{0.5\linewidth}
\end{minipage}
\begin{minipage}{0.5\linewidth}
\end{minipage}
\begin{minipage}{0.8\linewidth}
\end{minipage}
\begin{minipage}{0.5\linewidth}
\end{minipage}
\begin{minipage}{0.5\linewidth}
\end{minipage}
\begin{minipage}{0.8\linewidth}
\end{minipage}
\begin{minipage}[t]{0.45\linewidth}
  \centering
  \begin{tikzpicture}
    \coordinate (Origin)   at (0,0);

   \clip (-2.9,-1.5) rectangle (4.1cm, 4.2cm);

   \foreach \x in {-1,0,...,3}{
      \foreach \y in {4}{
        \node[draw,circle,inner sep=1.5pt,fill][blue] at (\x,  \y) {};
                }
}
    \foreach \x in {-1,0,...,3,4}{
      \foreach \y in {-1}{
        \node[draw,circle,inner sep=1.5pt,fill][blue] at (\x,  \y) {};
                 }
    }
  \foreach \x in {-1,0,2,3,4}{
      \foreach \y in {0}{ 
        \node[draw,circle,inner sep=1.5pt,fill][blue] at (\x,  \y) {};
                }
    }
  \foreach \x in {-1,0,2,3,4}{
      \foreach \y in {1}{
        \node[draw,circle,inner sep=1.5pt,fill][blue] at (\x,  \y) {};
                 }
    }

  \foreach \x in {-1,0,1,3,4}{
      \foreach \y in {2}{
        \node[draw,circle,inner sep=1.5pt,fill][blue] at (\x,  \y) {};
                 }
    }

  \foreach \x in {-1,0,1,2,4}{
      \foreach \y in {3}{
        \node[draw,circle,inner sep=1.5pt,fill][blue] at (\x,  \y) {};
                 }
    }
  \foreach \y in {-1,0,1,2,3,4}{
      \foreach \x in {-2}{
        \node[draw,circle,inner sep=1.5pt,fill][blue] at (\x,  \y) {};
                 }
    }

\node[draw,circle,inner sep=1.5pt,fill][black] at (0,1){};
\node[draw,circle,inner sep=1.5pt,fill][black] at (4,4){};
\node[draw,circle,inner sep=1.5pt,fill][black] at (0,0){};
\node[draw,circle,inner sep=1.5pt,fill][black] at (1,0){};

\node[draw,circle,inner sep=1.5pt,fill][blue] at (1,1){};
\node[draw,circle,inner sep=1.5pt,fill][blue] at (2,2){};
\node[draw,circle,inner sep=1.5pt,fill][blue] at (3,3){};

\draw[thick, black] (0,0)-- (1,0) node [below] {\small{(1,0)}}-- (4,4) node [left] {\small{$(p,p)$}}-- (0,1) node [left,above] {\small{($p$-$q$ -1,$p$-$q$)~~~~~~~~~~~}}--(0,0) node[below] {\small{(0,0)}};
 \draw[thick,black]   ; 
  \end{tikzpicture}
  \caption{Toric diagram for $Y^{p,q}$. 
  The figure is  representative of the choice $p=4$ and $q=3$. }
  \label{figure:toricdata}
  \end{minipage}
  \end{center}
\end{figure}


Finally, we may read off the toric data from the matrix $\Lambda$: the four vertices are given by the rows of $\Lambda$, namely
\be
\begin{pmatrix}
1\\
0\\
0
\end{pmatrix}~,~~~
\begin{pmatrix}
1\\
0\\
\podd
\end{pmatrix}~,~~~
\begin{pmatrix}
1\\
1\\
0
\end{pmatrix}~,~~~
\begin{pmatrix}
1\\
-1\\
\podd-\qodd
\end{pmatrix}~.
\ee
By an additional $SL_3\Z$ transformation the vectors take the form 
\be
\begin{pmatrix}
1\\
0\\
0
\end{pmatrix}~,~~~
\begin{pmatrix}
1\\
\podd\\
\podd
\end{pmatrix}~,~~~
\begin{pmatrix}
1\\
1\\
0
\end{pmatrix}~,~~~
\begin{pmatrix}
1\\
\podd -\qodd -1\\
\podd-\qodd
\end{pmatrix}~,
\ee
which agree formally  with the ones for the $Y^{p,q}$ Calabi--Yau singularity \cite{Benvenuti:2004dy}. Notice however that because $\qodd >\podd$,  this no longer defines a convex polytope.
For comparison, we contrast two examples of toric diagrams in the two cases in the figures \ref{figure:toricdata1} and \ref{figure:toricdata}.


\section{Summary of the 4d $Y^{p,q}$ Field Theories }
\label{ypqappendix}

\begin{figure}
\begin{center}
  \includegraphics[height=5cm]{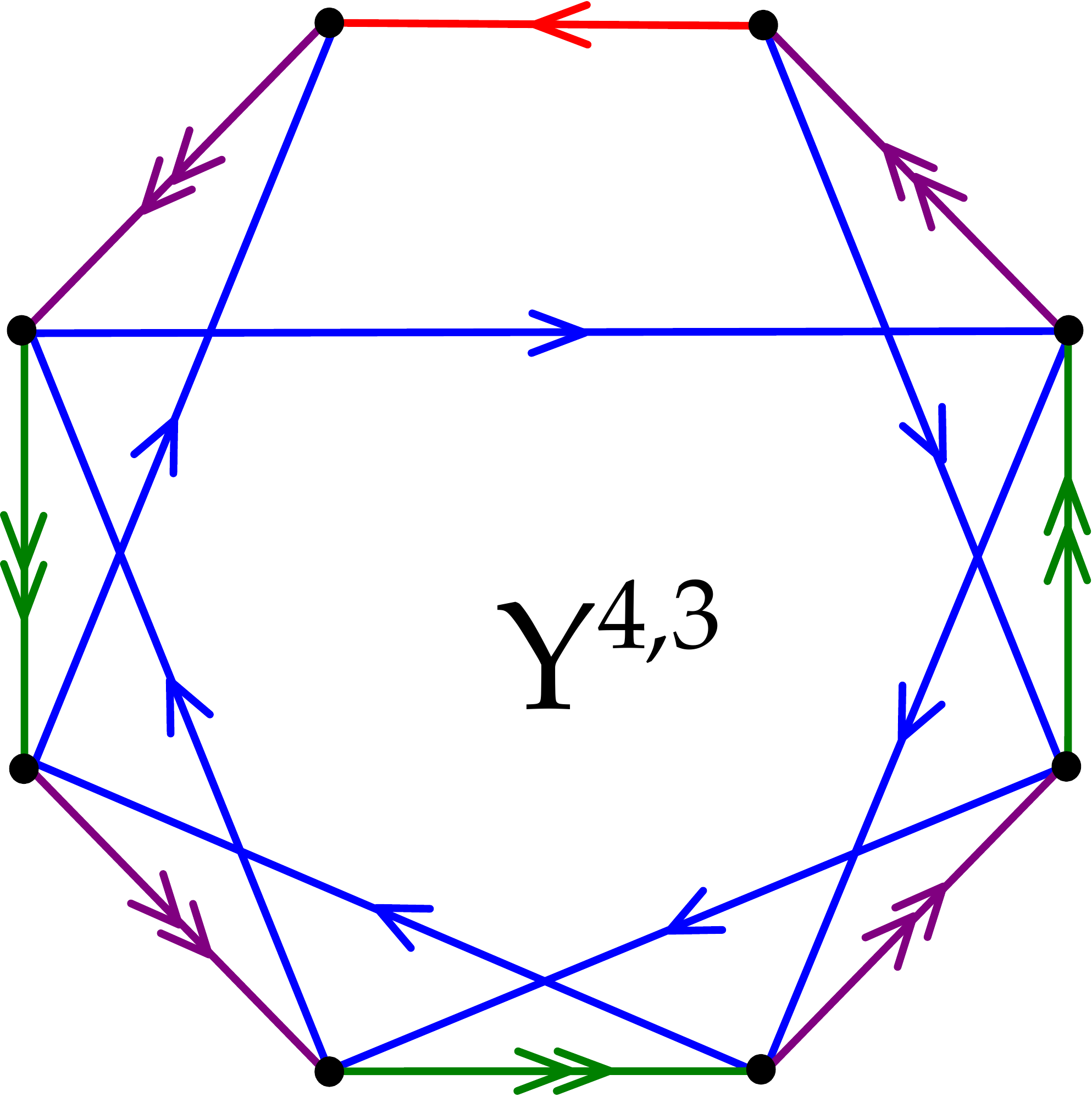}
 \caption{\label{fig:Ypqquiver} The $Y^{4,3}$ quiver diagram. 
 The quiver diagram for $Y^{4,3}$. The fields have been colour-coded as follows: the $Y$ fields are shown in \textcolor{blue}{blue}, $Z$ fields in \textcolor{red}{red}, $U_{i}$ fields are  \textcolor{purple}{purple} and $V_{i}$ fields {\color [rgb]{.1,.8,.1} {green}}.
 }
\end{center}
\end{figure}

In the paper we use  the    ${\cal N}=1$ four-dimensional  theories   \cite{Benvenuti:2004dy}  compactified on a Riemann surface  as an example of AdS$_3/$CFT$_2$ duality for which we can obtain an F-theory embedding. 
 For clarify of notation, 
below  we present  a summary of useful facts about the 4d field theories and the related AdS$_5/$CFT$_4$ duality, before compactifying on a Riemann surface.
These  are an infinite family of quiver gauge theories specified by the gauge group $G=SU(N)^{2p}$ and by a set of $4p+2q$ chiral multiplets, transforming in bi-fundamental representations of pairs of $SU(N)$ factors. Here $p$ and $q$ are two positive integers satisfying $p>q$. The precise representation can be conveniently  encoded in a quiver diagram, as in Figure \ref{fig:Ypqquiver}.

The bi-fundamentals are  grouped in four types of fields, denoted $Y,Z,U_\alpha,V_\alpha$, each with different global charges. There is a superpotential $W$, whose detailed form we will not need here. The theories have global symmetries $SU(2)_1\times U(1)_{2} \times U(1)_R\times U(1)_B$, but for many purposes it is convenient to consider the charges with respect to the Cartan generator $U(1)_{1} \subset SU(2)_1$. There are two categories of global symmetries, referred to as flavour and baryonic symmetries, respectively.  For the convenience of the reader, the charges of the fields, together with their multiplicities,  are summarised in Table \ref{Ypqcharges}.
\begin{table}
\begin{center}
\label{Ypqcharges}
\begin{tabular}{|c|c|c|c|c|c|}
\hline
Fields&Multiplicity&$U(1)_{1}$&$U(1)_{2}$& $U(1)_{B}$&$U(1)_{R}$\\
\hline
$Y$&$p+q$&$0$&$-1$&$p-q$&$R_{Y}$\\
$Z$&$p-q$&0&1&$p+q$& $R_{Z}$\\
$U_{1}$&$p$&$1$&$0$&$-p$&$R_{U}$\\
$U_{2}$&$p$&$-1$&$0$&$-p$&$R_{U}$\\
$V_{1}$&$q$&1&1&$q$&$R_{V}$\\
$V_{2}$&$q$&$-1$&1&$q$&$R_{V}$\\
$\lambda$&$2p$&$0$&$0$&$0$&$1$\\ 
\hline
\end{tabular}
 \caption{\label{tab:YpqCharges}The charges of the various fields in the 4d $Y^{p,q}$ theories.}
\end{center}
\end{table}
The $R_X$ in the last column  denote the $R$-charges of the (scalar) fields ($X=Y,Z,U_\alpha,V_\alpha$) under the true $R$-symmetry of the SCFTs at their IR fixed point. 
The $R$-charges of the fermions in the chiral multiplets are given by $R_X-1$ and  we have included the $2p$ gauginos $\lambda$ for reference.
The superconformal $R$-symmetry in the IR can mix with the abelian global symmetries of the theory and can be determined uniquely by employing $a$-maximization 
\cite{Intriligator:2003jj}. In fact, it turns out that the baryonic symmetry $U(1)_B$ does not participate to this mixing, and the result of the extremization 
 \cite{Bertolini:2004xf,Benvenuti:2004dy} provides the R-charges of the theory, which read
\begin{align}
R_Y &=  \frac{3 q^2 +2 p q -4 p^2+(2 p-q) \bbcw}{3 q^2}~,\qquad  R_U ~= \frac{2p (2 p- \bbcw)}{3 q^2}~,\nn\\
R_Z &= \frac{3 q^2 -2 p q -4 p^2+(2 p+q) \bbcw}{3 q^2}~,\qquad R_V ~= \frac{3 q -2 p +\bbcw}{3 q}~,
\label{R2D2}
\end{align}
where we have defined $\bbcw= \sqrt{4 p^2 -3 q^2}$. For generic values of the parameters $p$ and $q$ these numbers are famously irrational, which corresponds to the fact that the $R$-symmetry is not compact. 

There are two related ways to think about the gravity duals of these field theories. On one hand, one can show that there is a branch of the  mesonic vacuum moduli space of these theories that contains a copy of a Calabi--Yau three-fold singularity, denoted $C(Y^{p,q})$ \cite{Martelli:2004wu}. It follows that the field theories may be thought of as arising from $N$ D3 branes transverse to this conical singularity. On the other hand, in the large $N$ limit, the Type IIB geometry near the branes (``near horizon'') is AdS$_5\times Y^{p,q}$, where $Y^{p,q}$ are the  five-dimensional Sasaki--Einstein manifolds \cite{Gauntlett:2004yd}.  The integers $p,q$ characterising these manifolds can be consistently identified with the $p,q$ characterising the field theories.

There are several checks that can be performed on this conjectured duality. The most basic check consists in matching the central charges on the two sides.  Moreover, one can also compare successfully the charges under the global symmetries on the two sides. To this end, it is useful to consider certain baryonic operators ${\cal B}_X$,  which correspond to particles moving in AdS$_5$, arising from D3-branes wrapping supersymmetric three-manifolds in the Sasaki--Einstein manifold  \cite{Berenstein:2002ke}.  This establishes a map between baryonic operators ${\cal B}_X$ and supersymmetric three-manifolds $\Sigma$, in particular the $R$-charges of the former may be computed in terms of volumes of the latter by using the formula  \cite{Berenstein:2002ke}
\bea
R [X]  & = & \frac{N \pi}{3} \frac{\mathrm{vol} (\Sigma)}{\mathrm{vol} (Y^{p,q})}~.
\label{offer}
\eea

The baryonic charges of the fields can also be inferred from the string duals of the baryonic operators. Recall that the RR four-form potential $C_4$ gives rise upon Kaluza-Klein reduction to five dimensions to the background gauge field $A_B$ associated to the baryonic symmetry through the ansatz $C_4={\cal H}\wedge A_B$, 
where ${\cal H}$ is a (harmonic) representative of
 $H^{3}(Y^{p,q};\Z)\simeq \Z$. From this it follows that the
 baryonic charges of the baryonic operators can be computed in the supergravity solution by integrating  ${\cal H}$
  on the various sub-manifolds $\Sigma$ \cite{Franco:2005sm}, so that 
 \bea
Q_B (X) = \int_{\Sigma_X} {\cal H}~.
 \eea  
Indeed this was explicitly done in \cite{Herzog:2004tr}, obtaining agreement 
with the field theory $U(1)_B$ charges written in Table  \ref{Ypqcharges}. Finally, the multiplicities of the fields written in the second column of  \ref{Ypqcharges} can be reproduced by calculating 
$\pi_1 (\Sigma_X)$ in the geometry $\Sigma$ \cite{Franco:2005sm}. 

Notice that all the comparisons that we have recalled here can be done without using techniques of toric geometry. Indeed, the computations that we presented in section \ref{sec:NewSol} mimic these results, in the context of the 
$\Yodd^{\podd,\qodd}$ manifolds, that as we have explained are not toric, in the sense of symplectic toric geometry.  


\bibliographystyle{JHEP}

\providecommand{\href}[2]{#2}\begingroup\raggedright\endgroup


\end{document}